\documentclass[12pt]{amsart}
\usepackage{graphicx}
\usepackage{amsfonts}
\usepackage{amsmath}
\usepackage{amssymb,latexsym}
\usepackage{amsthm}
\usepackage{mathrsfs}
\usepackage{bm}
\usepackage{float}
\usepackage{enumerate}
\usepackage{caption}
\usepackage{subcaption}
\DeclareCaptionSubType*{figure}

\usepackage{color}

\usepackage[a4paper,left=1in,right=1in]{geometry}
\begin{document}
\title[Numerical simulations of nonlinear modes in mica]{Numerical
  simulations of nonlinear modes in mica: past, present and future}

\author{J. Bajars}
\address{Maxwell Institute and School of Mathematics\\
  University of Edinburgh\\ James Clerk Maxwell Building, The King's
  Buildings, Mayfield Road, Edinburgh EH9 3JZ, UK}
 \email{J.Bajars@ed.ac.uk} 
\author{J. C. Eilbeck}
\address{Maxwell Institute and
  Department of Mathematics, Heriot-Watt University, Riccarton,
  Edinburgh EH14 4AS, UK}
\email{J.C.Eilbeck@hw.ac.uk} 
\author{B~Leimkuhler}
\address{Maxwell Institute and School of Mathematics, The
  University of Edinburgh, James Clerk Maxwell Building, The King's
  Buildings, Mayfield Road, Edinburgh EH9 3JZ, UK}
  \email{b.leimkuhler@ed.ac.uk}

\newcommand{\D}{\textrm{d}}
\newcommand{\I}{\textrm{i}}
\newcommand{\E}{\textrm{e}}
\newcommand{\cm}{\textrm{cm}}
\newcommand{\nm}{\textrm{nm}}
\newcommand{\seg}{\textrm{s}}
\newcommand{\DB}{{\textrm{DB}}}
\newcommand{\ILM}{{\textrm{ILM}}}
\date{}

\begin{abstract} 
  We review research on the role of nonlinear coherent phenomena (e.g
  breathers and kinks) in the formation linear decorations in mica
  crystal.  The work is based on a new model for the motion of the
  mica hexagonal K layer, which allows displacement of the atoms from
  the unit cell.  With a simple piece-wise polynomial inter-particle
  potential, we verify the existence of localized long-lived breathers
  in an idealized lattice at 0$^\circ$K.  Moreover, our model allows
  us to observe long-lived localized kinks.  We study the interactions
  of such localized modes along a lattice direction, and in addition
  demonstrate fully two dimensional scattering of such pulses for the
  first time.  For large interatomic forces we observe a spreading
  horseshoe-shaped wave, a type of shock wave but with a breather
  profile.
\end{abstract}

\maketitle

\section{Introduction}
The role of nonlinear localized coherent phenomena for the formation
of anomalous structures in crystalline materials remains unclear,
despite a number of efforts over the last two decades.  In this
article, we begin with a short survey of the state of the art in
research on this topic.  This serves to introduce a number of relevant
issues in relation to atomistic models, including the work of
Mar\'{\i}n et al. on breathers \cite{mer98} in the K layer in mica.

\begin{figure}
\centering
\includegraphics[scale=0.75]{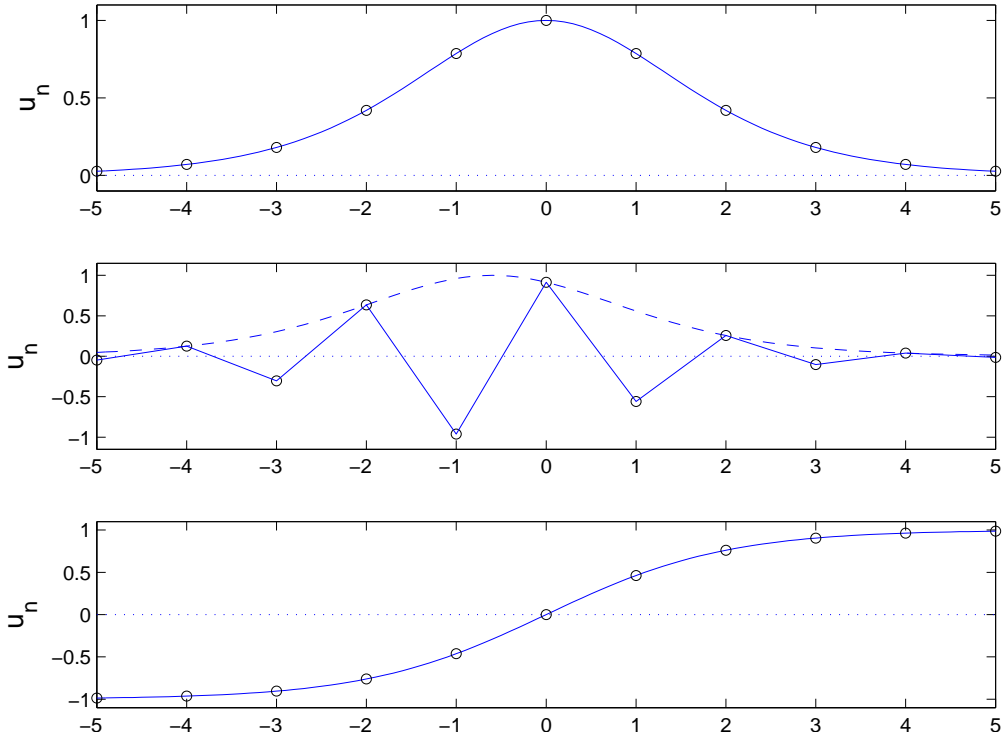}
\caption{Solitons, breathers and kinks, in 1D discrete lattices.}
\label{fig:Soliton}
\end{figure}

From a heavily simplified perspective, there are three types of
localized excitations in dispersive nonlinear systems. These are (in
1D) solitons, kinks, and breathers, as illustrated in
Fig. \ref{fig:Soliton}.

\begin{itemize}
\item {\bf Soliton}. Strongly localized package (lump) of energy, can
  move large distances with no distortion, very stable even under
  collisions or perturbations.
\item {\bf Kink}. Similar to a soliton, but with different boundary
  conditions as $x \rightarrow \pm \infty$. May be even more stable
  due to topological conservation laws.
\item {\bf Breather}. A more complicated form of nonlinear wave. It
  looks like a soliton modulated by an internal carrier wave. Not
  common in continuous systems but more frequently seen in discrete systems.
\end{itemize}

Note that breathers are also known as {\it Intrinsic Localized Modes}
(ILMs). M. Russell's {\it quodon} discussed in this article is now
believed to be a breather.

Breathers in discrete systems were first studied by Ovchinnikov
\cite{ov70}, but this pioneering paper was overlooked for many
years. Ovchinnikov also considered the mobility of such objects.

Independently in the early '80s, breathers were studied in the
Discrete Nonlinear Schr\"odinger (DNLS) equation.
\[
{\rm i}\frac{\D A_j}{\D t} + (A_{j-1}-2A_j+A_{j+1}) + \gamma |A_j|^2A_j=0 \;,
\]
where $A_j(t)$ is the {\em complex} oscillator amplitude at the $j$th
lattice site.  An early application of the DNLS equation was as a
simple model for so-called Davydov solitons on a protein molecule.
Arguably the first paper on the single breather in the system was due
to Scott and MacNeil \cite{sm83} (although such states were still
called solitons in the early papers).
\begin{figure}
\begin{center}
\includegraphics[scale=0.75]{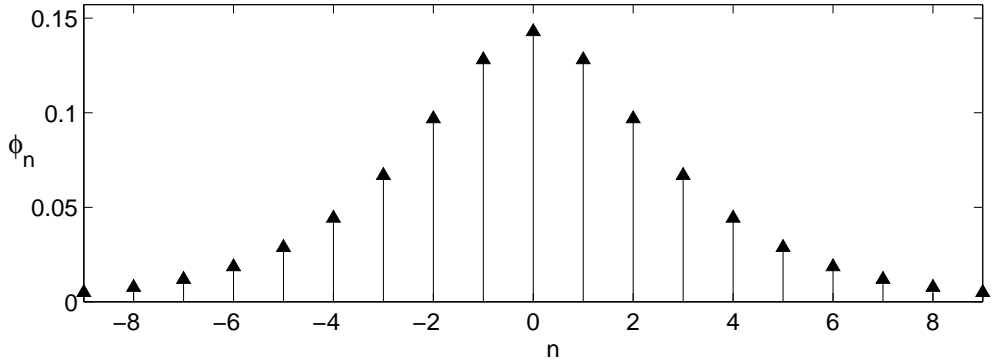}
\caption{Breather, DNLS equation.}\label{fig:DNLS}
\end{center}
\end{figure}

\begin{figure}
\centering
\includegraphics[width=0.75\textwidth]{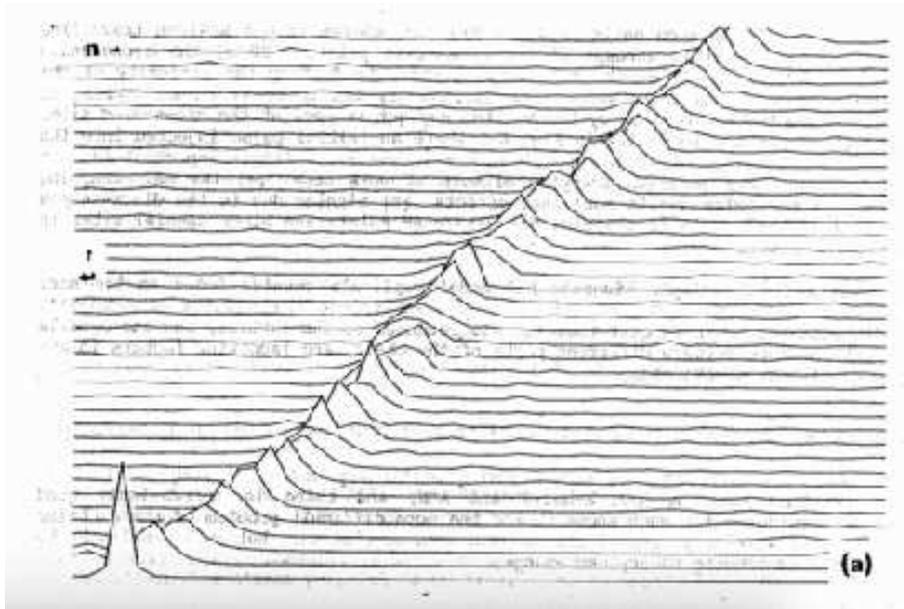}
\caption{Moving breather in a simple model system (DNLS). Here {\bf
    energy} is plotted rather than complex amplitude.}
\label{fig:mov_bre}
\end{figure}

Fig. \ref{fig:DNLS} shows a stationary breather on the DNLS
lattice. The time dependence in the DNLS model for {\em stationary}
solutions is extremely simple: $A_n(t)=\phi_n\exp(i\omega t)$. The
amplitude goes to zero exponentially as $|n|\rightarrow \infty$.
Eilbeck, Lomdahl and Scott took the first tentative step towards a 2D
theory of breathers in the DNLS model by considering two {\em coupled}
chains in a study of a crystal called Acetanilide (ACN) which modelled
protein structure \cite{els84}. This work found examples of staggered
breathers (i.e.~breather energies spread over two or more sites) and
the use of path-following from what is now called the anti-continuum
limit.  They also considered more complex non-chain geometries,
finding many exact solutions on small graphs \cite{els85}.  In the
course of work in this area, a relatively long-lived example of a
moving breathers in a 1D discrete systems was found \cite{ei86}, see
Fig.\ \ref{fig:mov_bre}.

Many workers found other examples of discrete breathers in various
systems (see \cite{fw98,fl12} for reviews).  In 1994, MacKay and Aubry
found a general mathematical proof for the existence of {\em
  stationary} breathers in a quite general class of systems
\cite{ma94}.  For {\em mobile} breathers in the DNLS equation,
Feddersen found a very accurate numerical description of travelling
wave solutions in 1991 \cite{fe91,defw93}.

The study of kinks in continuum and discrete models is a large subject
in its own right.  A good early paper by Peyrard and Kruskal
\cite{pk84}, on kinks in a highly discrete sine-Gordon model, is a
nice introduction.  Some results on the numerical studies of solitons
in discrete systems will be found in \cite{defw93}.

\subsection{Solitons, kinks and breathers in 2D}
\label{sec:Solitons}
\begin{figure}
\centering
\includegraphics[scale = 0.45]{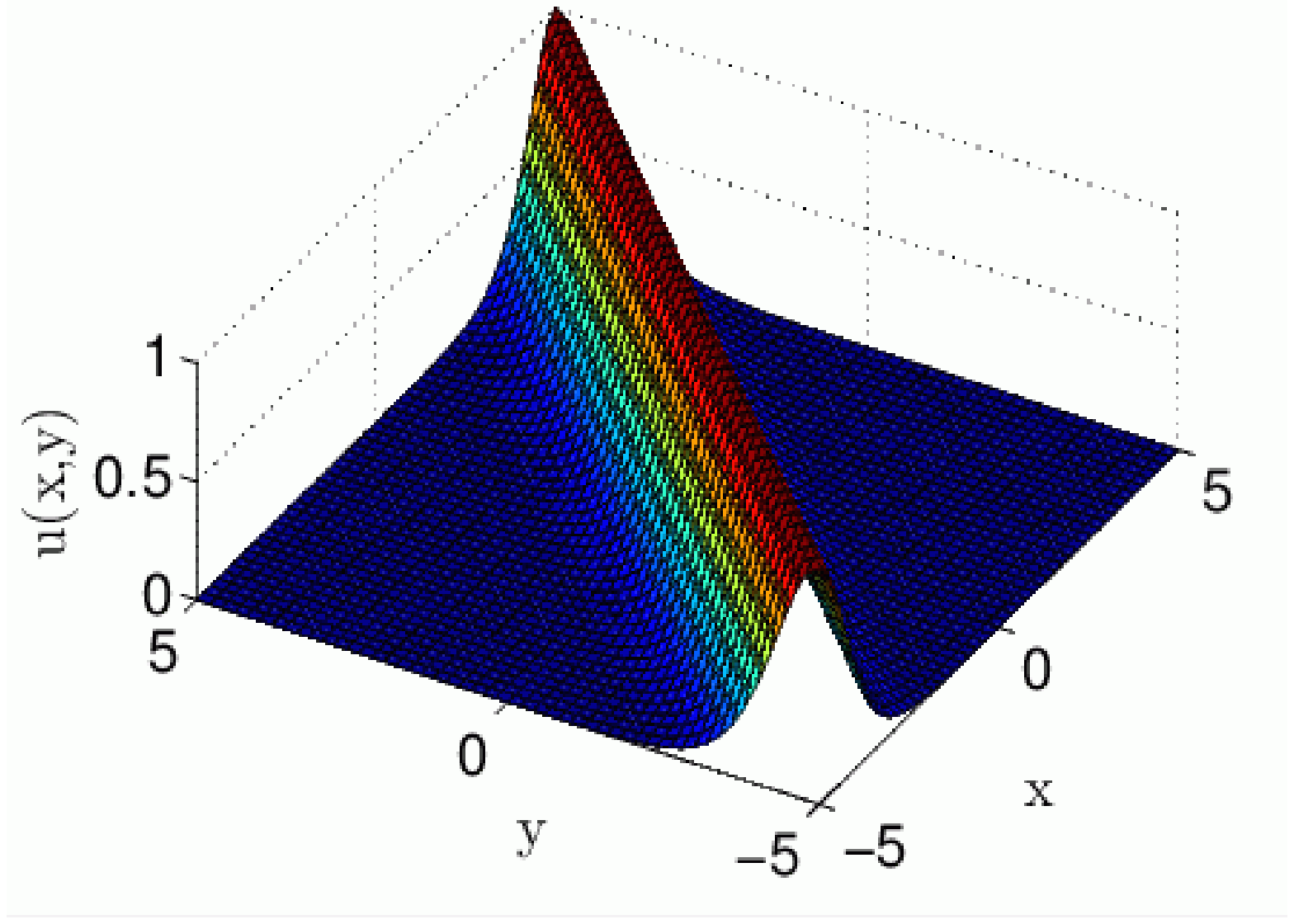}
\includegraphics[scale=0.45]{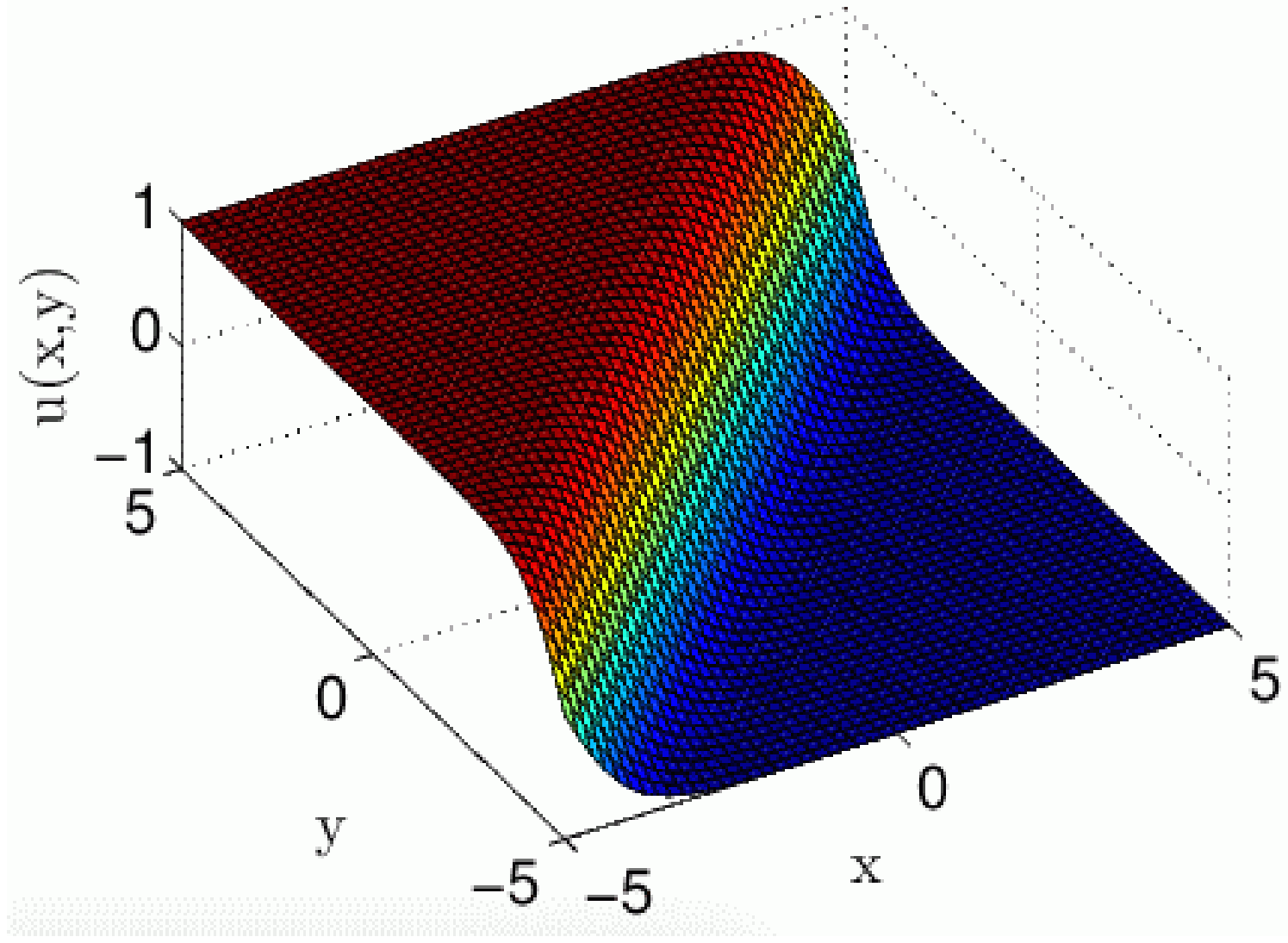}
\caption{Soliton on the left, kink on the right.}
\label{fig:SolKink2D}
\end{figure}

It is not difficult to generalise soliton or kink equations to give
models which have {\it plane wave} solutions in 2D, see Fig.\
\ref{fig:SolKink2D}.
However there is a problem - the wave front has a finite energy
density so the infinite wave front has infinite energy. What we
need is a {\em localized} pulse with finite energy.

\begin{figure}
\centering
\includegraphics[scale=0.45]{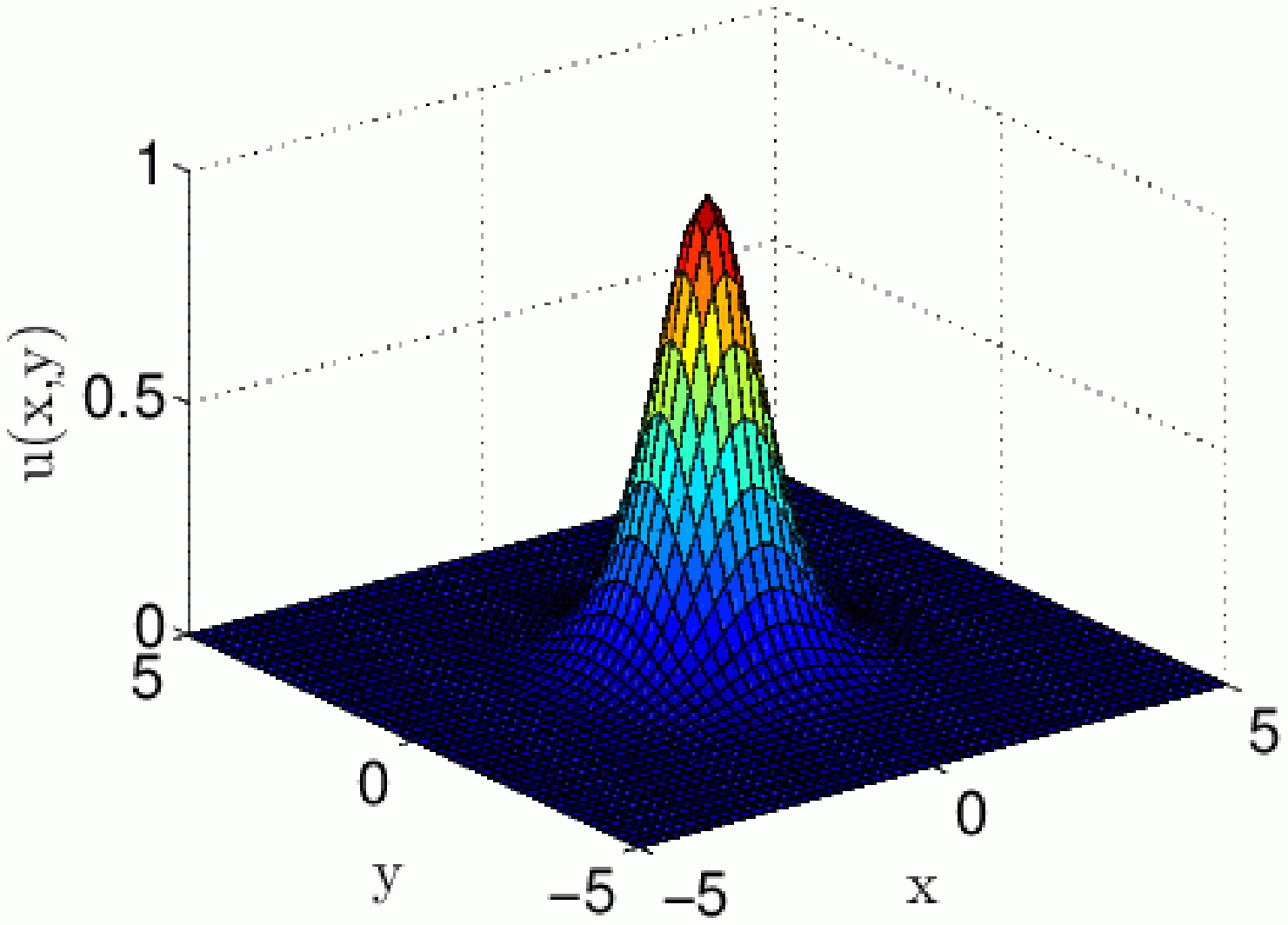}
\includegraphics[scale=0.45]{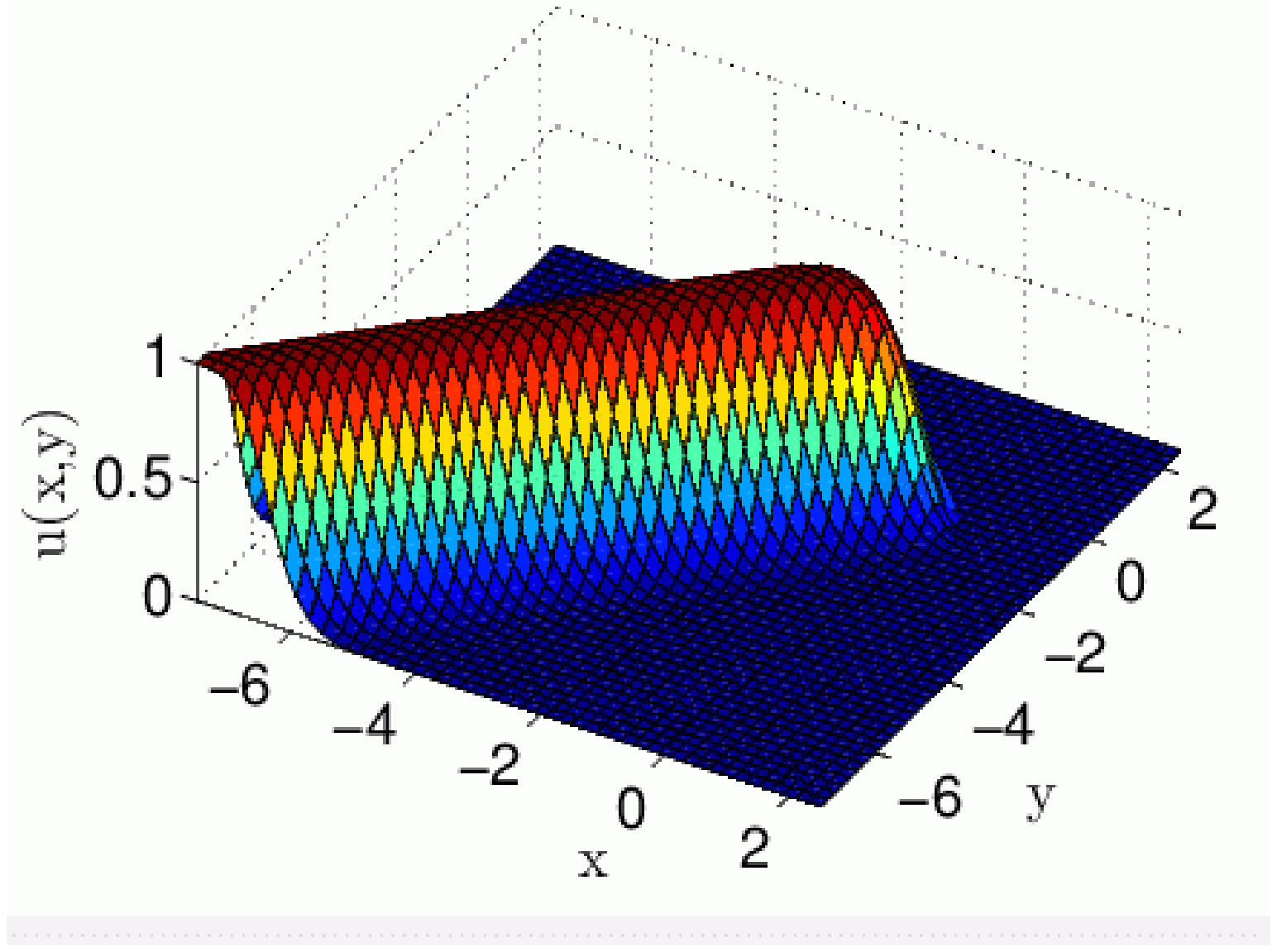}
\caption{Localized soliton (left), candidate for a kink on the right,
  travelling from left to right.}
\label{fig:loc}
\end{figure}

Schematically we can envisage pulses such as that shown in Fig.\ \ref{fig:loc}.

The soliton looks reasonable, but for topological reasons the kink has
an infinite ``side wall'' dislocation which may lead to infinite
energy. The tail can be truncated--but this brings us back to a
soliton-type wave.  The challenge then is to develop a suitable model
for a kink or soliton solution in a 2D system, or failing that to find
breather solutions.

\subsubsection{Derrick's theorem}\label{sec:Derrick}
In a simple single component homogeneous scalar {\bf continuum} field
theory, we have a {\bf non-existence} proof for stationary solitons
due to Derrick (see \cite{ms04}). The simple idea is to start by
supposing that, for example, our $n$-dimensional Hamiltonian is
\begin{eqnarray*}
  E(\phi) &=& \int\left(W(\phi)\nabla(\phi)\cdot\nabla(\phi) 
    +U(\phi)\right) \,\D^n x\\
  &&\equiv E_2+E_0 \;.
\end{eqnarray*}
Consider scaling the spatial variable
$\vec{x}\rightarrow\mu\vec{x}$. It is easy to show that
\[
E(\phi(\mu \vec{x}))=\mu^{2-n}E_2+\mu^{-n}E_0 \;.
\]
If the soliton solution $\phi(\vec{x})$ is a stable minima, then
$\D{E}/\D\mu=0$. The solution for $n=1$ is $\mu=\sqrt{E_0/E_2}$, but
there is {\bf no solution} for $n=2,3$.

Note that the theorem does not apply when we have a discrete system
which does not have a continuum limit - here breathers/ILMs/quodons
may play a part. The theorem (and the arguments given following the
figures above) give an indication that problems may arise if we try to
generalise in a naive way from 1D to higher dimensions.

\subsection{The work of Mar\'{\i}n et al.\ on breathers in 
   the K layer of mica}
\label{sec:MarinSim}
\begin{figure}
\centering
\includegraphics[width=0.5\textwidth]{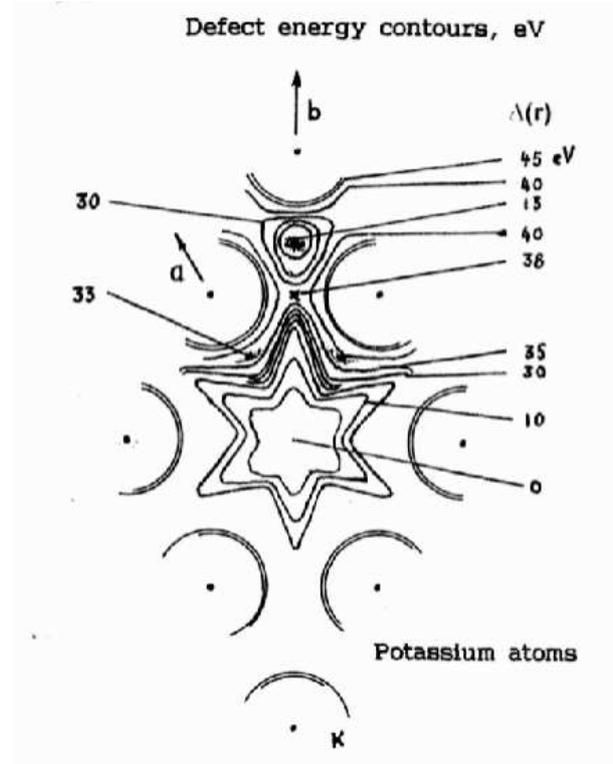}
\caption{Energy levels in the K layer in Mica.}
\label{fig:EnK_layer}
\end{figure}

\begin{figure}
\centering
\hspace{-2.5cm}\includegraphics[scale=0.73]{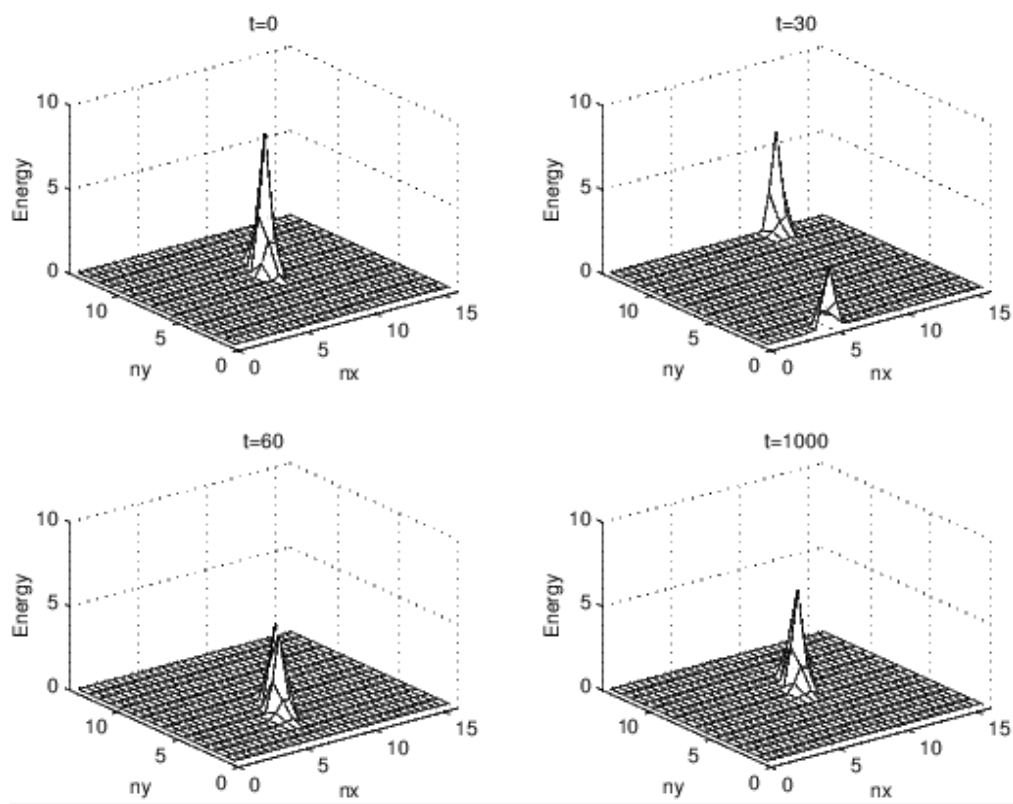}
\caption{Breather motion in model hexagonal lattice.}
\label{fig:MarinSim}
\end{figure}

Russell's work on mica led to Collins preparing a potential energy
plot on the Potassium layer - the energy of moving one $\mathbf{K}$
atom with all the others being fixed \cite{rc95}, see
Fig. \ref{fig:EnK_layer}.
In 1998, Mar\'{\i}n et al.\ \cite{mer98} used the quantitative features of
this plot to make a careful numerical study of a simple 2D model of
the K layer in mica. His program \texttt{hexlatt} simulates the motion
of a classical 2D hexagonal lattice, with displacements in the
plane. It includes both a nonlinear nearest-neighbour coupling ($W$)
between the $\mathbf{K}$ and some type of nonlinear ``on-site''
potential ($V$) \cite{mer98}. The Hamiltonian is
\begin{equation}\label{eq:MHam}
  H= \sum_{i,j} \frac12 \| \dot{\vec{u}}_{i,j}\|^2 + V(\vec{u}_{i,j}) 
  + \frac12 \lambda \sum_{i',j'} 
  W(\vec{u}_{i,j}, \vec{u}_{i',j'}) \;,
\end{equation}
where $\vec{u}_{i,j}$ is the $(i,j)$th atom's displacement from its
equilibrium state, and $\dot{\vec{u}}_{i,j}$ is the displacement's time
derivative. For the on-site potential (mimicking the effect of the
$\mathbf{O}$ atoms above and below the $\mathbf{K}$ plane, assumed
fixed) he used 6 atoms interacting in a Morse potential:
\begin{equation}\label{eq:MOnSite}
V_{\mbox{Morse}}(s)= \frac12 (1-\exp(-s))^2 \;,
\end{equation}
where $s$ is the distance between potassium and fixed oxygen
atoms. For interatomic potential ($\mathbf{K}$-$\mathbf{K}$) he used a
scaled classical 6-12 Lennard-Jones
\[
W_{\mbox{LJ}}(r)= 1 + \left(\frac{\sigma}{r}\right)^{12} -
2\left(\frac{\sigma}{r}\right)^6 \;,
\]
where $r$ is the distance between neighbouring potassium atoms and
$\sigma$ is a lattice constant.  The best results were found when both
the on-site and interatomic potentials have similar strengths.  Fig.\
\ref{fig:MarinSim} shows one typical simulation on a 16x16 lattice.
Note that we plot local energy density at various times on the
lattice.  At $t=0$ we give three atoms in the centre an asymmetric
kick, to mimic the radioactive decay of a K atom in the mica.  At $t=30$
the breather resulting from this kick has moved to the edge of the
small lattice and is beginning to reappear on the opposite side due to
the imposed periodic boundary conditions.  At $t=60$ it has continued
in the same direction and has almost reached the starting point.  The
final frame is at a much later time, $t=1000$, and shows the breather
after it has traversed the lattice about ten times.

Mar\'{\i}n's study showed stable breathers propagating up to around
$\le$ $10^4$ lattice constants before breaking up.  This is
encouraging, but to demonstrate tracks in mica of centimetres, we need
an extra factor of $10^5$ in the lifetime.  Mar\'{\i}n's 2D
calculation also included a brief study of inline breather-breather
collisions \cite{mer00}.  Most simulations were performed on a 16x16
lattice due to CPU speeds at the time, but a few were done using 32x32
lattices.  The K atoms were constrained to stay within the unit
cell--with no hopping to other sites (hence no kinks).  All
simulations were carried out at zero temperature.  Similar results
were also observed for cubic lattices \cite{mre01}.

A key feature in the model is that the forces have the so-called
quasi-one-dimensional property--that is, if an atom is moved along one
of the crystallographic directions, the restoring forces is exactly in
the same line, but with a negative sign.  Technically this is a $C_2$
symmetry.  Note that we also use ``quasi-one-dimensional'' in a
different sense, to describe the fact that a localized breather or
kink is observed travelling along a crystallographic direction with
very little disturbance in a transverse direction.  The two concepts
are conjectured to be closely related, although no formal proof of
this exists.

\clearpage
\newpage
{\small \noindent
\rule[0.5ex]{0.4\linewidth}{2pt}
{\em Historical Anecdote.} 
\rule[0.5ex]{0.4\linewidth}{2pt}
The second author's
involvement in this problem began in 1995, when he was contacted for
the first time by Mike Russell.  Mike was interested in attending the
soliton and nonlinear waves meeting, photo shown in
Fig. \ref{fig:Recr}, that JCE was organising in Edinburgh that summer.
He was keen to discuss nonlinear effects in mica crystal.
\begin{figure}[h]
\centering
\includegraphics[width=0.39\textwidth]{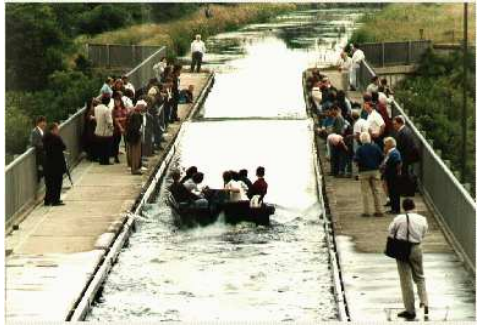}
\caption{Recreation of the soliton on the Union canal in 1995.}
\label{fig:Recr}
\end{figure}
Mike was studying the tracks in mica as seen in Fig. \ref{fig:Mica}
and believed that these could provide evidence for some sort of
nonlinear wave like a soliton in a 2D crystal - no linear theory
seemed to fit the data.
\begin{figure}[h]
\centering
\includegraphics[width=0.39\textwidth,angle=90]{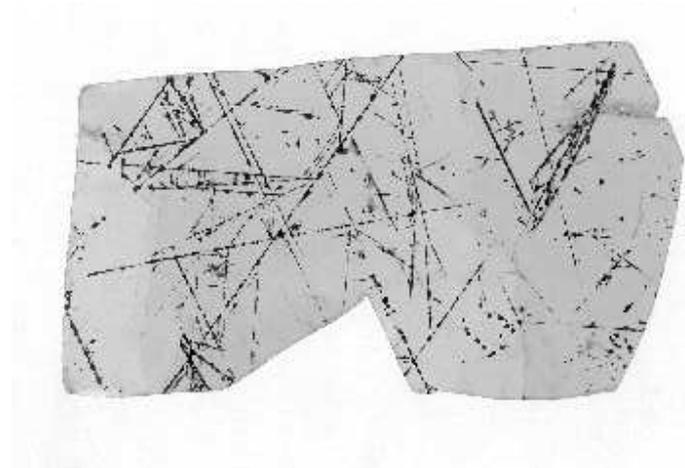}
\vspace{-0.5cm}\caption{Tracks in mica, several cm long.}\label{fig:Mica}
\end{figure}
JCE had long been interested in nonlinear waves, initially in
continuous systems, but more recently in discrete systems.  JCE, at
that time, was especially interested in breathers in
lattices. Subsequent collaboration of the two led to a series of
papers aimed at understanding the theoretical underpinnings of the
mica tracks (among other phenomena).
 
Although the consequent papers of Mar\'{\i}n, Eilbeck and Russell
received some attention, the calculations have never been replicated.
The Altea meeting in Spain in 2013 provided an excellent opportunity
to revisit and extend these calculations.  What follows is a more
extensive examination of results based on the simple model we
presented there.

\noindent
\rule[0.5ex]{\linewidth}{2pt}
} 
\newpage

\section{Preliminary results from numerical 
   experiments}
\label{sec:Present}
In the main part of this section, we describe a new 2D mathematical
model used for the present study of long-lived propagating breather
and kink solutions in mica at 0$^\circ$K.  With this model we allow
atoms in the lattice to be displaced out of the unit cells compared to
the nearest neighbour interactions considered in Mar\'in's model from
Sect.\ \ref{sec:MarinSim}.  Thus we can now allow the possibility of
kink solutions in our 2D lattice model.  In addition, in the choice of
potentials we take a more academic point of view and explore
alternative approaches besides Lennard-Jones.  Current research raises
new and not yet fully understood questions, and motivates further,
more intensive study.

In the present work we are concerned with the Hamiltonian dynamics of
$N$ potassium atoms $\mathbf{K}$ in a 2D $\mathbf{K}$-$\mathbf{K}$
sheet of mica crystal lattice.  Equivalently to (\ref{eq:MHam}) the
Hamiltonian of the system is
\begin{equation}\label{eq:Hamilt}
  H = K + V + U = \sum_{n=1}^{N} \left( \frac{1}{2}
    \|\vec{u}_{n}\|^2
    + \sum_{n'=1,\,n'\neq{n}}^{N} V(\vec{r}_{n},\vec{r}_{n'}) 
    + U(\vec{r}_{n})\right) \;,
\end{equation}
where $\vec{r}_{n}\in\mathrm{R}^2$ is the 2D position vector of
$n^{th}$ $\mathbf{K}$ atom in $(x,y)$ coordinates,
$\vec{u}_{n}=\dot{\vec{r}}_{n}$ is momentum, $K$ is kinetic energy,
$V$ is the interaction potential energy and $U$ is the on-site
potential energy.  All masses of atoms are normalized to one. The
system of equations is
\begin{eqnarray}
  \dot{\vec{r}}_{n} &=& \vec{u}_{n} \;, \label{eq:r}\\
  \dot{\vec{u}}_{n} &=& -\partial_{\vec{r}_{n}} \left(
    \sum_{n'=1,\,n'\neq{n}}^{N} 
    V(\vec{r}_{n},\vec{r}_{n'}) + U(\vec{r}_{n}) \right) \;, 
\label{eq:u}
\end{eqnarray}
for all $n=1,\dots,N$.

\subsection{On-site potential}
\label{sec:OnSitePot}
\begin{figure}
\centering
\begin{subfigure}[t][]{6cm}
\centering
\includegraphics[width=1.2\textwidth]
    {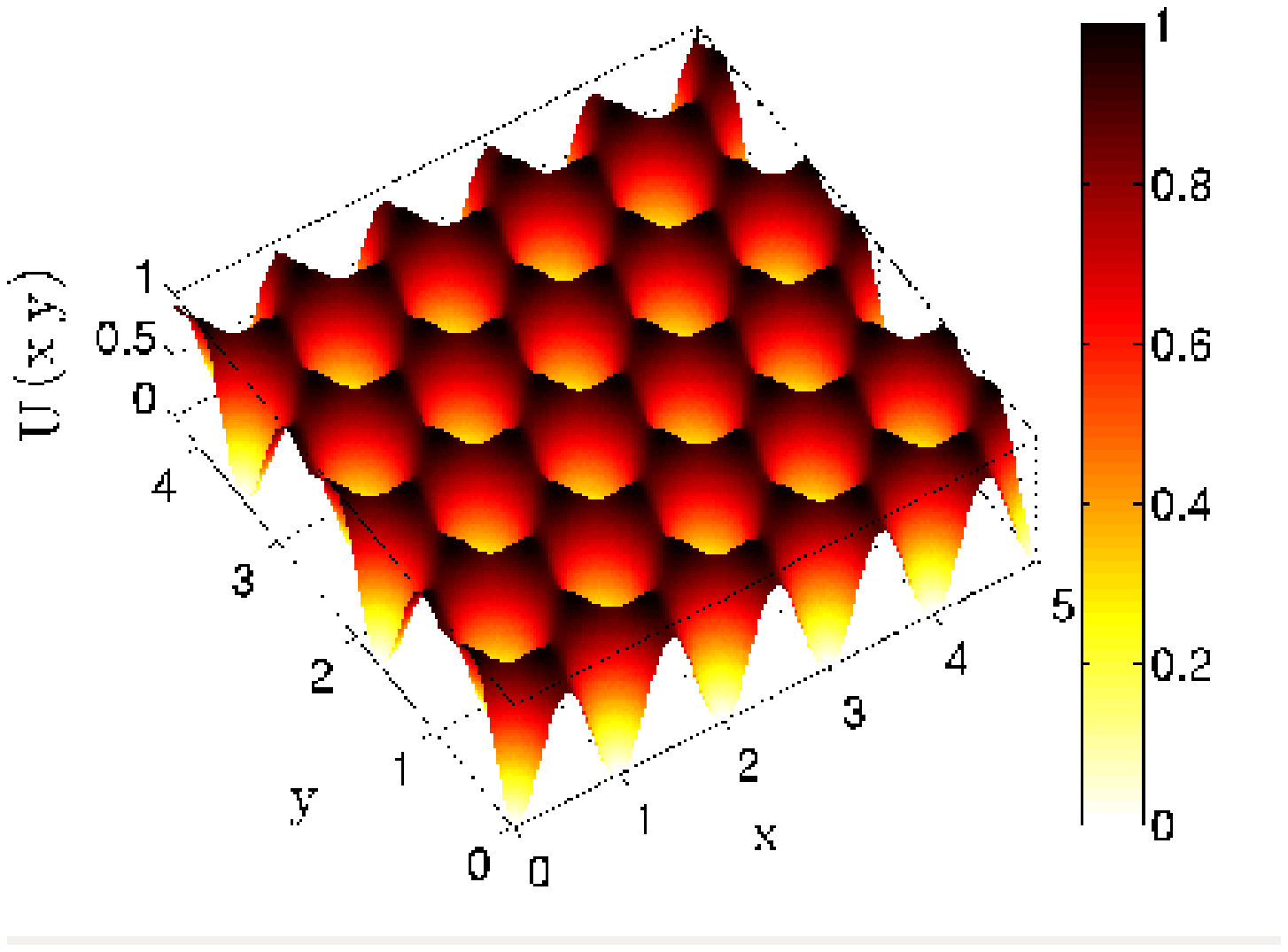}
\subcaption{Egg-box carton on-site potential with $\sigma=1$ and
  $U_{0}=1$.}
\label{fig:PotA}
\end{subfigure}
\qquad\qquad
\begin{subfigure}[t][]{6cm}
\centering
\includegraphics[width=1.2\textwidth]
     {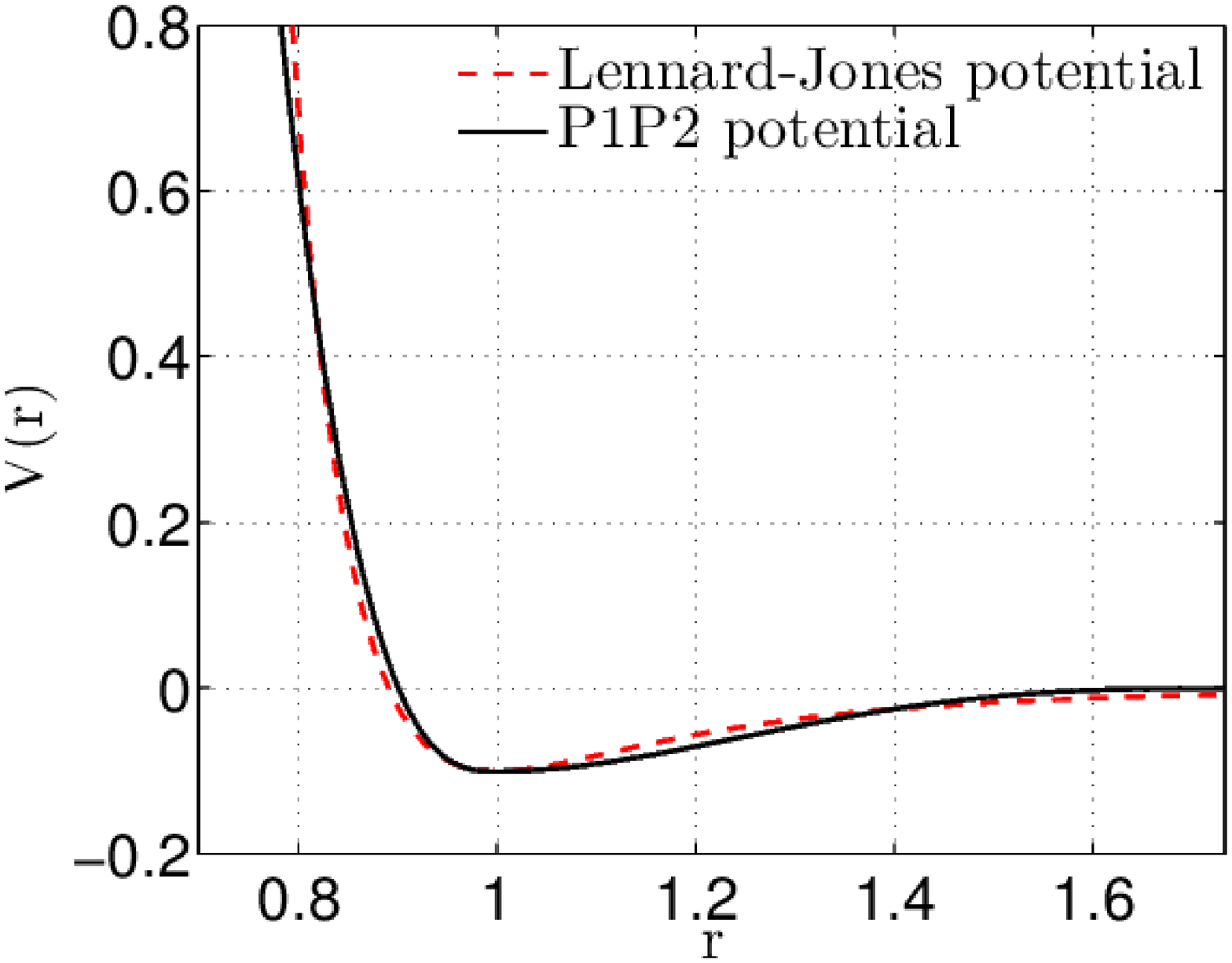}
\subcaption{Comparison between the Lennard-Jones potential and the
  polynomial potential ($P_{1}P_{2}$) with $M=25$, $\epsilon=0.1$,
  $\sigma=1$ and $r_{cut}=\sqrt{3}\sigma$.}\label{fig:PotB}
  \end{subfigure}
\end{figure}

In contrast to the on-site potential (\ref{eq:MOnSite}) considered by
Mar\'in et al. \cite{mer98}, but with the same assumptions of the fixed
upper and lower layers of oxygen atoms, we consider a smooth periodic
function with hexagonal symmetry \cite{yaduyachli11}, i.e.~a function
resembling an egg-box carton
\begin{eqnarray}\label{eq:OnSite}
  U(x,y) &=& \frac{2}{3} U_{0} \Bigg( 1-\frac{1}{3} 
    \Bigg( \cos{\Bigg( \frac{4\pi{y}}{\sqrt{3}\sigma} \Bigg)} 
  \nonumber\\
  &&  +\cos{ \Bigg( \frac{2\pi(\sqrt{3}x-y)}
          {\sqrt{3}\sigma}\Bigg)} + 
      \cos{ \Bigg( \frac{2\pi(\sqrt{3}x+y)}{\sqrt{3}\sigma} 
        \Bigg) } \Bigg) \Bigg) \;,
\end{eqnarray}
where $x = (\vec{r}_{n})_1$, $y=(\vec{r}_{n})_2$, $\sigma$ is the
lattice constant and $U_{0}>0$ is the maximal value of the on-site
potential, see Fig. \ref{fig:PotA}.  This model has the same
quantitative features as Fig.\ \ref{fig:EnK_layer}.

Note that a simple product of cosine functions would not provide the
required hexagonal symmetry.  Also, in a 1D approximation,
i.e.~$y=\mbox{const.}$, the on-site potential (\ref{eq:OnSite})
reduces to the cosine function which is an on-site potential of the
discrete sine-Gordon equation and the periodic potential of a 1D model
considered in \cite{dcer11}. The model in \cite{dcer11} can be thought
as a 1D approximation of the 2D model (\ref{eq:Hamilt}) in any of
three crystallographic lattice directions which can be prescribed by
the direction cosines, that is, with vectors: $(1,0)^T$ and
$(1/2,\pm\sqrt{3}/2)^T$.  Without periodic boundary conditions, a
smooth cut-off of potential (\ref{eq:OnSite}) can be imposed.

\subsection{Interaction potential $\mathbf{(P_{1}P_{2})}$}
\label{sec:InterPot}
There are very well known and much used empirical interaction
potentials from the molecular dynamics community such as Lennard-Jones
12-6, Morse and Buckingham potentials, among others.  Essentially all
of these interaction potentials model repulsive and attractive forces
of particles.  The detailed structure of these potential energy
functions may strongly influence the behaviour observed in simulations,
particularly dynamical properties.

All potentials mentioned above are built from completely monotone
functions with a possible singularity at vanishing interparticle
distance.  For example, the Lennard-Jones 12-6 potential has been
extensively used in molecular dynamics models, on account of its good
representation of van der Waals attraction forces and its efficient
implementation in numerical codes.  In this article, we use a simple
family of interaction potentials, defined by piecewise polynomials,
which allow for easy adjustment of modelling features such as well
depth and which do not have a singularity at the origin.  Importantly
we have found that these simplified potentials lead to interesting
properties of the numerical solutions for our lattice model compared
to those obtained using more conventional interaction potentials (in
particular, we observe kinks in certain simulations, see below.)  We
refer the interested reader to \cite{bel14} where the authors have
performed a numerical study of propagating localized modes in a 2D
hexagonal lattice, by considering conventional Lennard-Jones potential
for the interparticle interactions and the same on-site potential
(\ref{eq:OnSite}).

The numerical results observed in this article suggest the need for
deeper analytical investigations, particularly where these may lead to
the design of completely new materials \cite{GeGr13}.  In addition,
the use of piecewise polynomial potentials may provide additional
freedom to better match the material properties in consideration,
while excluding singularities and directly incorporating smooth
cut-offs; such potentials can be constructed to different orders of
regularity.
  
In this paper, for the interaction potential $V$, we consider a
distance dependant potential of two joint $4^{th}$ order polynomials
$P_{1}(r)$ and $P_{2}(r)$, that is
\[ V(r) = \left\{
  \begin{array}{l l}
    P_{1}(r)\;, & \quad 0 \leq r \leq \sigma \;, \\
    P_{2}(r)\;, & \quad \sigma < r \leq r_{cut} \;, \\
    0\;, & \quad \mbox{otherwise} \;,
  \end{array}
\right.
\]
where $r=\|\vec{r}_{n}-\vec{r}_{n'}\|$ for all $n$ and $n'$,
$n\neq{n'}$.  The parameter $\sigma$ is the lattice constant and
$r_{cut}$ is the cut-off radius of the potential.  The coefficients of
the polynomials $P_{1}(r)$ and $P_{2}(r)$ are found from the following
constraints:
\begin{eqnarray*}
  && P_{1}(0) = M\;, \quad \partial_{r}P_{1}(0) = 0\;, \quad M >0\;, \\
  && P_{1}(\sigma) = P_{2}(\sigma) = -\epsilon\;, \quad \epsilon > 0\;,\\
  &&\partial_{r}{P}_{1}(\sigma) = \partial_{r}{P}_{2}(\sigma) = 0\;, \\
  &&\partial_{rr}{P}_{1}(\sigma) = \partial_{rr}{P}_{2}(\sigma)\;,\\
  && P_{2}(r_{cut}) = 0\;, \quad \partial_{r}{P}_{2}(r_{cut}) = 0\;, 
  \quad \partial_{rr}{P}_{2}(r_{cut}) = 0\;,
\end{eqnarray*}
such that $V(0)=M$, $V(\sigma)=-\epsilon$, $V(r_{cut})=0$,
$\partial_{r}V(0)=0$, $\partial_{r}V(\sigma)=0$,
$\partial_{r}V(r_{cut})=0$ and $\partial_{rr}V(r_{cut})=0$.

For small atomic displacements from the mechanical equilibrium state,
which we will consider as our initial conditions, the particular
choice of the cut-off radius $r_{cut}=\sqrt(3)\sigma$ leads to the
closest representation of the nearest neighbour interaction model,
i.e.~Hamiltonian dynamics of atoms with only nearest neighbour
interactions, such as the model by Mar\'in et al. \cite{mer98}.
Importantly, there is no formal restrictions to consider larger values
of the cut-off radius $r_{cut}$.

For our computations we choose $\sigma=1$, $M=25$ and $\epsilon=0.1$
to approximately match the scaled Lennard-Jones potential
\[ 
V_{LJ}(r) = \epsilon \left( \left( \frac{\sigma}{r}\right)^{12} - 2
  \left( \frac{\sigma}{r}\right)^{6} \right)
\]
with $\sigma=1$ and $\epsilon=0.1$, where $\sigma$ is the same lattice
constant and $\epsilon$ is the potential well depth value, see
Fig. \ref{fig:PotB}.  Additional motivation for the particular choice
of parameter values will be given in Sec. \ref{sec:ParVal}.

\subsection{Time integration method}
\label{TimeInt}
In simulations of Hamiltonian systems,
e.g.~(\ref{eq:r})--(\ref{eq:u}), it is essential to use a symplectic
time integration procedure.  In our simulations, we employed the
Verlet method, a second order, explicit symplectic scheme
\cite{alti89,lr04}.  The method is known for its good energy
conservation properties in long time simulations where energy stays
bounded in time and is conserved up to second order with respect to a
time step.  For a Hamiltonian of the form $H=\frac{1}{2}\|\vec{p}\|^2+V(\vec{q})$, the
Verlet timestep approximates Newtonian dynamics by the steps:
\begin{eqnarray*}
\vec{q}^{n+1/2} &=& \vec{q}^{n} + \frac{1}{2}\tau \vec{p}^n \;, \\
\vec{p}^{n+1} &=& \vec{p}^{n} - \tau \nabla_{\vec{q}} V(\vec{q}^{n+1/2}) \;, \\
\vec{q}^{n+1} &=& \vec{q}^{n+1/2} + \frac{1}{2}\tau \vec{p}^{n+1} \;,
\end{eqnarray*}
where $\tau$ is the time step, $\vec{q}^{n}\approx\vec{q}(t^{n})$ and
$\vec{p}^{n}\approx\vec{p}(t^{n})$ at time level $t^{n}=n\tau$ where
$n=0,1,\dots$.  As mentioned above, the method preserves the
symplectic property of Hamiltonian dynamics, i.e.\
$\D\vec{q}^{n+1}\wedge\D\vec{p}^{n+1} =
\D\vec{q}^{n}\wedge\D\vec{p}^{n}$, where $\wedge$ is a wedge product
of two differential 1-forms in vector representation.  Thus the model
is also volume preserving in phase space. A valuable feature of
symplectic integrators is that they may, under certain circumstances,
be interpreted as being essentially equivalent to the exact
propagation of a modified Hamiltonian ($\tilde{H}_{\tau} = H +
O(\tau^{k})$) (for a $k^{th}$ order scheme) meaning that we may
reinterpret the trajectories generated by our numerical method as
dynamical paths for a perturbed system.  Interested readers in
numerical methods for Hamiltonian dynamics are referred to
\cite{lr04}.

\subsection{Parameter values}
\label{sec:ParVal}
To proceed with the numerical study of propagating localized modes of
system (\ref{eq:r})--(\ref{eq:u}) we must select system parameter values.  
Without loss of generality, we set the lattice constant $\sigma$
equal to one.  Once the interaction potential parameter values are
fixed, we are left with one parameter value to consider, that is, the
strength of the on-site potential parameter $U_{0}$.  Thus with the
parameter $U_{0}$ we can control the relative strengths of forces in
the system.  With very small values of $U_{0}$, the system will
be dominated by the forces of the interaction potential, {\em and vice
  versa}.

As was noted by Mar\'in et al.\ \cite{mer98}, the best conditions to
observe propagating discrete breathers seemed to be when both
potentials are of roughly equal strength.  We find that for given
interaction potential parameter values $r_{cut}=\sqrt{3}\sigma$,
$\sigma=1$, $M=25$ and $\epsilon=0.1$, and with value $U_{0}=2$, both
potentials agree well for the small displacements of the potassium
$\mathbf{K}$ atom from its mechanical equilibrium state while the
neighbouring $\mathbf{K}$ atoms have been fixed in their positions.
For the comparison of potentials we consider an atom with the six fixed
neighbouring atoms in their equilibrium states.  

\begin{figure}
\centering
\includegraphics[width=0.45\textwidth]{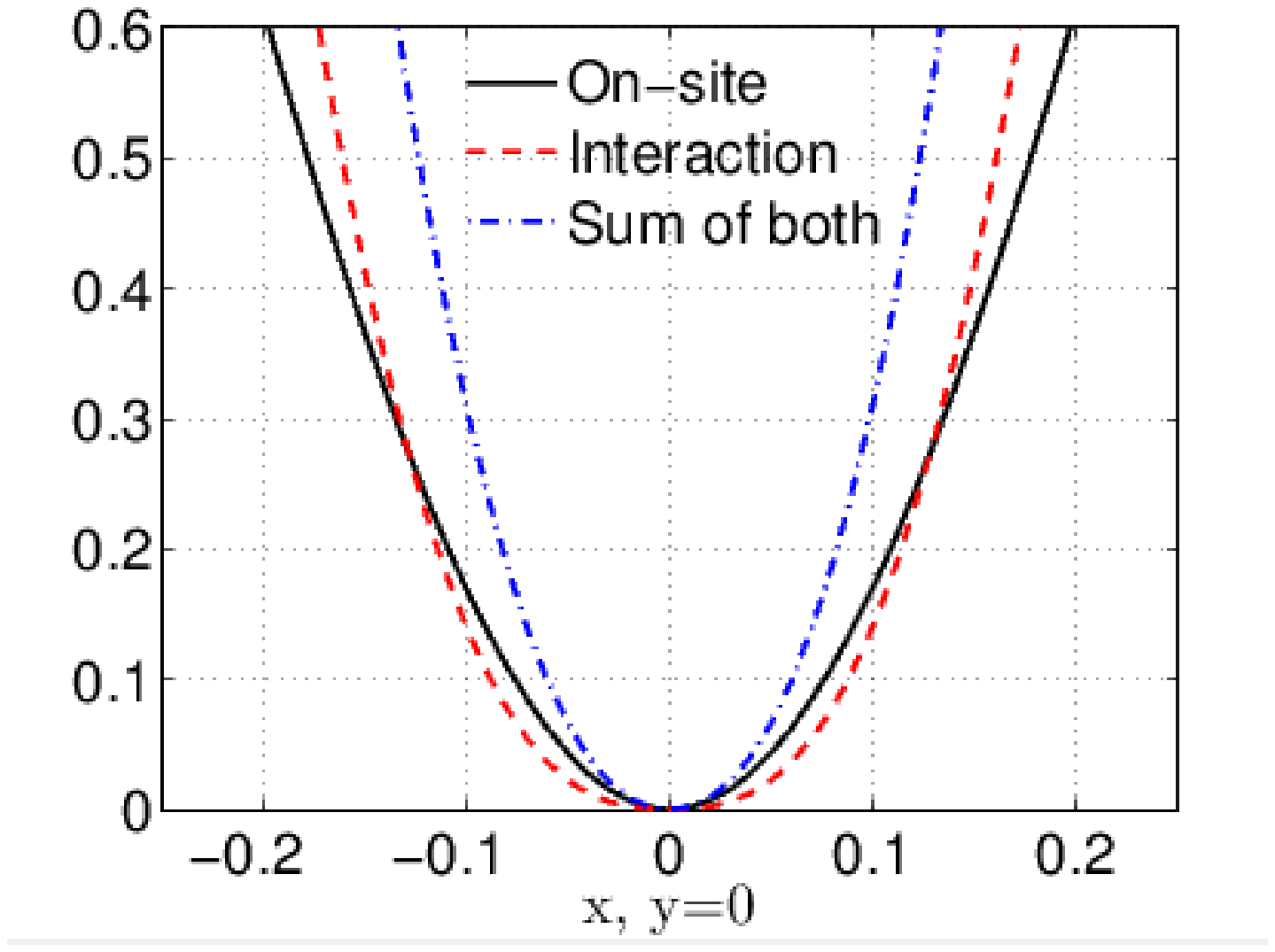}
\includegraphics[width=0.45\textwidth]{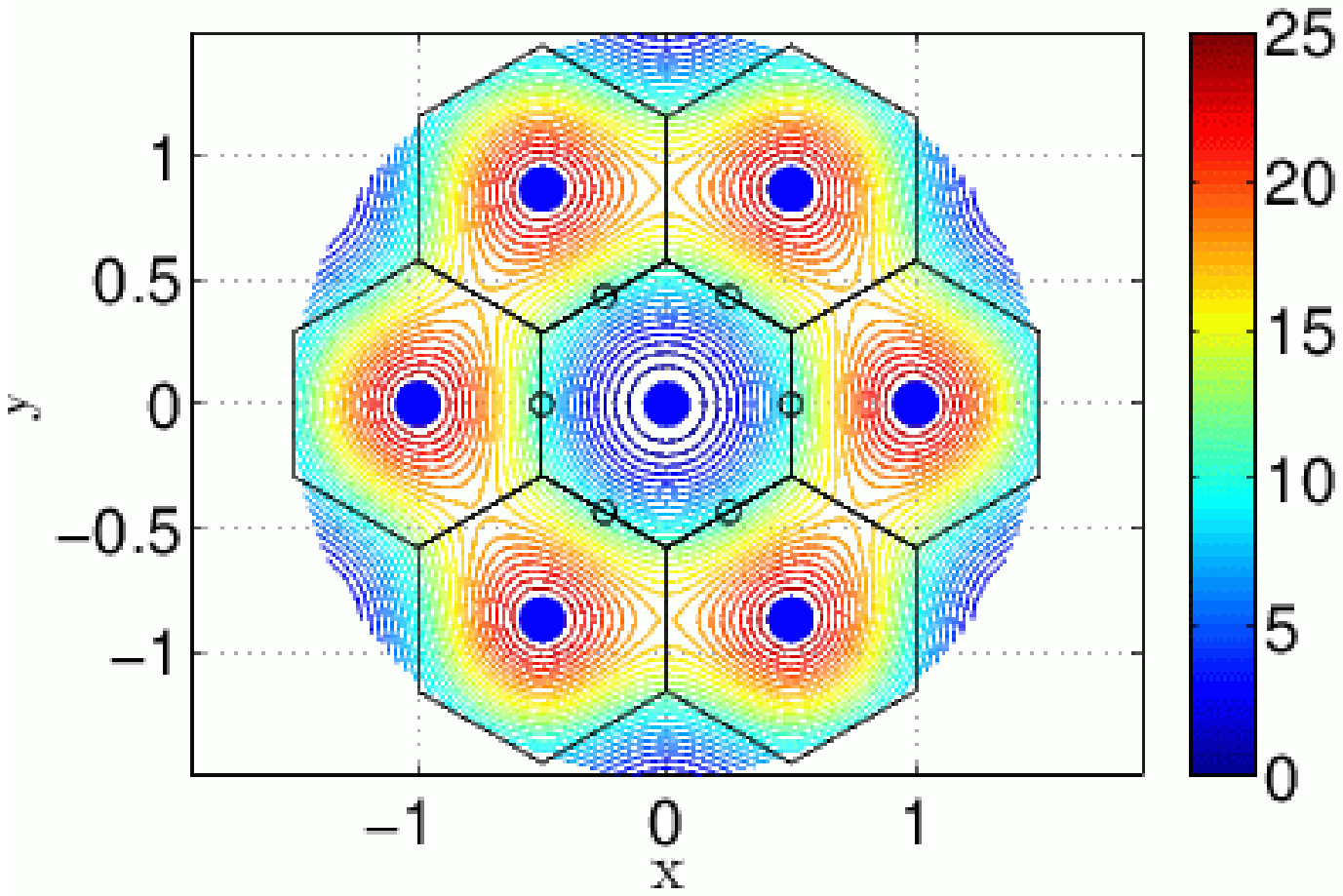}
\caption{Unrelaxed potential computations. Parameter
    values: $r_{cut}=\sqrt{3}\sigma$, $\sigma=1$,
    $M=25$,$\epsilon=0.1$ and $U_{0}=2$. Left: unrelaxed potential
    functions as seen by a $\mathbf{K}$ atom moving in the $(1,0)^T$
    crystallographic direction in a 2D $\mathbf{K}$-$\mathbf{K}$ sheet
    of the mica crystal lattice model. Right: energy contour lines as
    seen by the $\mathbf{K}$ atom at the origin moving in a 2D
    $\mathbf{K}$-$\mathbf{K}$ sheet of the mica crystal lattice
    model.}\label{fig:Unrl}
\end{figure}

Results of unrelaxed potential computations are shown in
Fig. \ref{fig:Unrl} for the parameter values given above. In both
plots we normalize the interaction potential values such that
$V\ge{0}$.  In the left plot of Fig. \ref{fig:Unrl}, we compute
unrelaxed potentials as seen by an $\mathbf{K}$ atom moving in the
$(1,0)^T$ crystallographic direction in a 2D $\mathbf{K}$-$\mathbf{K}$
sheet of mica crystal lattice, while in the right plot of
Fig. \ref{fig:Unrl} we plot the total potential energy contour lines
as seen by the $\mathbf{K}$ atom at origin.  The colour axis agrees
with the location of the $\mathbf{K}$ atom in space.  When the atom is
at the origin, the total potential energy is equal to zero.  However
when the atom approaches any of the other potassium atoms, the on-site
potential approaches zero and thus there is mainly only one
contribution from the interaction potential at $r=0$.   For this reason the
potential energy becomes close to value $M$ where $M=25$ in our
example.

For the purposes of illustration we have indicated in the right plot
of Fig. \ref{fig:Unrl}, the lines of hexagonal lattice, six
neighbouring $\mathbf{O}$ atoms and seven $\mathbf{K}$ atoms in their
dynamical equilibrium states.  Compare the energy levels of the right
plot of Fig. \ref{fig:Unrl} to the energy levels of
Fig. \ref{fig:EnK_layer}.

\subsection{Numerical results}\label{sec:NumResults}
In this section we describe numerical results showing propagating
discrete breather, kink and horseshoe wave solutions in an open
lattice.  Periodic boundary conditions can also be imposed.  Open
lattice simulation allows atoms to be ejected by the propagating waves
at the edge of the lattice, which has possibly relevance to the experiment by
Russell \cite{re07}.

With zero initial velocities (momentum) and all $\mathbf{K}$ atoms
being placed at the cell centres of the hexagonal lattice, the system
(\ref{eq:r})--(\ref{eq:u}) is in mechanical equilibrium, i.e.~all
forces of the system are equal to zero.  The lattice is defined by
$N_{x}$ atoms in the $x$ axis direction and an even number $N_{y}$ of
atoms in the $y$ axis direction.  The first atom is always placed at
the origin $(0,0)$. The spacing between atoms in the $x$ direction is
equal to the lattice constant $\sigma=1$ and in the $y$ direction, the
lattice spacing between atoms is $h=\sqrt{3}/2\sigma$.  The total
number of atoms considered in the simulations is
$N=N_{x}N_{y}-[N_{y}/2]$. In all simulations we use the Verlet method,
as described above, with fixed time step $\tau=0.01$.

For the initial conditions, we consider imparting a non-zero velocity
to one of the atoms while the rest of the lattice is kept at rest.
The initial velocities of this atom in the $x$ and $y$ axis directions
are indicated by $u_{x}^{0}$ and $u_{y}^{0}$, respectively.  With
different initial velocity kicks and with different parameter values
$U_{0}$, we are able to observe different phenomena as discussed in
the following sections.  In addition, we will refer to the horizontal
chain of atoms as the main chain of atoms along which the breather or
kink solutions propagates, that is, the most of their energy has been
localized on this chain.  The final computational time is indicated by
$T_{end}$.

By assigning half of the interaction potential energy to each atom in
an interacting pair, while adding also the kinetic and potential
energy values from the on-site potential, we can define an energy
density function for each atom.  Since the interaction potential may
take negative values, we can normalize it.  To see small scales
better, we take the logarithm of the energy density function, that is
\[
 H_{log} = \log ( H + |\min\{H\}| + 1),
\]
such that $H_{log}\ge{0}$. In all energy plots we plot $H_{log}$ and
interpolate its values on uniform meshes for plotting purposes only.

\subsubsection{Numerical results: propagating breather solutions}
\label{sec:BrSol}
\begin{figure}
\centering 
{\includegraphics[width=0.32\textwidth]{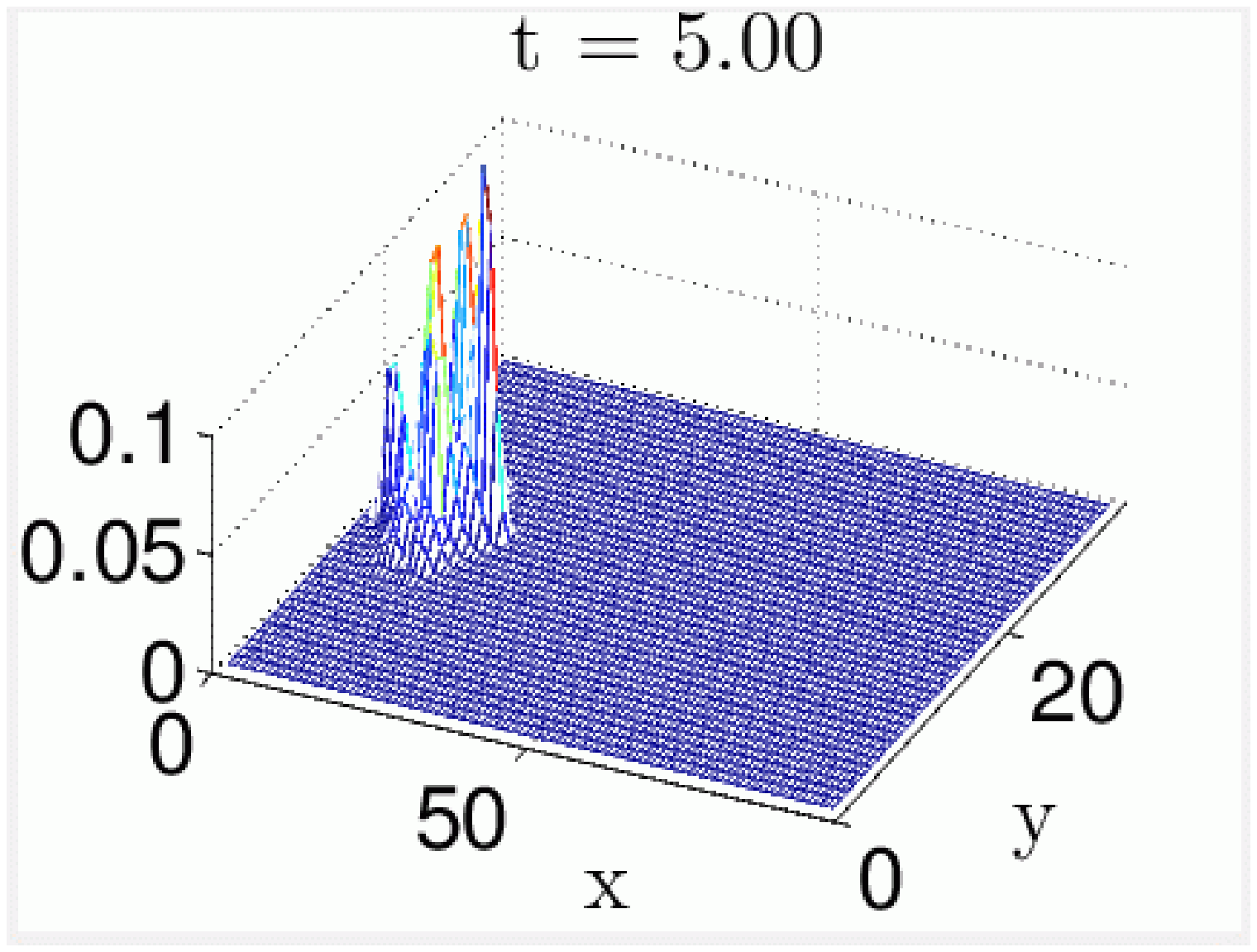}}
{\includegraphics[width=0.32\textwidth]{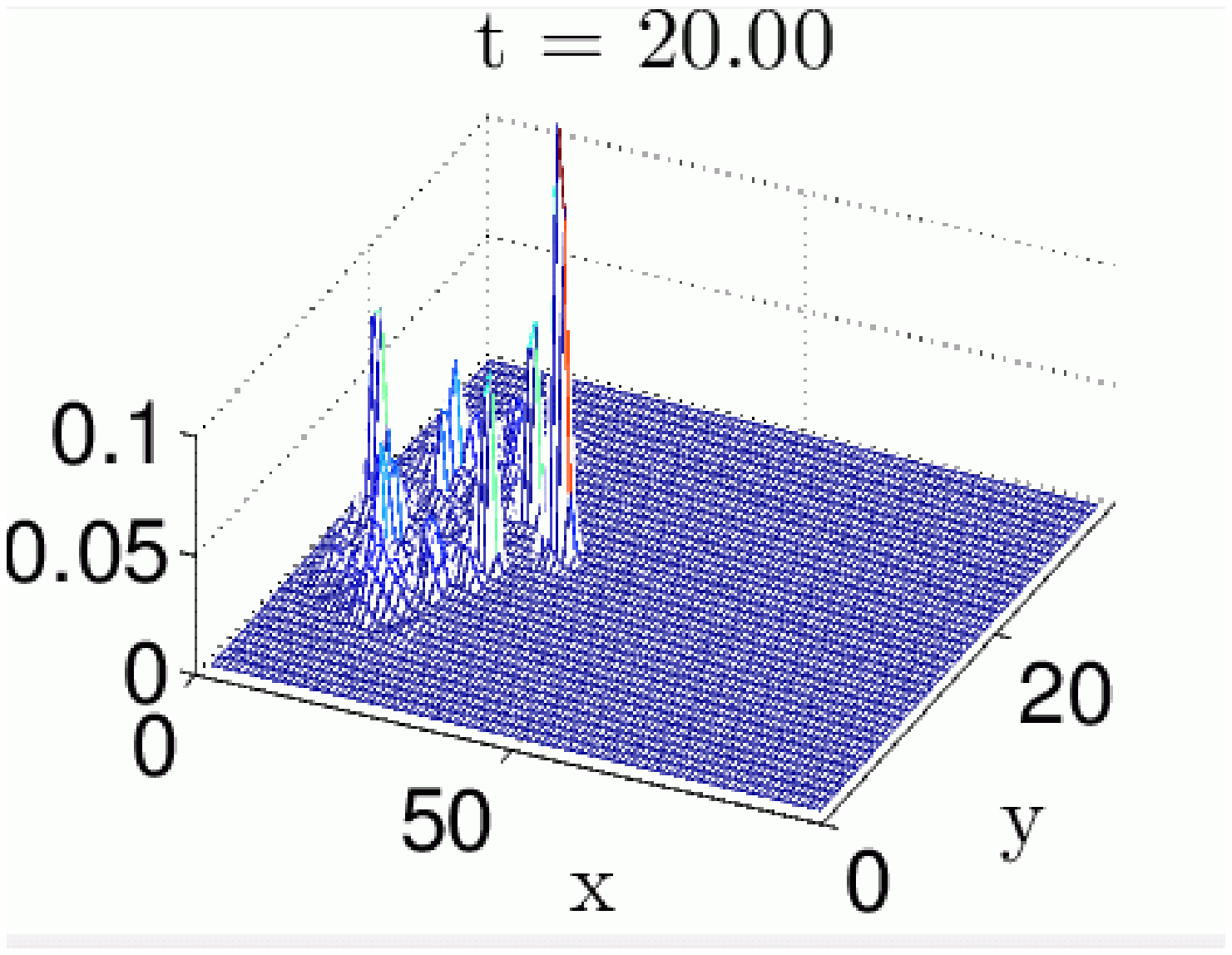}}\\
{\includegraphics[width=0.32\textwidth]{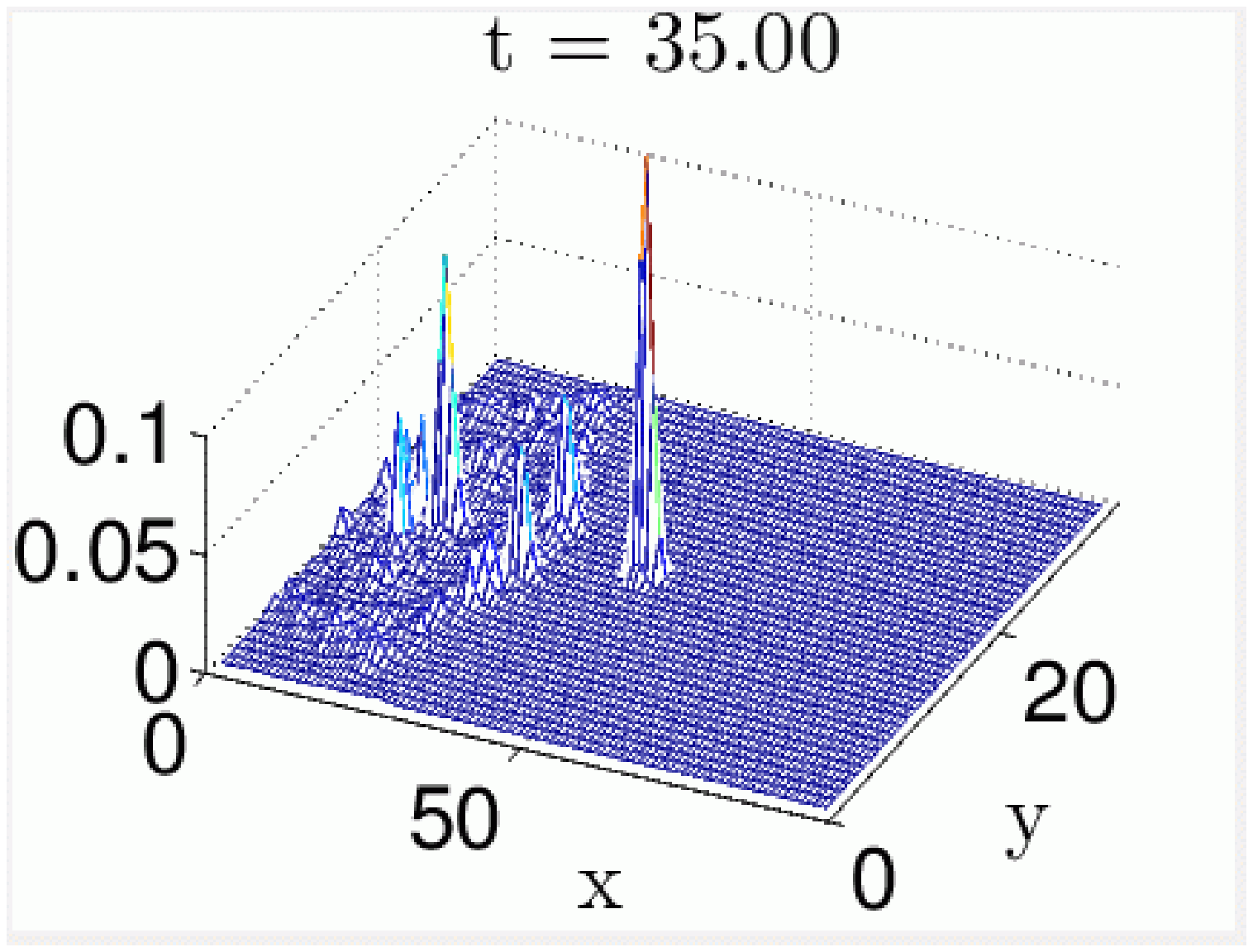}}
{\includegraphics[width=0.32\textwidth]{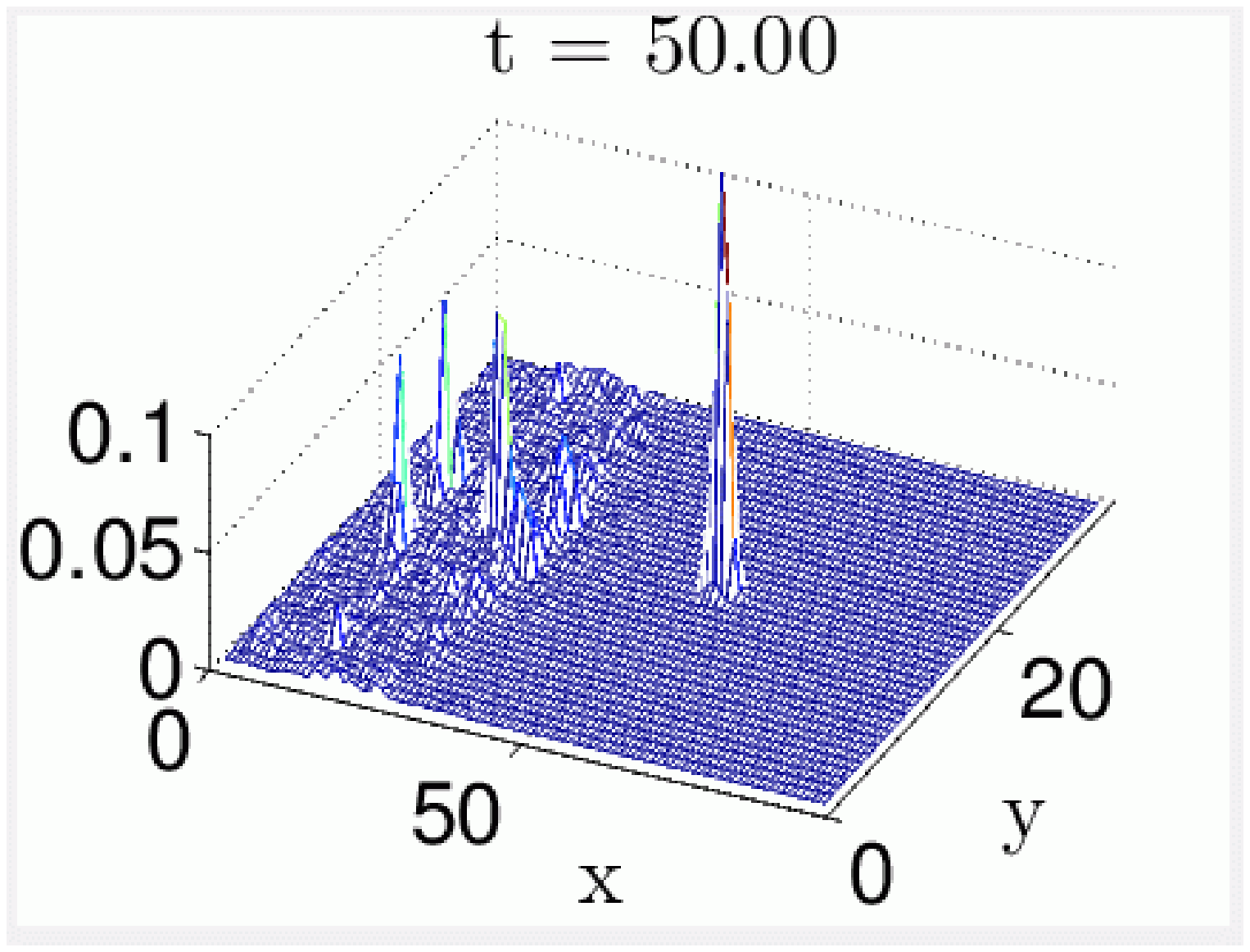}}\\
{\includegraphics[width=0.32\textwidth]{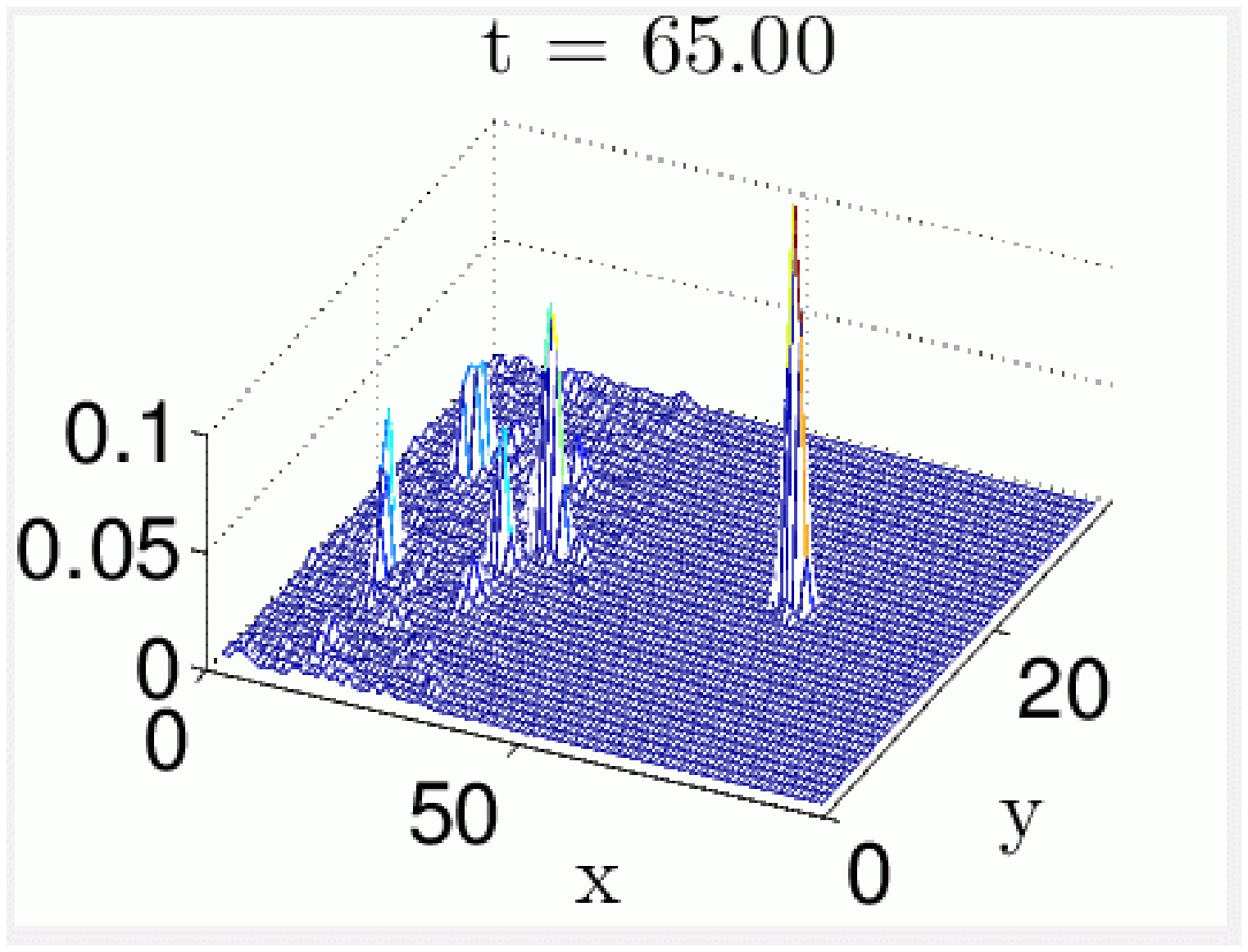}}
{\includegraphics[width=0.32\textwidth]{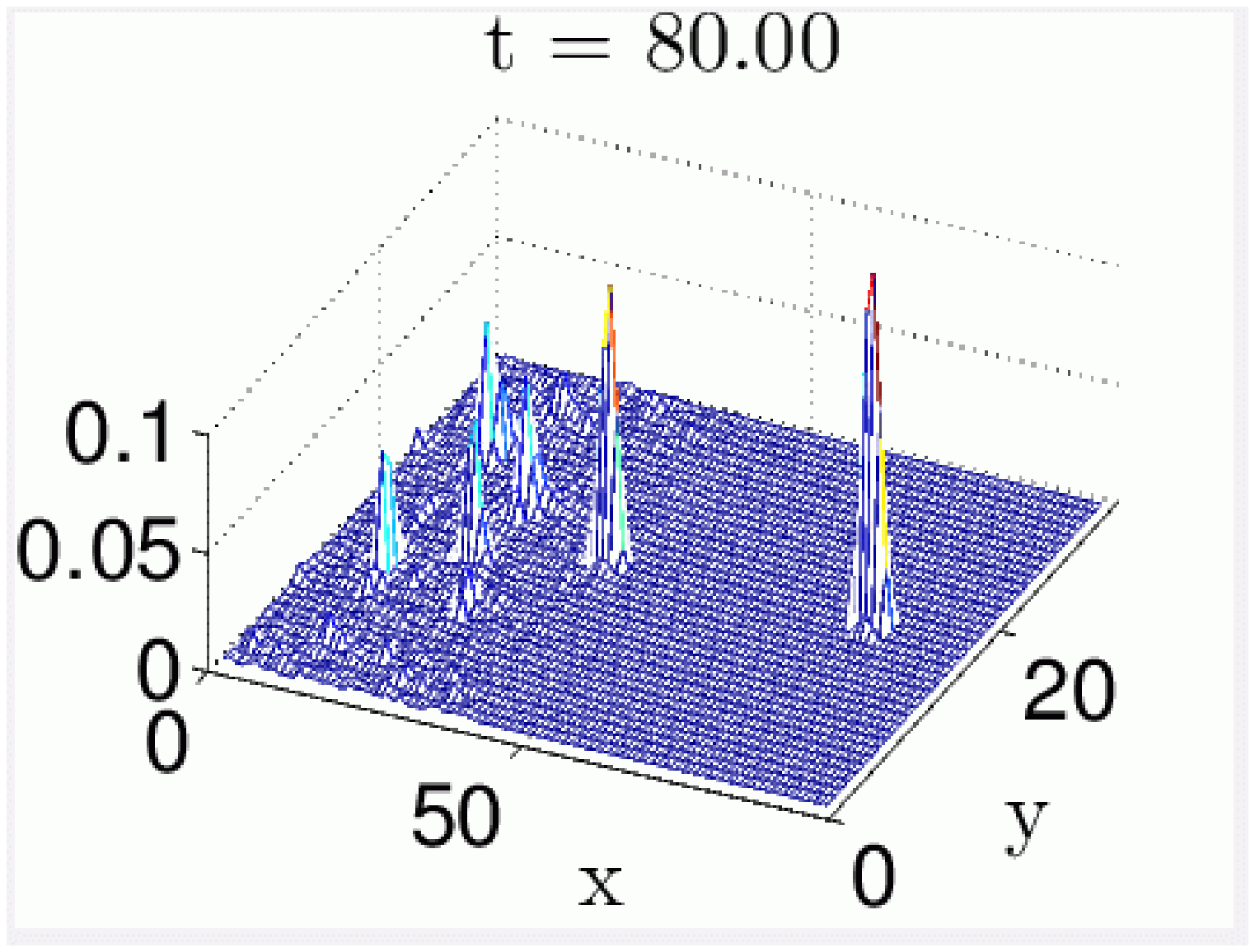}}
\caption{Evolution of the energy density function in time of the
  breather solution. $N_{x}=100$, $N_{y}=40$, $T_{end}=80$, $U_{0}=2$,
  $u_{x}^{0}=3$ and $u_{y}^{0}=0$.}\label{fig:PropBrEn}
\end{figure}

This subsection is devoted to the study of propagating discrete
breather solutions.  We perform numerical tests by exciting one atom
in the system, i.e.,\ by giving a single initial kick.  We provide the
impulse to the atom in the middle of the lattice with respect to the
$y$ axis.  Numerical results with $N_{x}=100$, $N_{y}=40$, $U_{0}=2$,
$u_{x}^{0}=3$ and $u_{y}^{0}=0$ are shown in Fig. \ref{fig:PropBrEn}.
We integrate in time until $T_{end}=80$.  Figure \ref{fig:PropBrEn}
illustrates the propagation of the breather energy in time on a
horizontal lattice chain in the $(1,0)^T$ crystallographic direction.
We have excluded atoms at the boundaries from the plots due to high
potential values at the boundaries.  The breather in the $x$ axis
direction is localized in space on about seven lattice sites and on
about three lattice sites in the $y$ axis direction.

\begin{figure}
\centering
\includegraphics[scale=0.5]{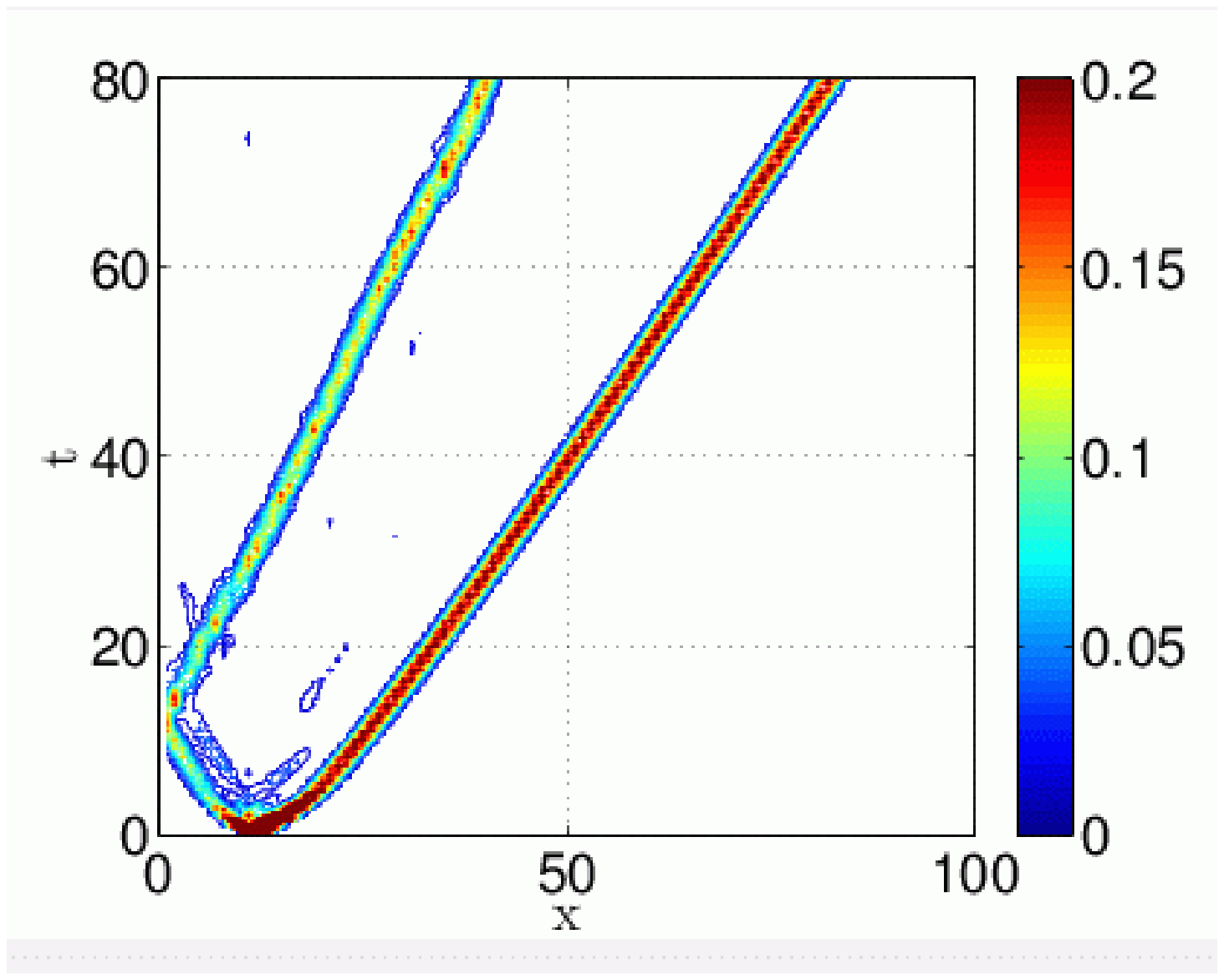}
\includegraphics[scale=0.5]{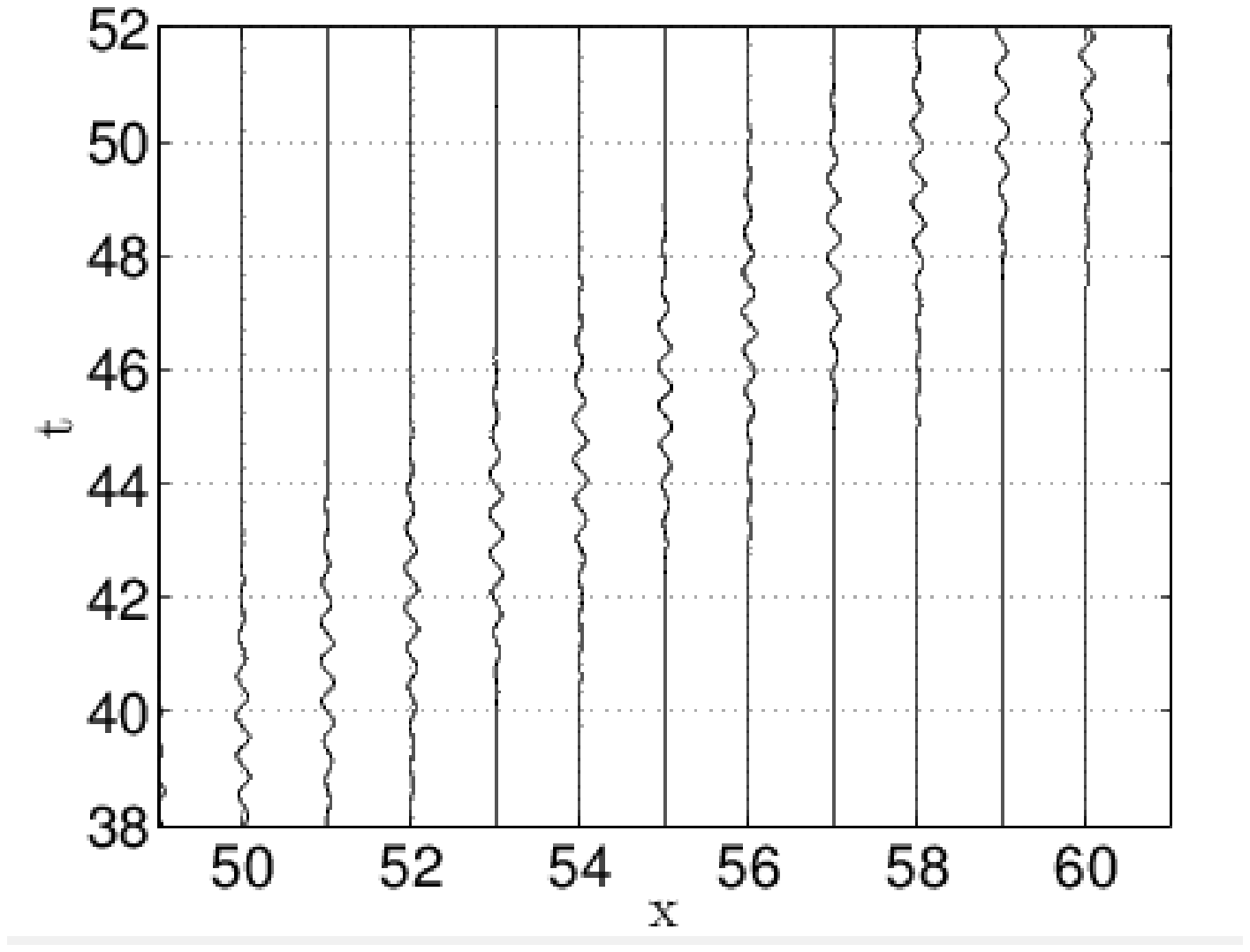}
\caption{Short time simulation of a propagating breather. $N_{x}=100$,
  $N_{y}=40$, $T_{end}=80$, $U_{0}=2$, $u_{x}^{0}=3$ and
  $u_{y}^{0}=0$. Left: contour plot of the atomic energy density
  function on a horizontal lattice chain. Right: atomic displacements
  in the $x$ axis direction from their equilibrium positions in time
  on a horizontal lattice chain.}
\label{fig:ShortSim}
\end{figure}

The initial kick has produced a highly localized quasi-one-dimensional
breather solution.  The excess energy of the kick produces phonons
which spread into the lattice at higher velocities than the breather.
In addition, the kick has produced a secondary breather solution with
a lower energy propagating in the opposite direction.  After some
time, this breather solution elastically reflects from the boundary
and follows the main breather solution.  To illustrate that, we plot
(in time after each $20$ time steps) the energy density function of
atoms on the main horizontal chain along which the breather
propagates, see the left plot of Fig. \ref{fig:ShortSim}.  We plot the
atomic displacements in the $x$ axis direction of the same lattice
chain in the right plot of Fig. \ref{fig:ShortSim}.  From the
displacement plot we can conclude that the localized mode is an
optical breather.

\begin{figure}
\begin{center}
{\includegraphics[width=0.48\textwidth]{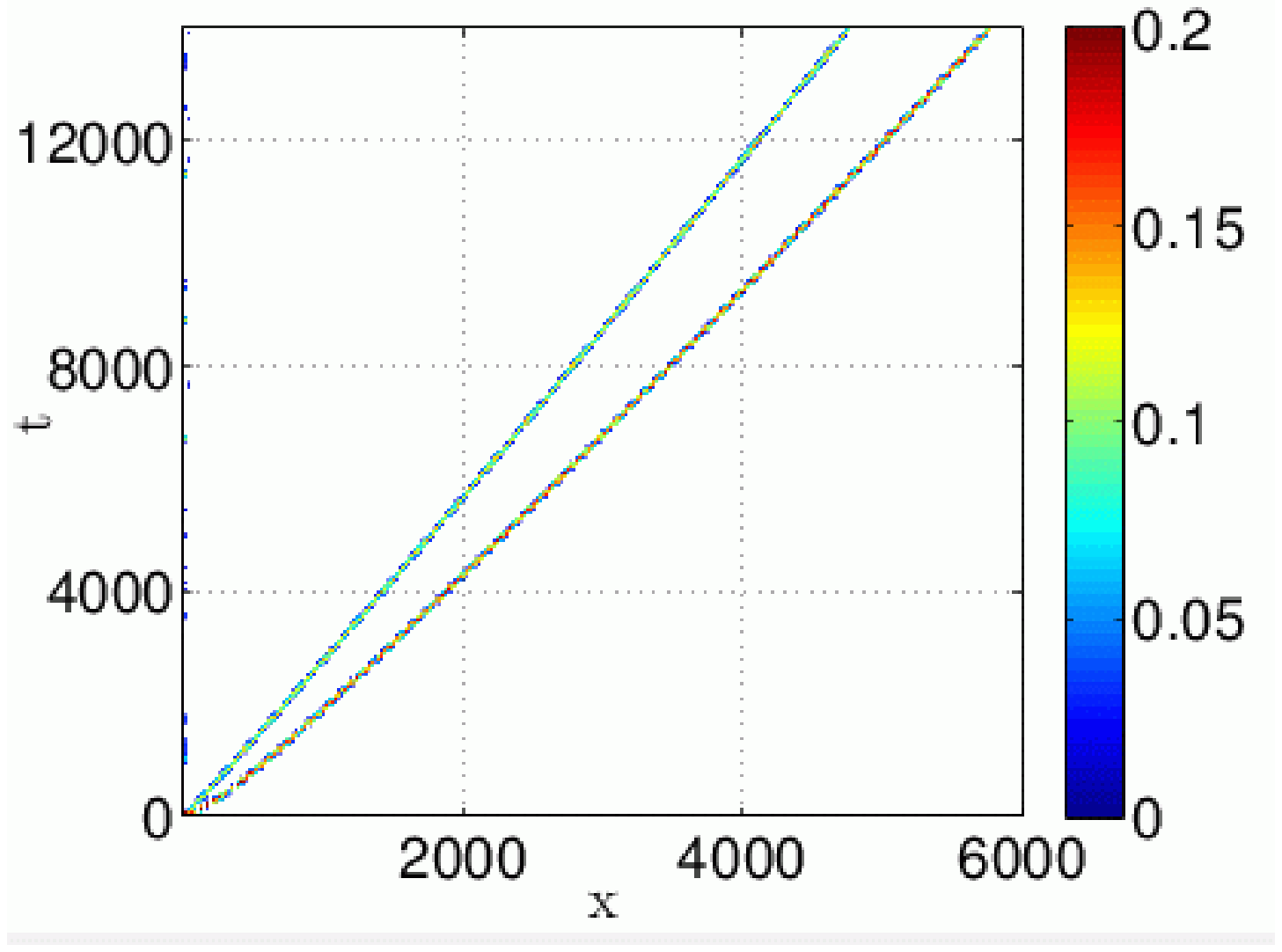}}
{\includegraphics[width=0.48\textwidth]{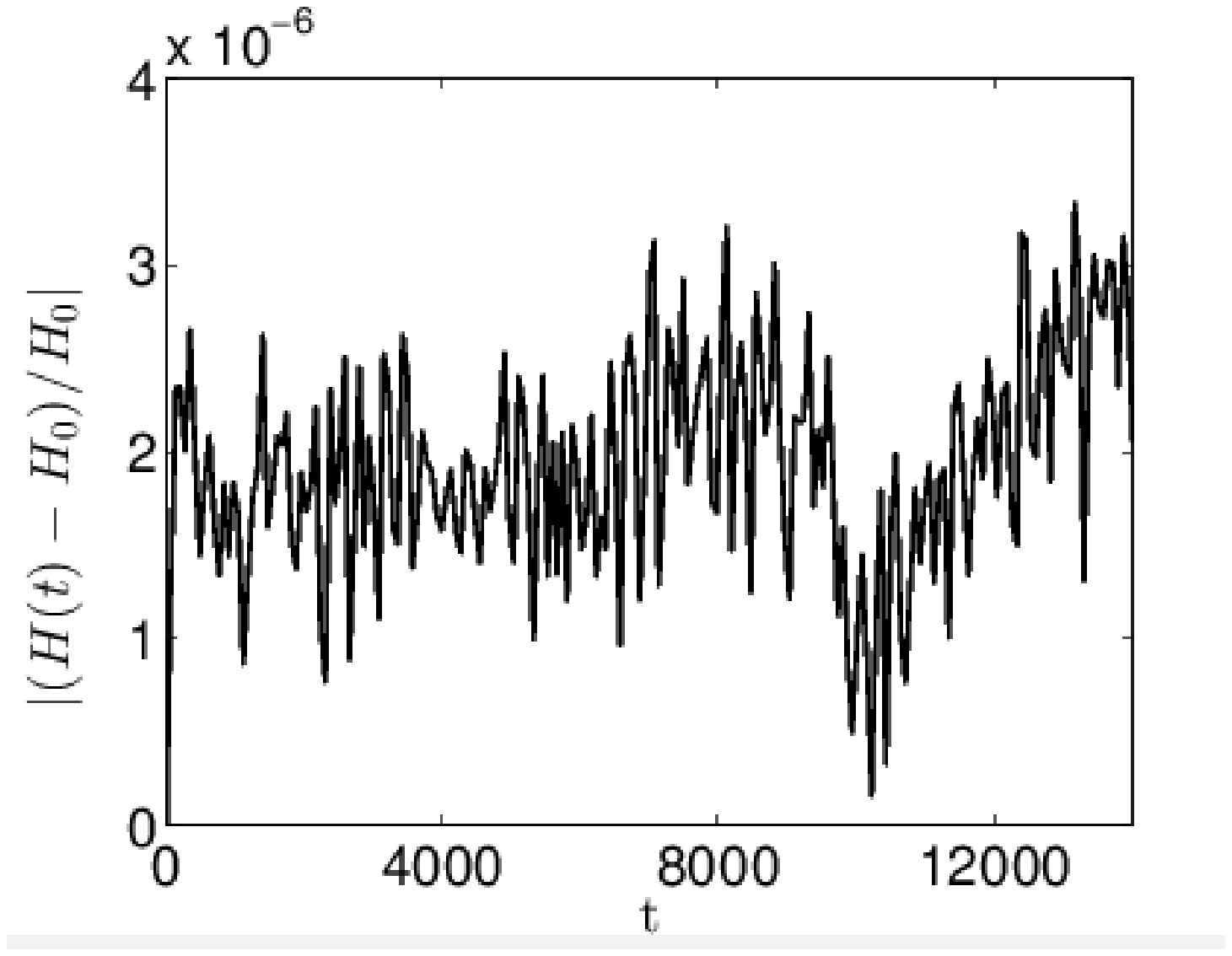}}
\end{center}
\caption{Long time simulation of propagating breather on a long strip
  lattice: $N_{x}=6000$, $N_{y}=40$, $T_{end}=14000$, $U_{0}=2$,
  $u_{x}^{0}=3$ and $u_{y}^{0}=0$. Left: energy density contour plot
  on a lattice line. Right: absolute relative error of total energy in
  time.}\label{fig:LongSim}
\end{figure}

\begin{figure} 
\centering 
{\includegraphics[width=0.48\textwidth]{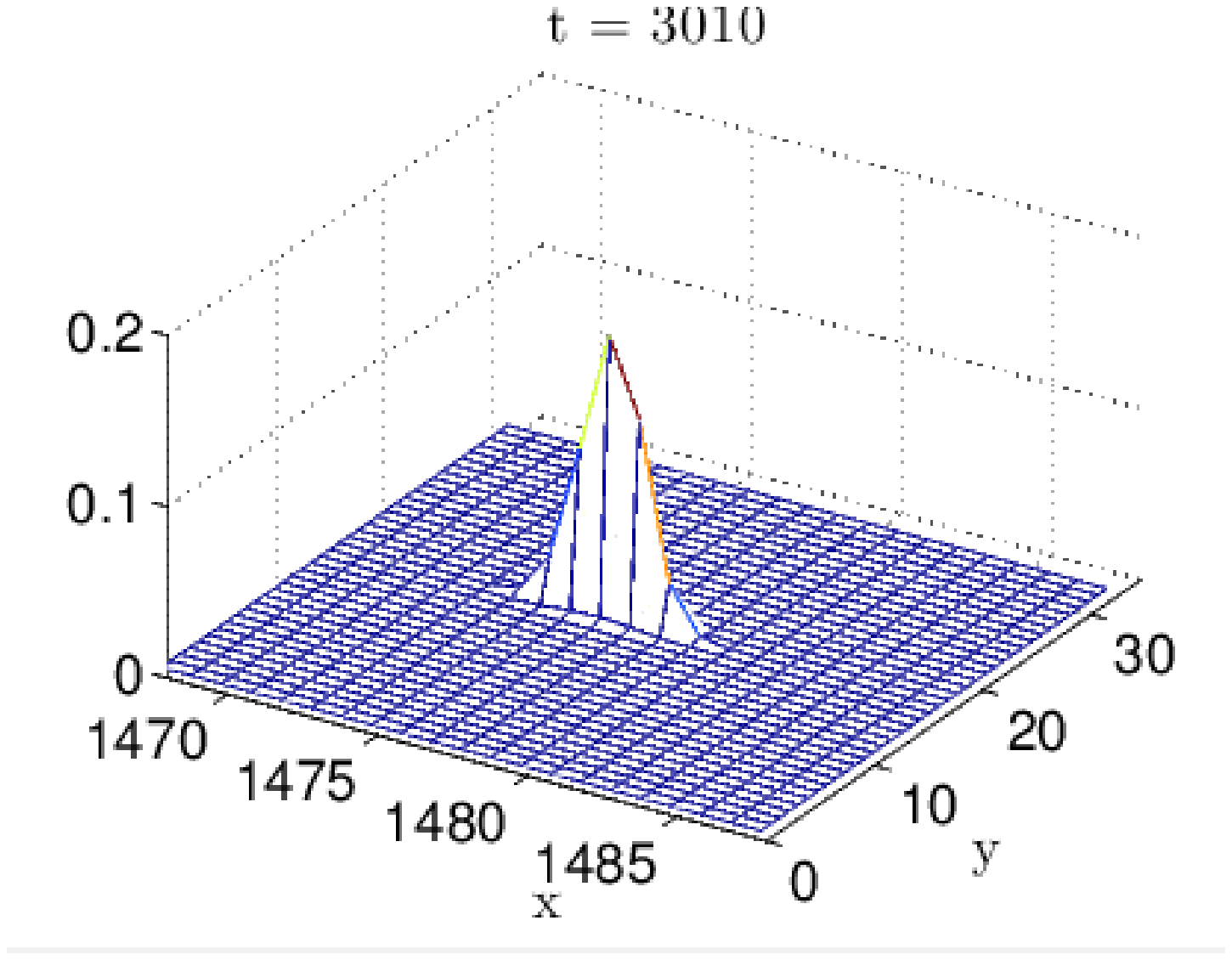}}
{\includegraphics[width=0.48\textwidth]{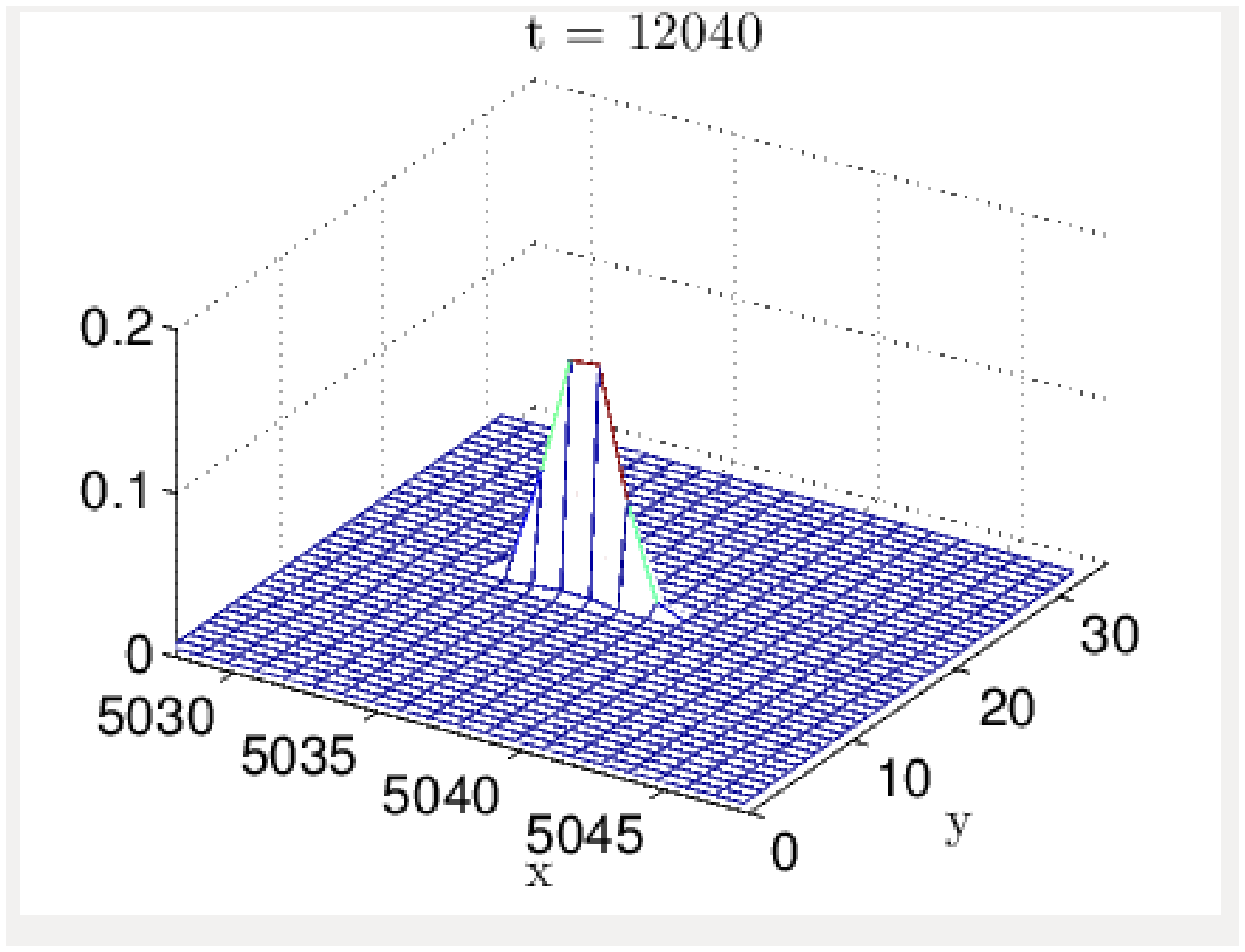}}
\caption{Snapshots of the energy density function of the propagating
  breather solution at two distinct times.  Long strip lattice
  simulation: $N_{x}=6000$, $N_{y}=40$, $T_{end}=14000$, $U_{0}=2$,
  $u_{x}^{0}=3$ and $u_{y}^{0}=0$.}\label{fig:LongSimEn}
\end{figure}

To test the lifespan of the breather solutions, we perform long time
simulations with the same initial conditions and parameter values on a
longer lattice, that is, on a long lattice strip: $N_{x}=6000$ and
$N_{y}=40$.  We integrate in time until $T_{end}=14000$.  In the left
plot of Fig. \ref{fig:LongSim}, we plot the energy density function of
atoms after each $7000$ time steps on the main lattice chain in time.
The result shows the long lifespan of propagating discrete breathers
in crystal model at 0$^\circ$K.  The breather has propagated more than $5000$
lattice sites.  The second curve in the left plot of
Fig. \ref{fig:LongSim} is due to the presence of the second
propagating breather, see the description above.  To see that these
localized energies in the left plot of Fig. \ref{fig:LongSim} are
associated with the discrete breathers, we take snapshots of the
energy density function at two distinct times from the simulation and
show them in Fig. \ref{fig:LongSimEn}.  To confirm the good energy
conservation properties of the Verlet method, we have included in the
right plot of Fig. \ref{fig:LongSim} a graph of absolute relative
error of the total energy in time.  The graph shows that the total
energy stays bounded for long integration times.

We can excite propagating discrete breathers for wide range of initial
kick values.  Taking smaller values for initial kicks leads to
stationary breather solutions.  For very small initial kick there is
no localization and only phonons are produced.  If we keep increasing
the initial kick values, the kink solutions appear, which are the
topic of next section.

{\bf Remark:} Numerical simulations showed that with the same initial
conditions but with larger values of $U_{0}$, the breather gets pinned
to the lattice, but with smaller values of $U_{0}$ very distinctive
horseshoe wave solutions appear, which we will discuss in
Sec. \ref{sec:FrontSol}. Recall that we control the relative strength
of the potentials in dynamics with the parameter value $U_{0}$.

\subsubsection{Numerical results: kink solutions}
\label{sec:KinkSol}
\begin{figure} 
\centering 
{\includegraphics[width=0.32\textwidth]{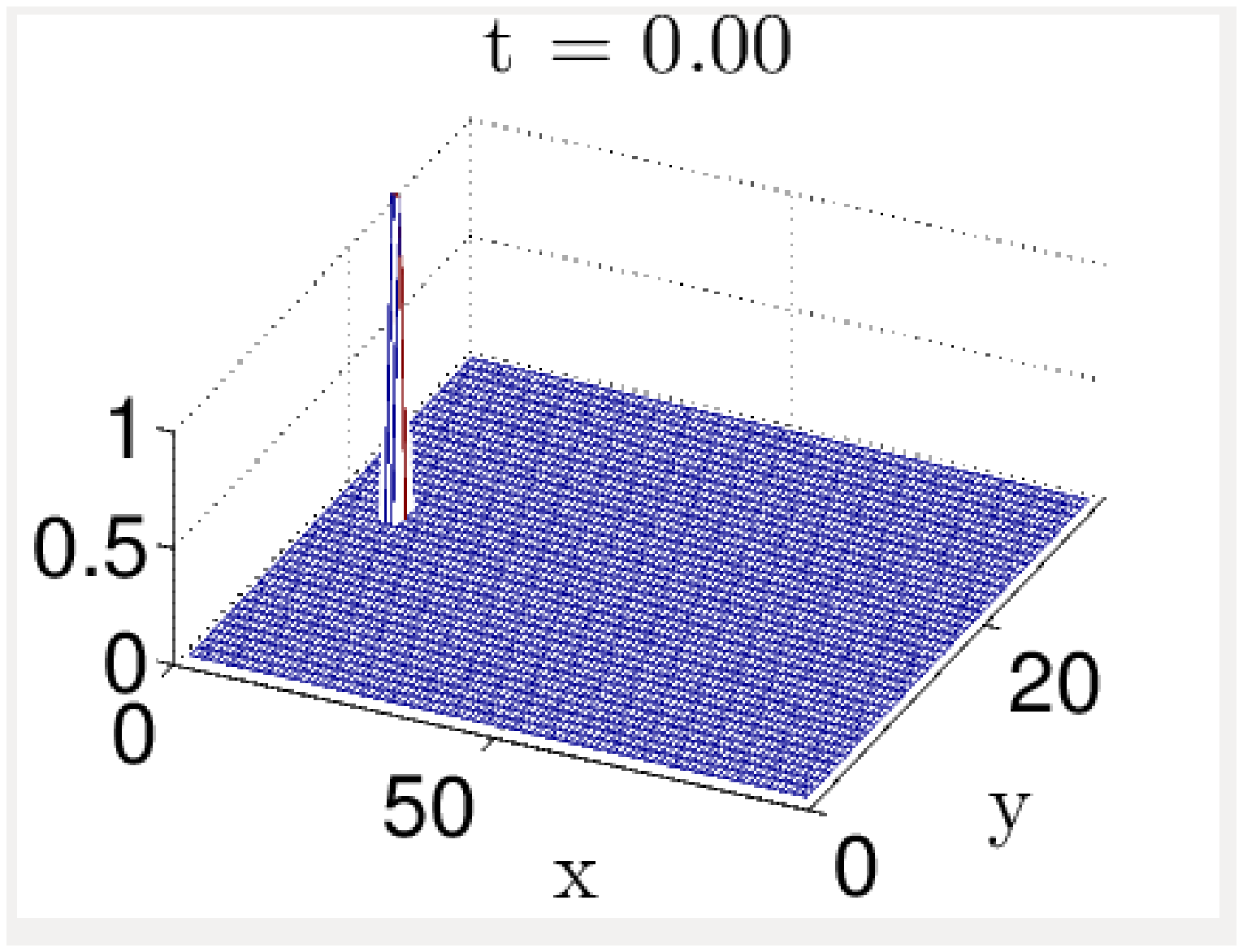}}
{\includegraphics[width=0.32\textwidth]{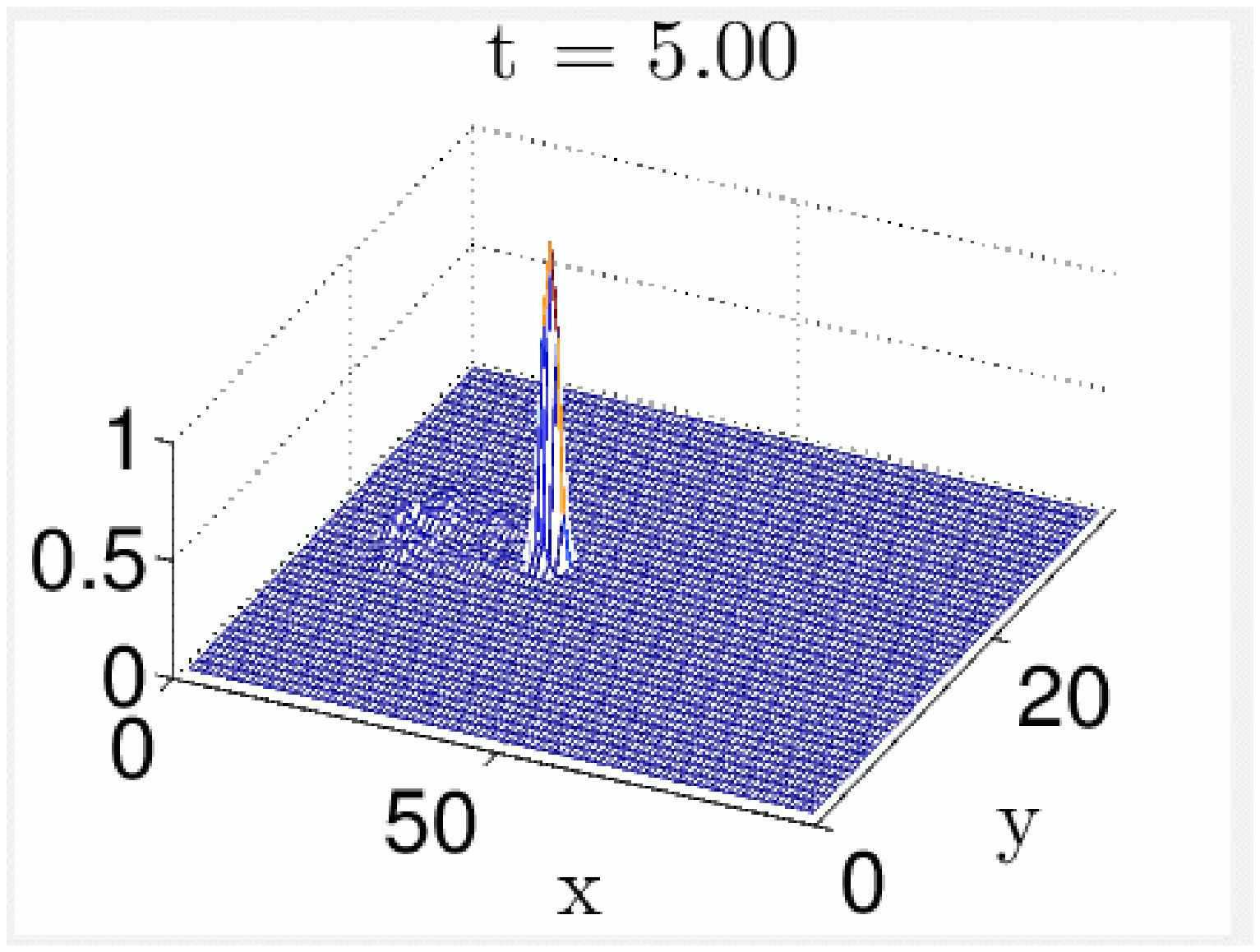}}
{\includegraphics[width=0.32\textwidth]{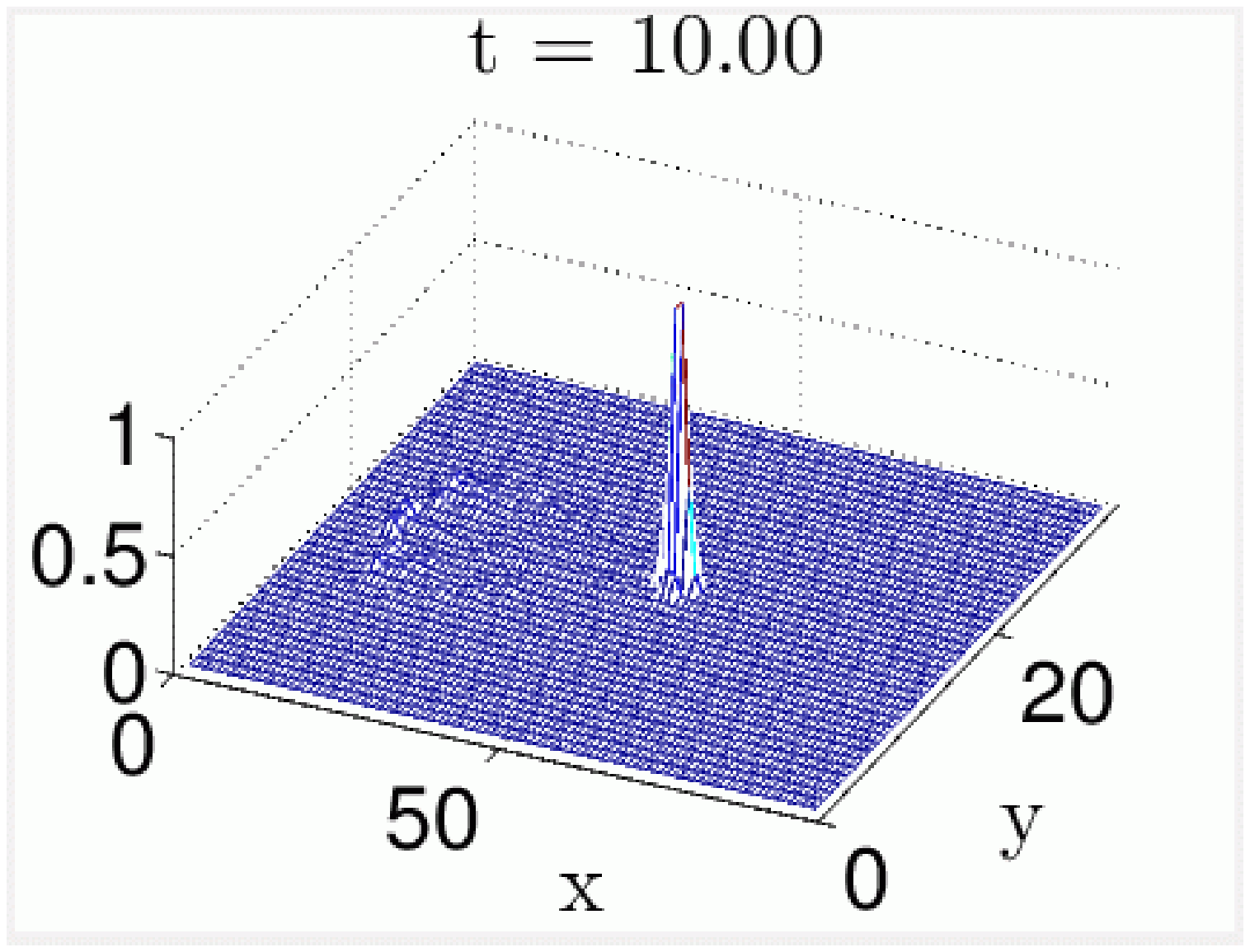}}
{\includegraphics[width=0.32\textwidth]{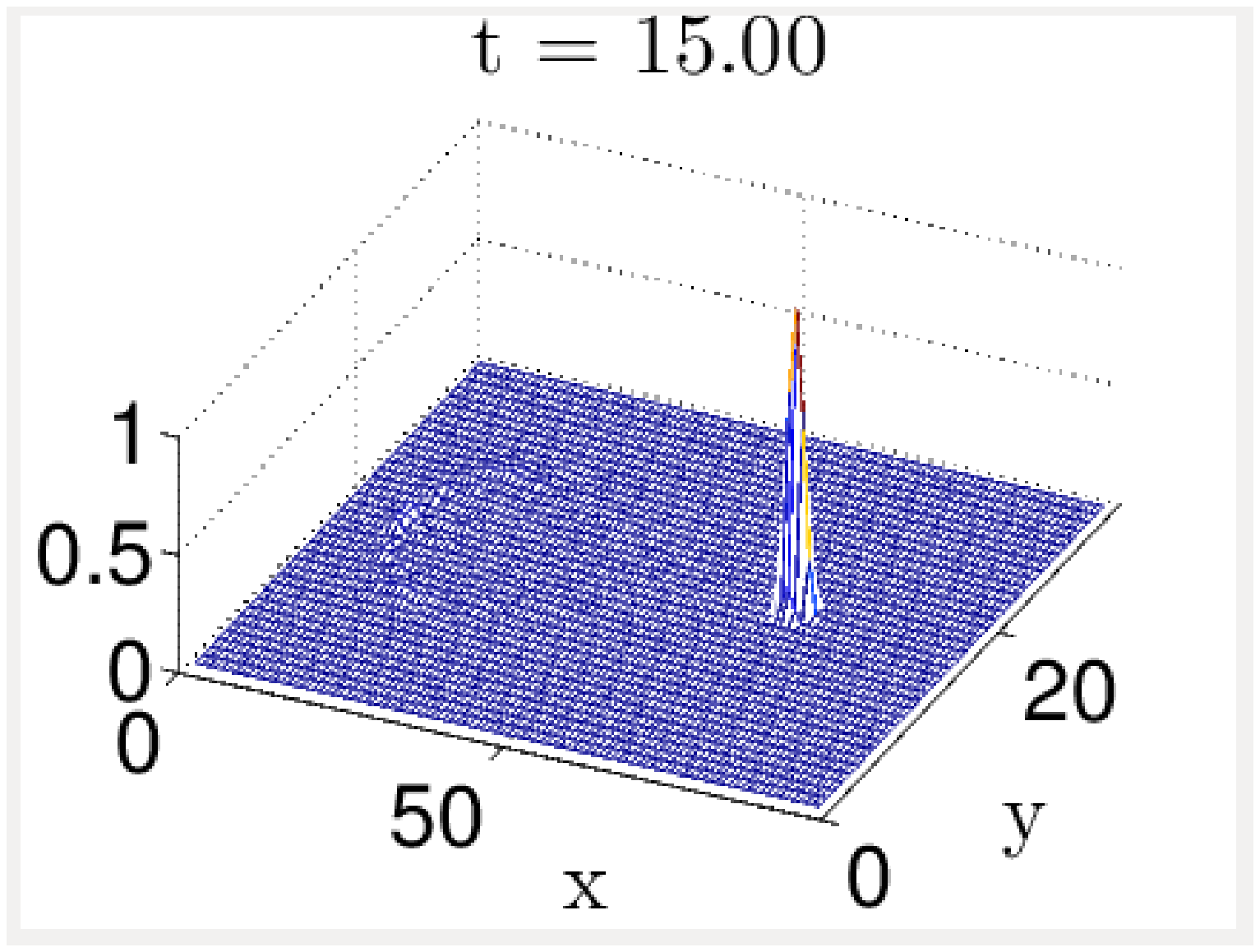}}
{\includegraphics[width=0.32\textwidth]{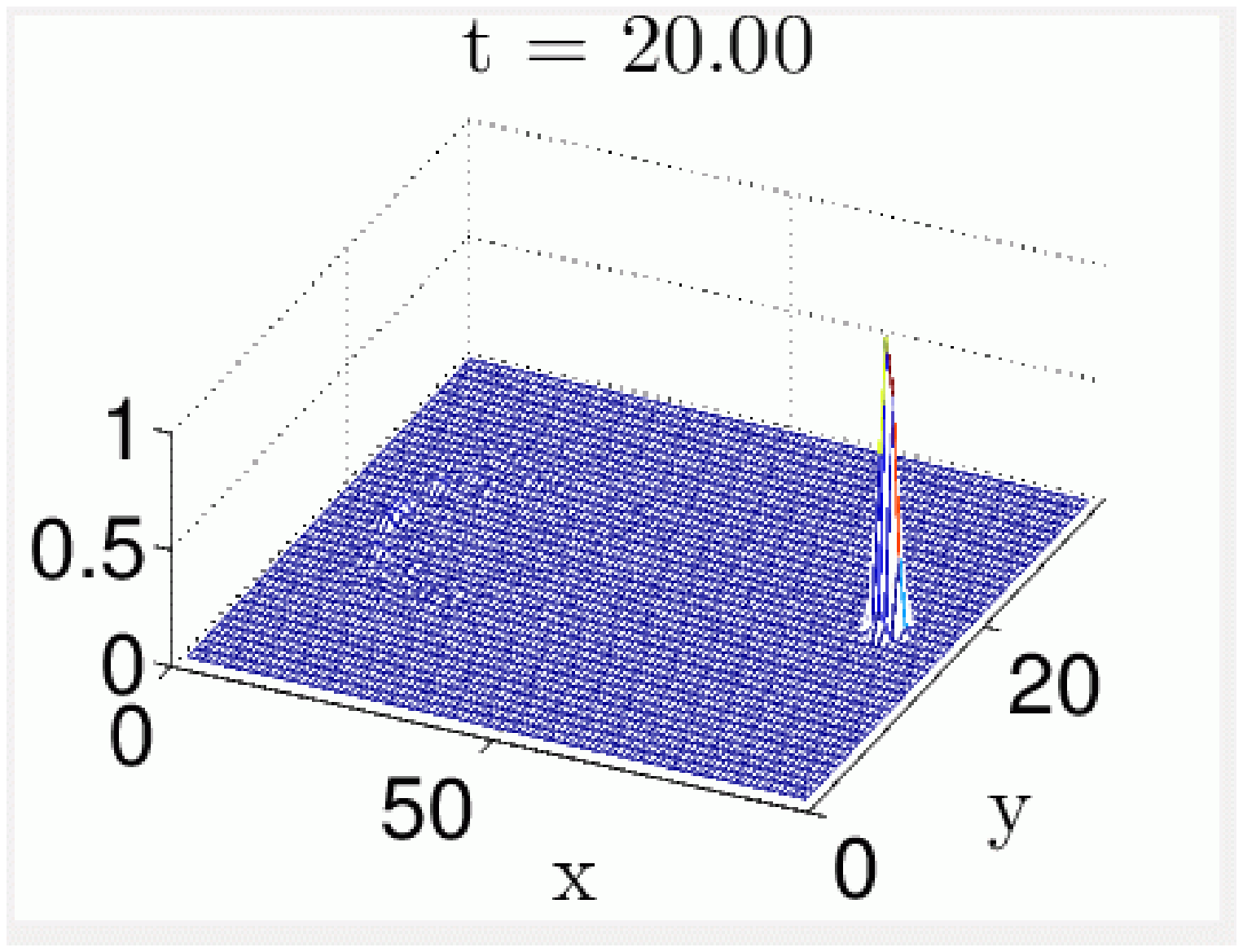}}
{\includegraphics[width=0.32\textwidth]{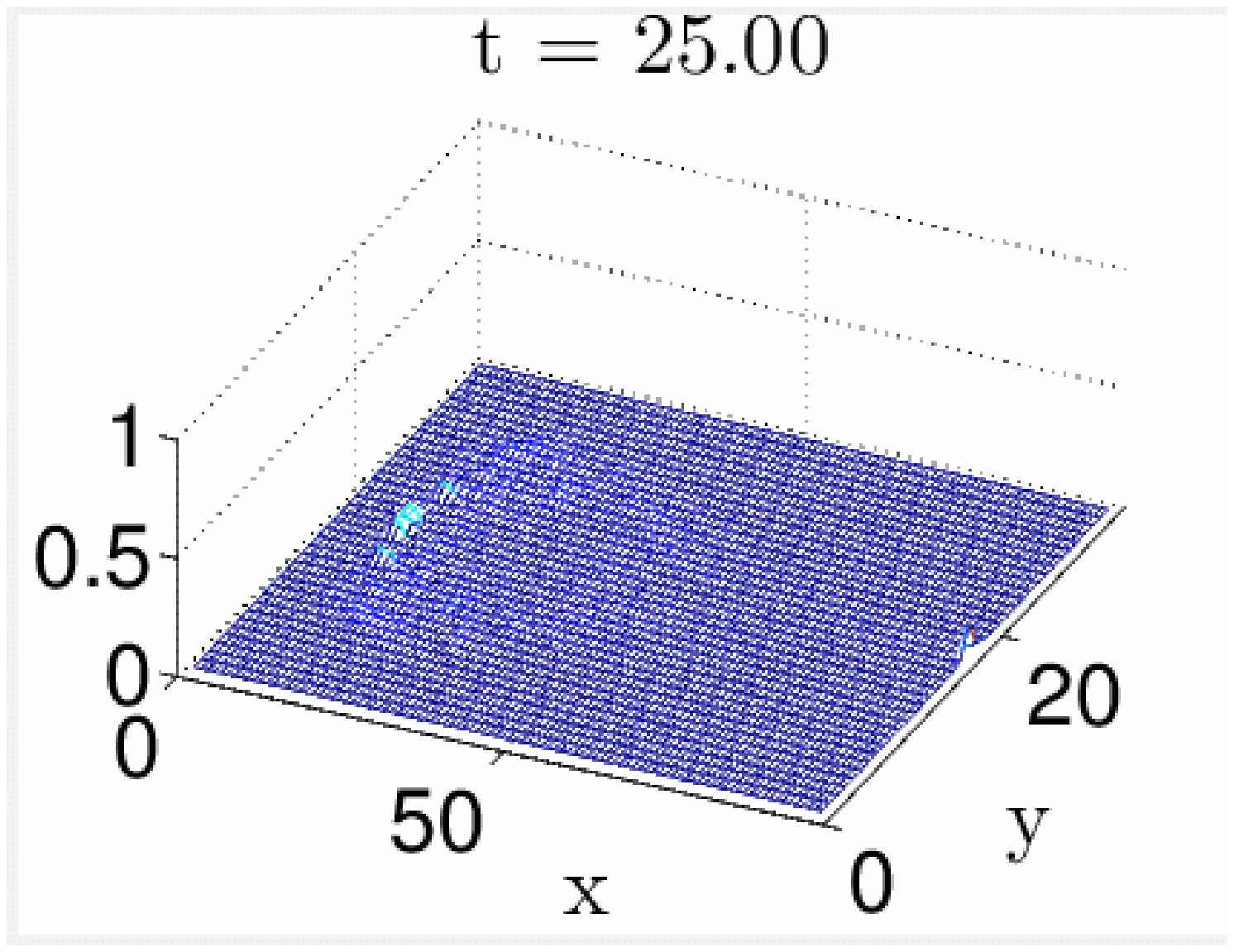}}
\caption{Evolution of the energy density function of the kink
  solution in time. $N_{x}=100$, $N_{y}=40$, $T_{end}=30$, $U_{0}=2$,
  $u_{x}^{0}=5.5$ and $u_{y}^{0}=0$.}\label{fig:KinkEn}
\end{figure}

\begin{figure} 
\centering 
\includegraphics[width=0.48\textwidth]{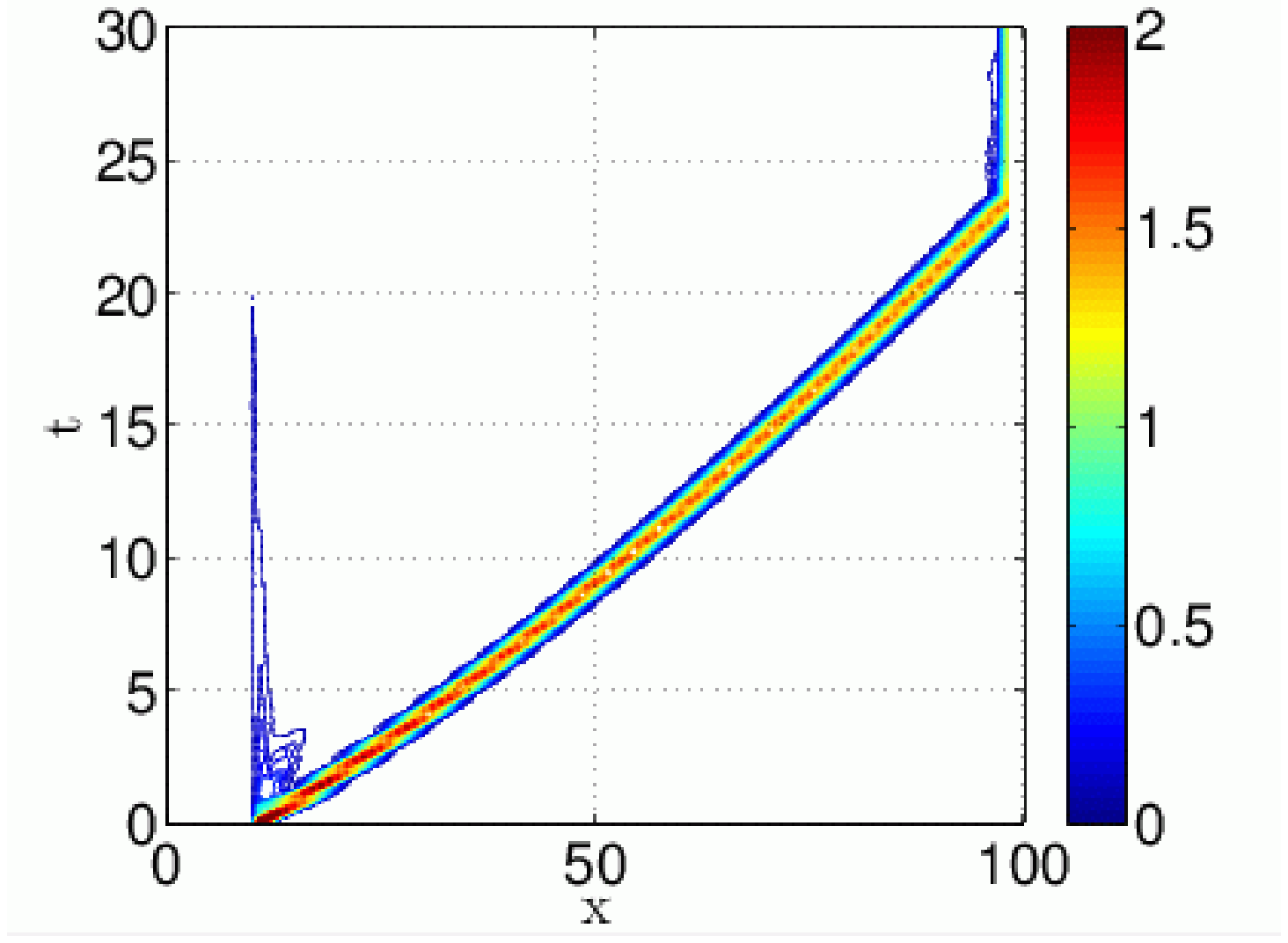}
\includegraphics[width=0.48\textwidth]{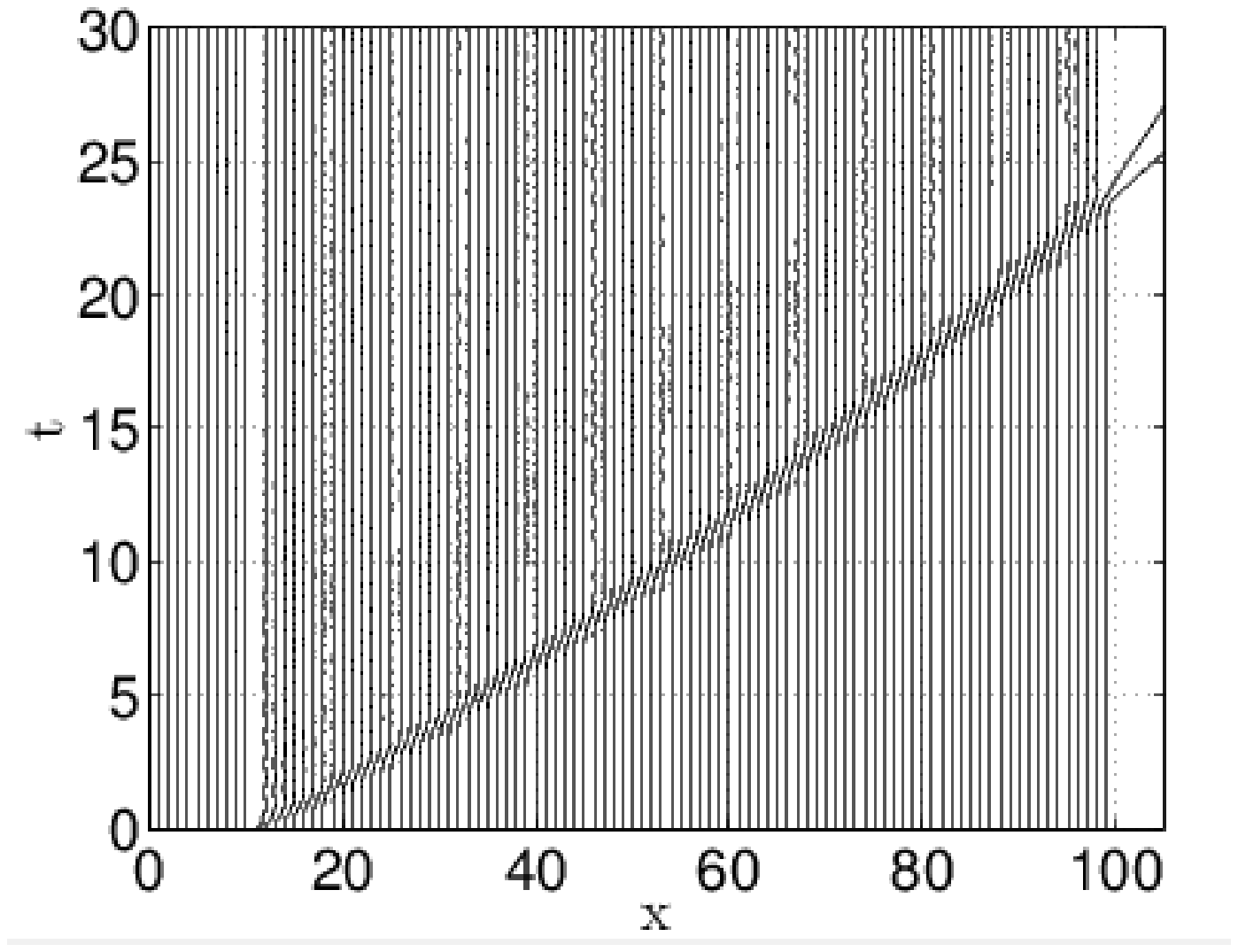}
\caption{Short time simulation of a kink solution, with $N_{x}=100$,
  $N_{y}=40$, $T_{end}=30$, $U_{0}=2$, $u_{x}^{0}=5.5$ and
  $u_{y}^{0}=0$. Left: contour plot of the atomic energy density
  function on a horizontal lattice chain.  Right: atomic displacements
  in the $x$ axis direction from their equilibrium positions in time
  on a horizontal lattice chain.}
\label{fig:Kinks}
\end{figure}

In this section we report on long lived kink solutions.  For fixed
value $U_{0}=2$ we keep increasing the initial velocity value of the
kick.  In the first numerical simulation, we consider a lattice with
$N_{x}=100$ and $N_{y}=40$.  The initial velocity kick values are
$u_{x}^{0}=5.5$ and $u_{y}^{0}=0$.  Such kicks produces a kink
solution propagating on a horizontal chain of atoms.  In Figure
\ref{fig:KinkEn} we show evolution of kink's energy in time.  We
integrate in time until $T_{end}=30$.  Shortly before $25$ time units,
the kink has approached the boundary and ejects two atoms from the
lattice.  That can be seen in the left plot of Fig. \ref{fig:Kinks}
where we plot the energy density function of atoms on the main chain
along which the kink propagates, after each $10$ time steps in time.
In the right plot of Fig. \ref{fig:Kinks} we plot atomic displacements
in the $x$ axis direction.  Note the fundamental difference between
breather and kink solutions.  The kink solution is carried by the
atoms from one unit cell to other, while a propagating breather passes
through the lattice without atoms leaving their unit cells.  Thus kink
solutions may form vacancies inside the lattice as evident from the
right plot of Fig. \ref{fig:Kinks}.

\begin{figure} 
\centering 
\includegraphics[width=0.48\textwidth]{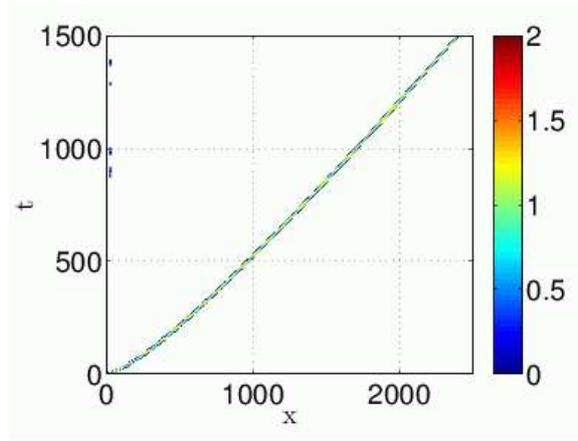}
\caption{Energy density function of a long time simulation of a
  propagating kink on a long strip lattice: $N_{x}=2500$, $N_{y}=40$,
  $T_{end}=1500$, $U_{0}=2$, $u_{x}^{0}=5.5$ and
  $u_{y}^{0}=0$.}\label{fig:LongKink}
\end{figure}

In Section \ref{sec:BrSol} we demonstrated the long lifespan of
propagating discrete breather solutions, see left plot of
Fig. \ref{fig:LongSim}.  We find that our model also supports
long-lived kink solutions.  For long-lived kink simulations, we
consider long strip lattice: $N_{x}=2500$ and $N_{y}=40$.  With the
same parameter values and initial conditions we integrate in time
until $T_{end}=1500$.  In Figure \ref{fig:LongKink}, we plot the
kink's energy in time after each $750$ time steps of the main lattice
chain. The kink has propagated over more than $2000$ lattice sites and
has not collapsed during the whole computational time window.

What about the argument above that a kink should not be able to
propagate in 2D?  In these solutions the essential feature is that the
``side wall'' of the kink has zero energy - the atoms on the main
chain have moved exactly $\sigma$ before and after the kink passes.
So displacements across the wall is zero. If the wall was wider, then
it would have finite energy, and we would {\em not} observe this
phenomena.

{\bf Remark:} If we keep the same initial condition but increase the
value of $U_{0}$, the kink disappears.  For a kink to appear again we
have to increase the initial velocity kick value $u_{x}^{0}$. On the
another hand if we keep the same initial condition but decrease the
value of $U_{0}$, the kink disappears too. Instead horseshoe wave
solutions appear, see Sec. \ref{sec:FrontSol}.

\subsubsection{Numerical results: horseshoe wave solutions}
\label{sec:FrontSol}
So far we have considered constant value of $U_{0}=2$. The parameter
$U_{0}$ controls the relative strength between two potentials
considered, i.e.~the atom-atom interaction and the on-site
potential.  In this section we perform numerical study with smaller
value of $U_{0}$, which lead to the observation of horseshoe wave
solutions.

\begin{figure} 
\centering 
{\includegraphics[width=0.32\textwidth]{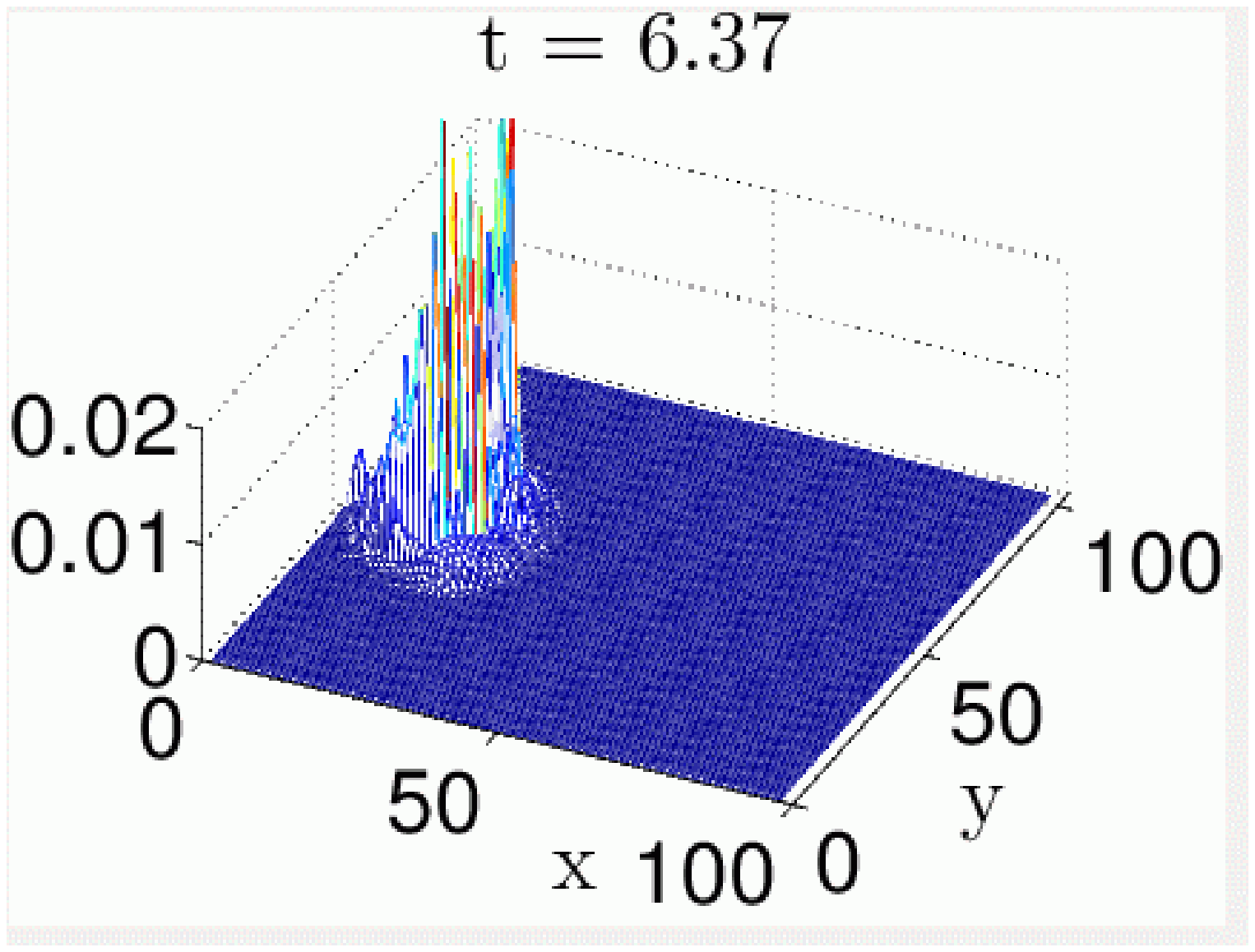}}
{\includegraphics[width=0.32\textwidth]{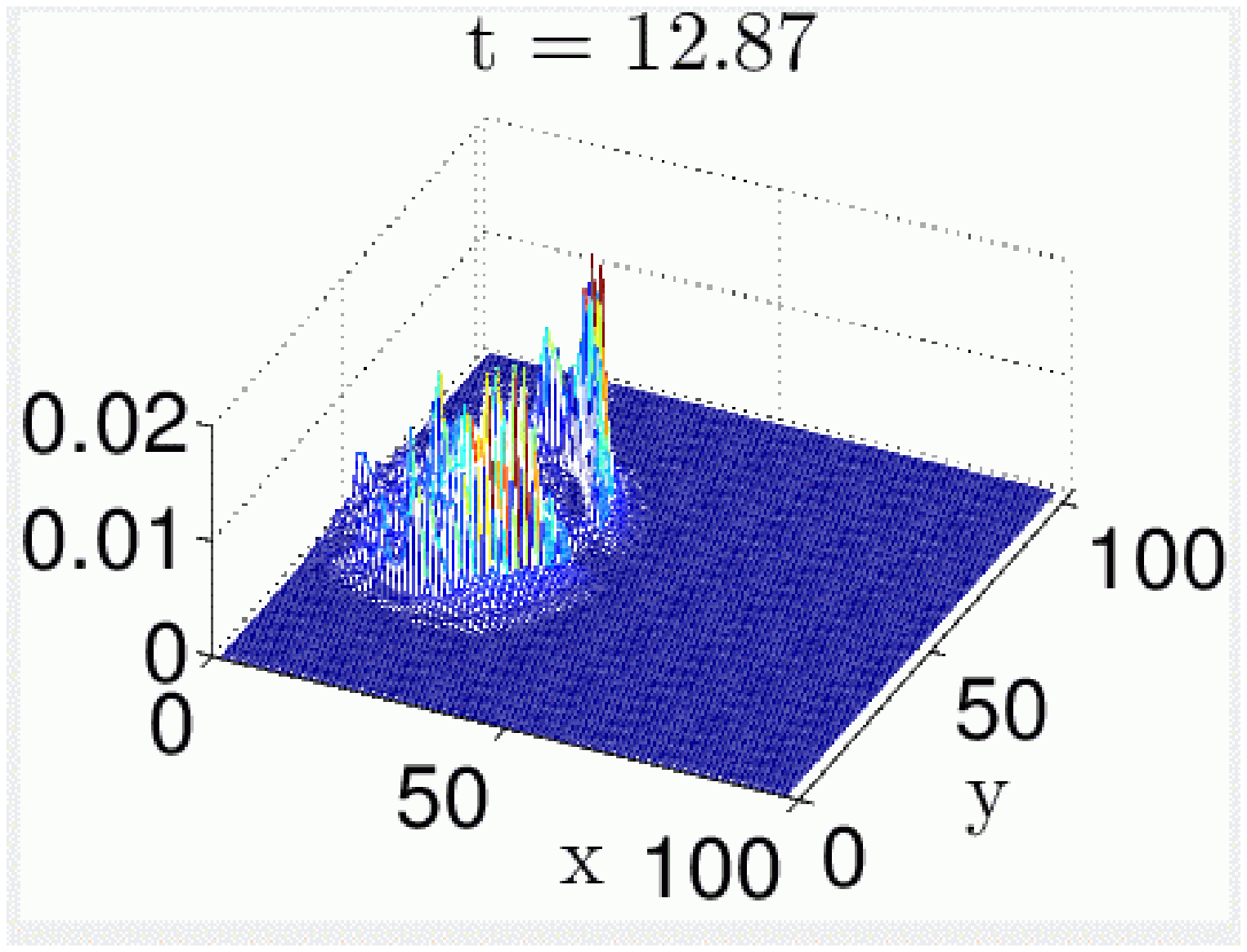}}
{\includegraphics[width=0.32\textwidth]{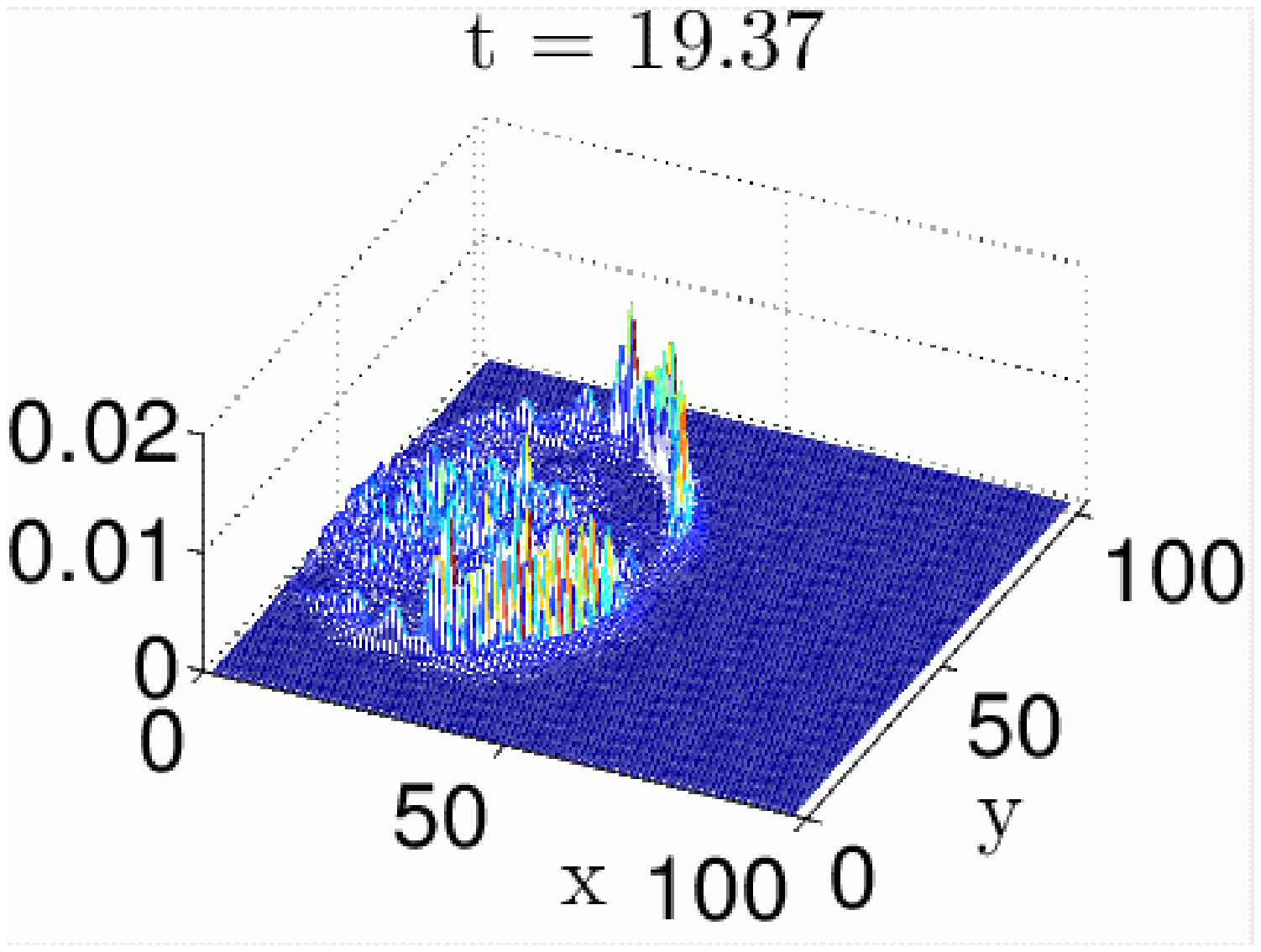}}
{\includegraphics[width=0.32\textwidth]{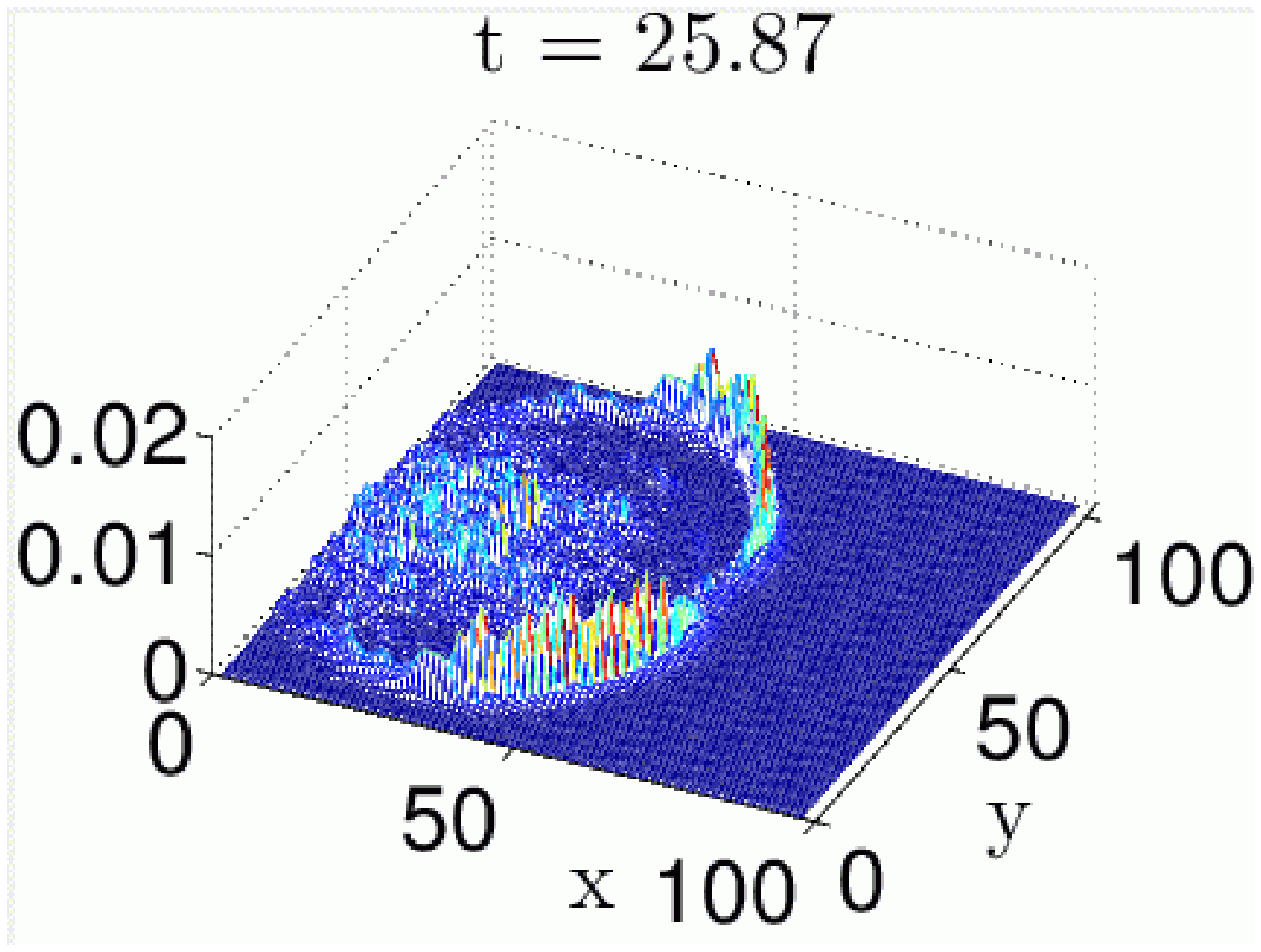}}
{\includegraphics[width=0.32\textwidth]{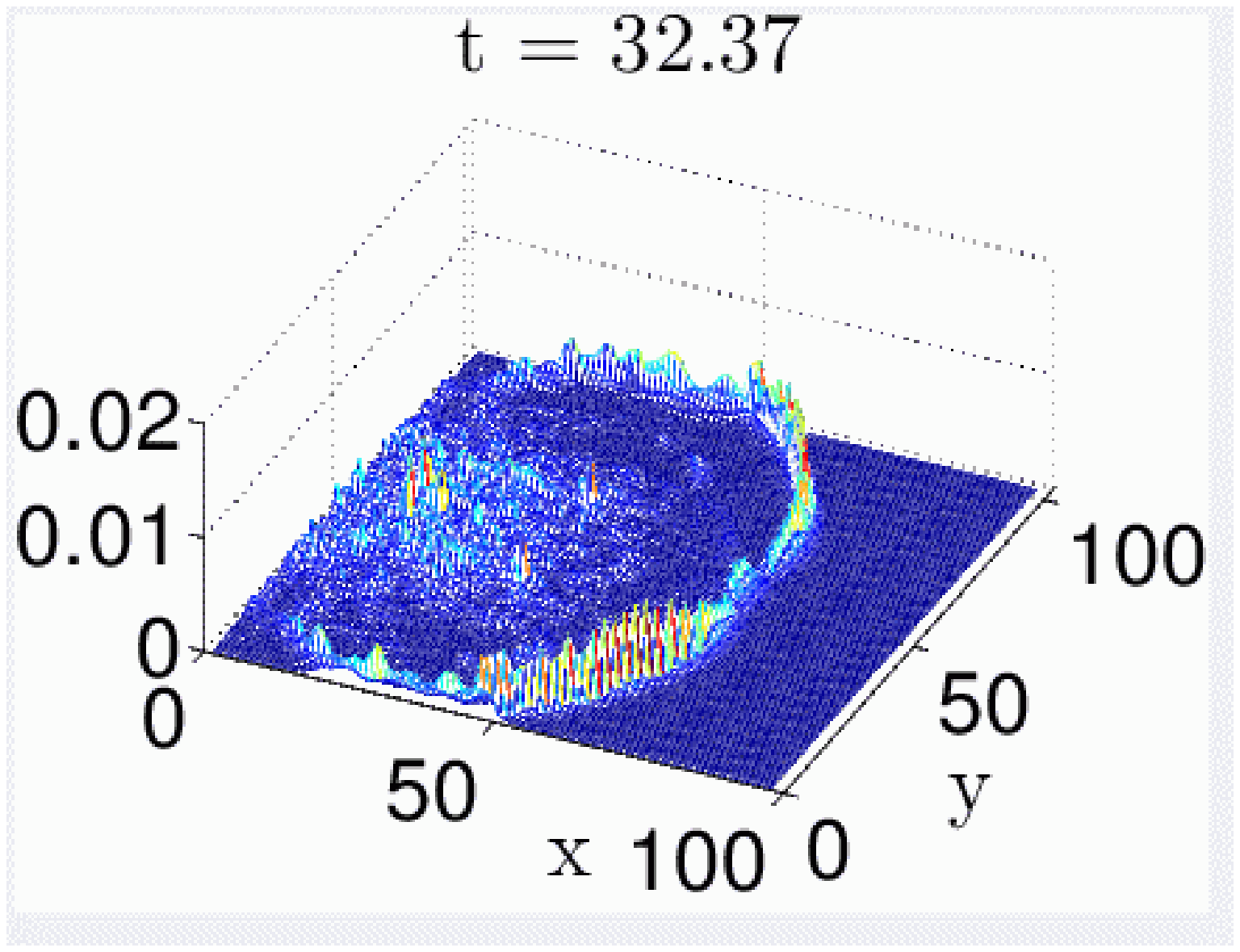}}
{\includegraphics[width=0.32\textwidth]{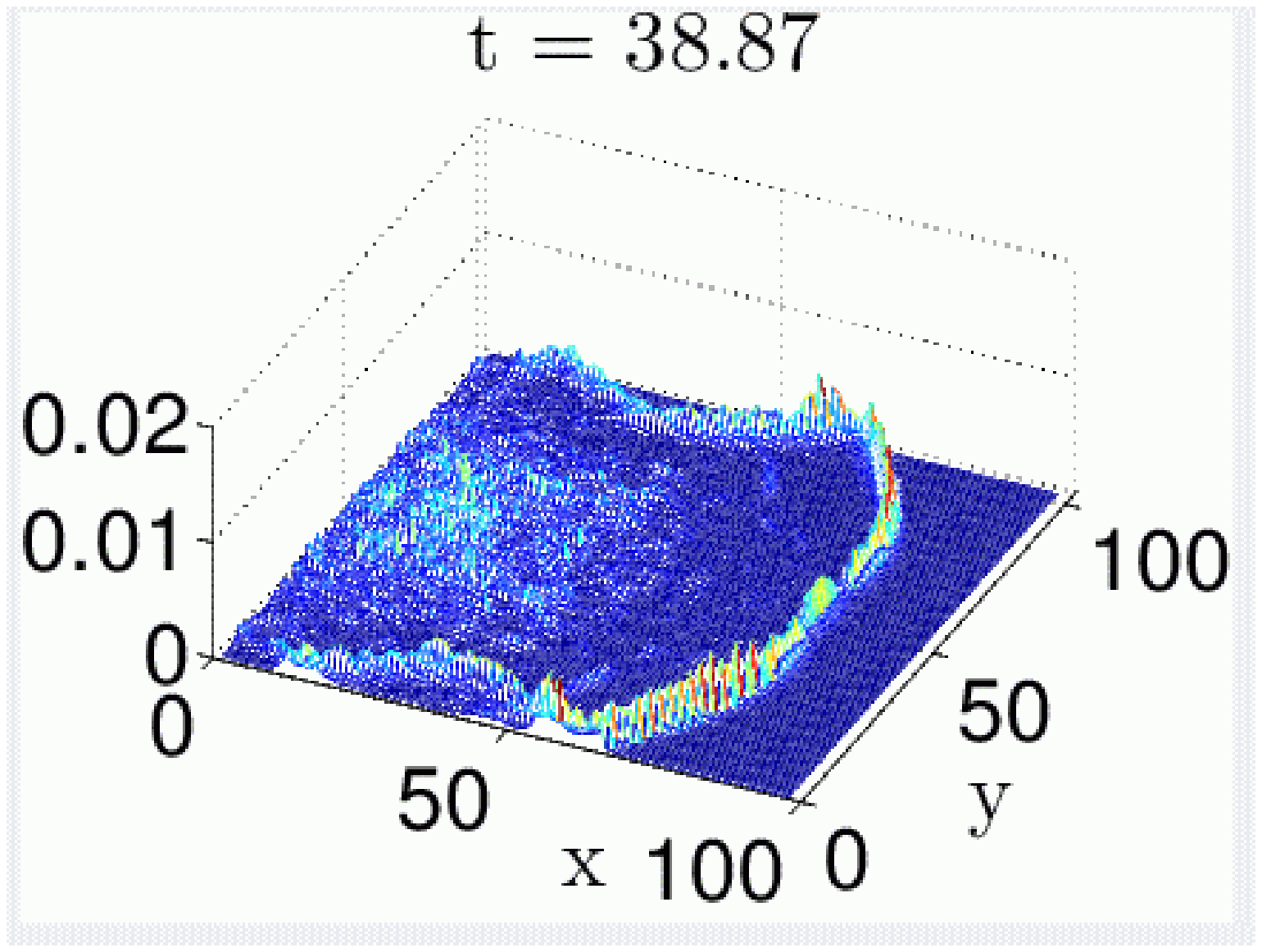}}
\caption{Evolution of the energy density function in time for the
  horseshoe wave solution. $N_{x}=100$, $N_{y}=120$, $T_{end}=52$,
  $U_{0}=0.1$, $u_{x}^{0}=3.0$ and
  $u_{y}^{0}=0$.}\label{fig:FrontWave}
\end{figure}

For this numerical test we consider a lattice: $N_{x}=100$ and
$N_{y}=120$, and the same initial kicks which led to the observation
of propagating breather solutions in Sec. \ref{sec:BrSol}, that is,
$u_{x}^{0}=3.0$ and $u_{y}^{0}=0$.  We integrate in time until
$T_{end}=52$ with $U_{0}=0.1$.  In Figure \ref{fig:FrontWave} we show
the evolution of the energy density function in time.  From the energy
plots, we observe circular propagating wave spreading in all directions
until it hits the boundaries.

\begin{figure}
\centering 
\includegraphics[width=0.48\textwidth]{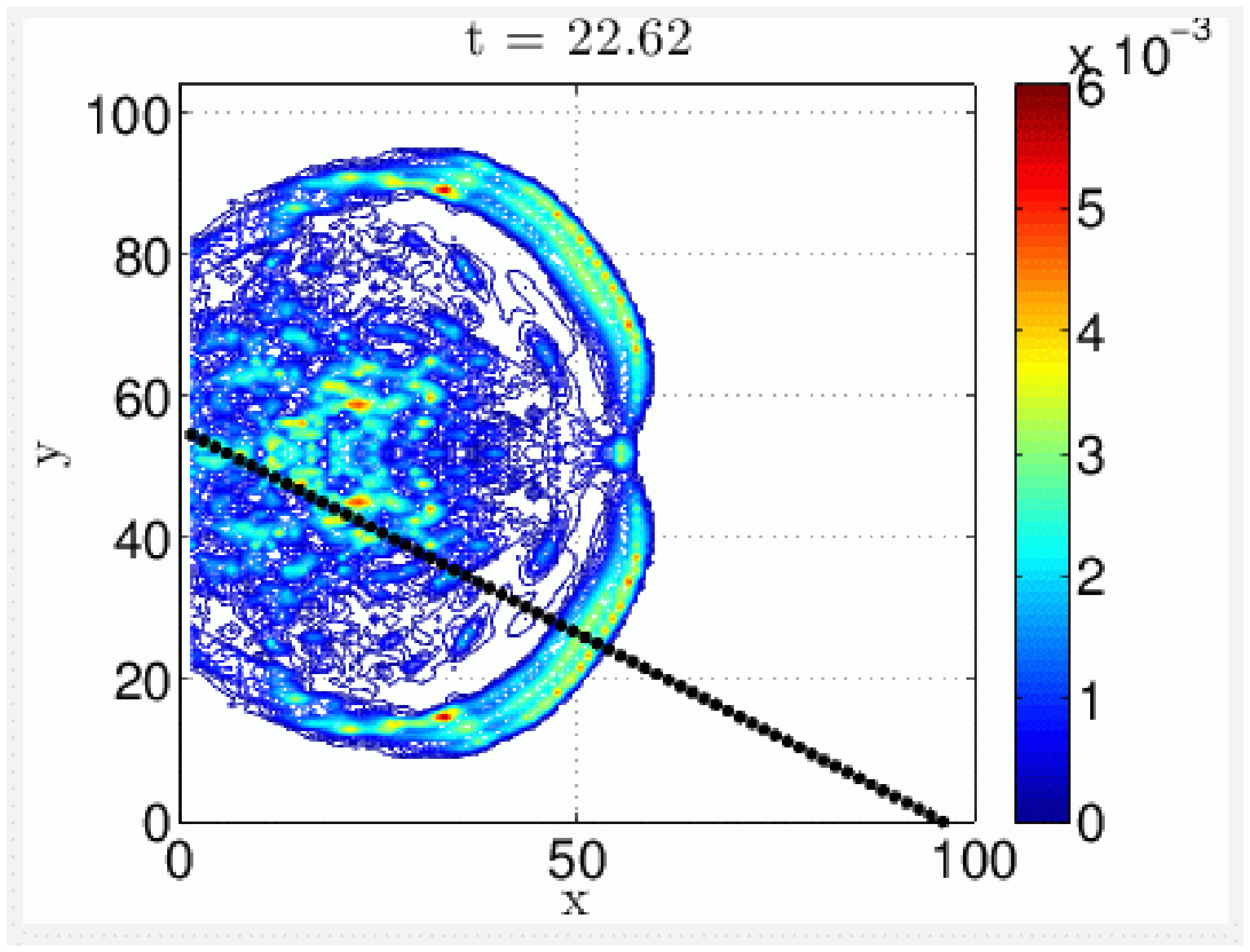}
\includegraphics[width=0.48\textwidth]{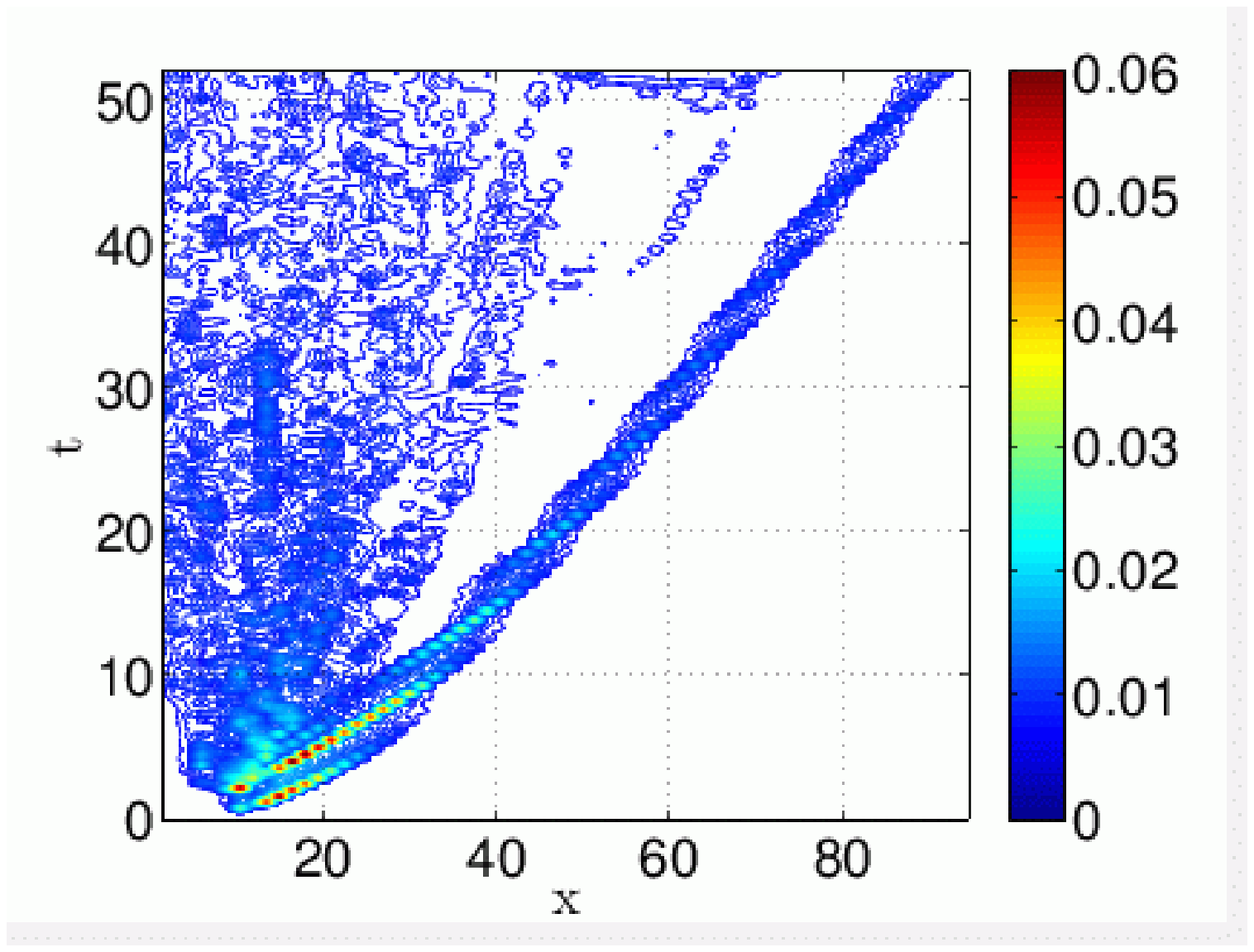}
\caption{Contour plots of the energy density function. Left: over the
  whole domain at time $t=22.62$. Dark dots indicate a chain of
  atoms. Right: over the cross-section of the front wave in time, that
  is, energy of the atoms on the chain shown in the left
  plot. $N_{x}=100$, $N_{y}=120$, $T_{end}=52$, $U_{0}=0.1$,
  $u_{x}^{0}=3.0$ and $u_{y}^{0}=0$.}\label{fig:FrontWaveEn}
\end{figure}

At fixed time we make a contour plot of the energy density function,
see the left plot of Fig. \ref{fig:FrontWaveEn}.  From this image it
becomes evident that the wave adopts a horseshoe shape.  We are
interested in understanding the properties of the front wave of the
horseshoe wave solutions.  We find that the cross-section of the front
wave is a breather solution.  We consider a chain of atoms (assuming
perpendicular to the front) shown by the dots in the left plot of
Fig. \ref{fig:FrontWaveEn} and show their energy density in time after
each $13$ time steps in the right plot of Fig. \ref{fig:FrontWaveEn}.
The particular chain of atoms is perpendicular to the
$(1/2,\sqrt{3}/2)^T$ crystallographic lattice direction and makes
$-30\,^{\circ}$ with the $x$ axis. The right plot of
Fig. \ref{fig:FrontWaveEn} confirms the propagating breather
characteristics of the front wave of the horseshoe wave solution.

\subsubsection{Numerical results: in-line collisions}
\label{sec:Coll}
In this section we study in-line breather-breather, kink-kink and
breather-kink collisions.  To initiate both types of wave
propagations, we excite two atoms in the lattice, that is, we give
initial velocity kicks to two atoms on the same lattice chain of
atoms.  The left atom initial velocity kick is $u_{x}^{0}$ and
$u_{y}^{0}$, and the right atom velocity kick is $u_{x}^{1}$ and
$u_{y}^{1}$.  We start with the rest of the lattice in its mechanical
equilibrium state.  In all the following numerical experiments,
$N_{x}=200$ and $N_{y}=40$, $U_{0}=2$ and $u_{y}^{0}=u_{y}^{1}=0$.

\begin{figure}
\centering 
\includegraphics[width=0.48\textwidth]{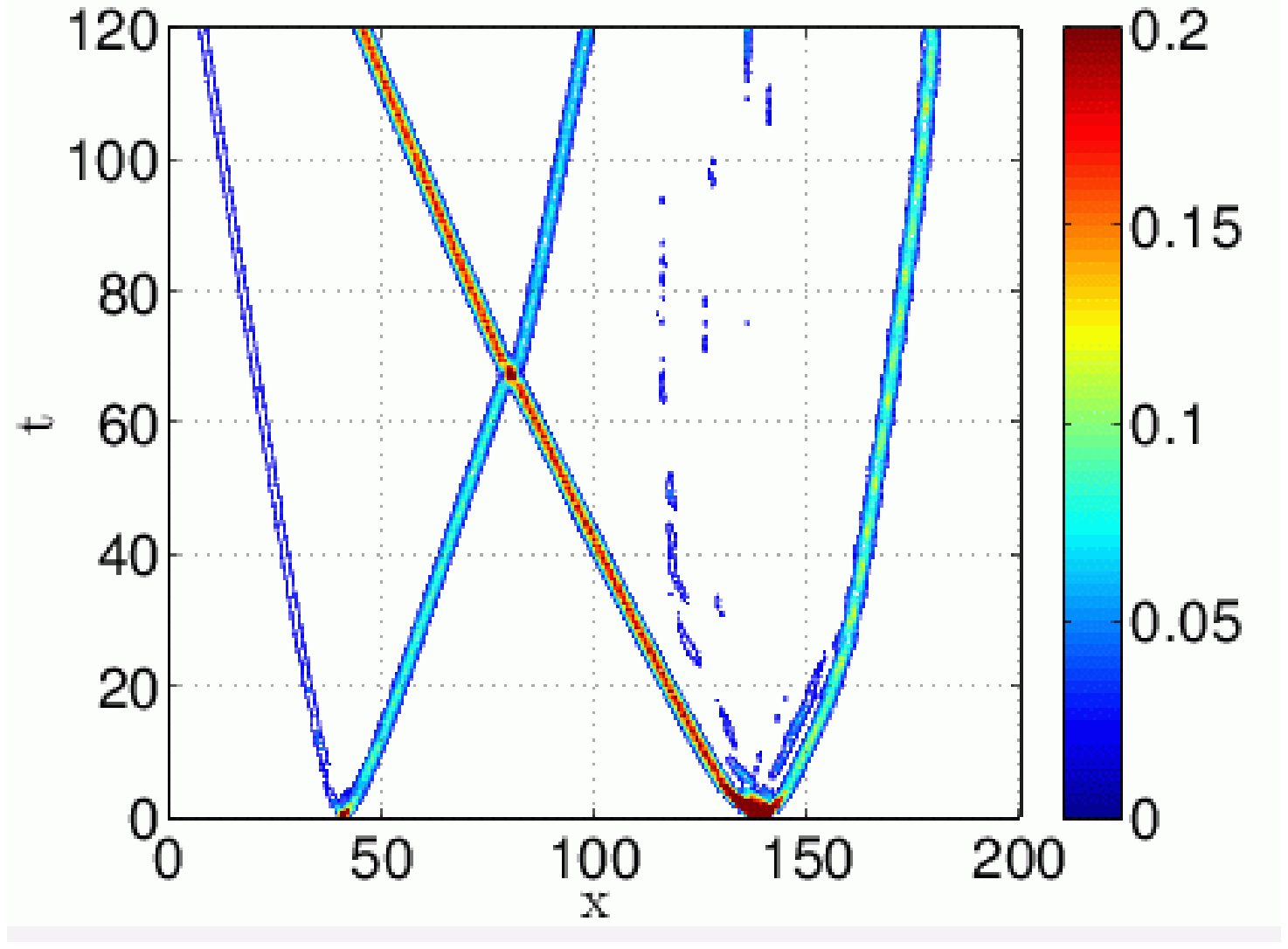}
\includegraphics[width=0.48\textwidth]{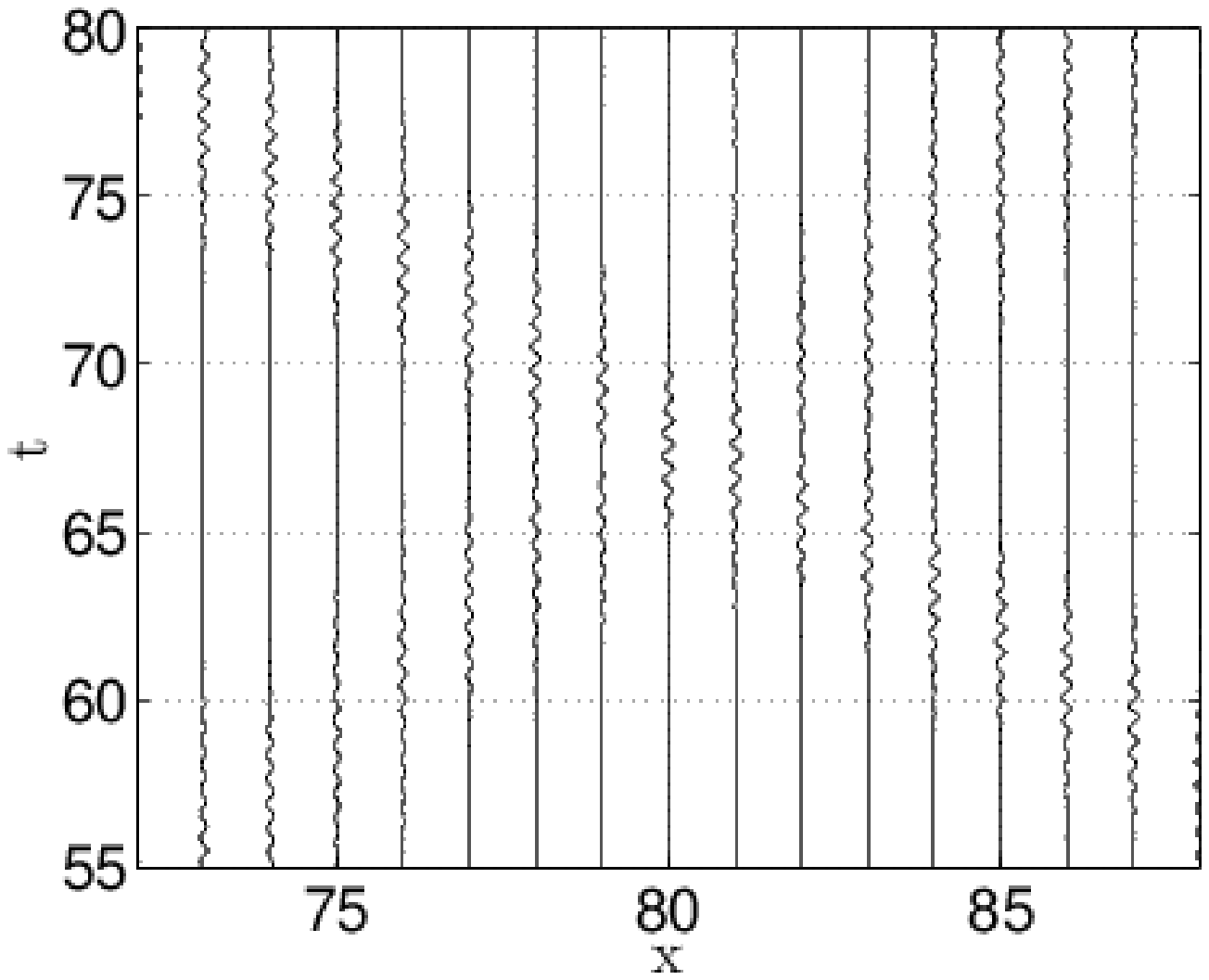} 
\caption{In-line collision of two propagating breathers. $N_{x}=200$,
  $N_{y}=40$, $T_{end}=120$, $U_{0}=2$, $u_{y}^{0}=u_{y}^{1}=0$,
  $u_{x}^{0}=1$ and $u_{x}^{1}=-3.5$. Left: contour plot of the energy
  density function in time on the lattice line. Right: atomic
  displacement plot in the $x$ axis direction on the lattice line
  during the collision.}\label{fig:BBcoll}
\end{figure}

For our first example we consider in-line breather-breather collision
with initial kicks: $u_{x}^{0}=1$ and $u_{x}^{1}=-3.5$.  Integration
in time is performed until $T_{end}=120$, see
Fig. \ref{fig:BBcoll}. In the left plot of Fig. \ref{fig:BBcoll}, we
show energy density function in time after each $20$ time steps on the
main chain of atoms. Both kicks have produced two propagating breather
solutions moving in opposite directions.  All four breathers have
different energies as can be seen by the colours.  After $60$ time
units, two middle breathers collide and pass through each other,
exchanging some energy in the process.  Evidently, the breather coming
from the left has lost some of its initial velocity and propagates
slower.  The displacement plot of atoms in the $x$ axis direction during
the collision can be seen in the right plot of Fig. \ref{fig:BBcoll}.

\begin{figure}
\centering 
\includegraphics[width=0.48\textwidth]{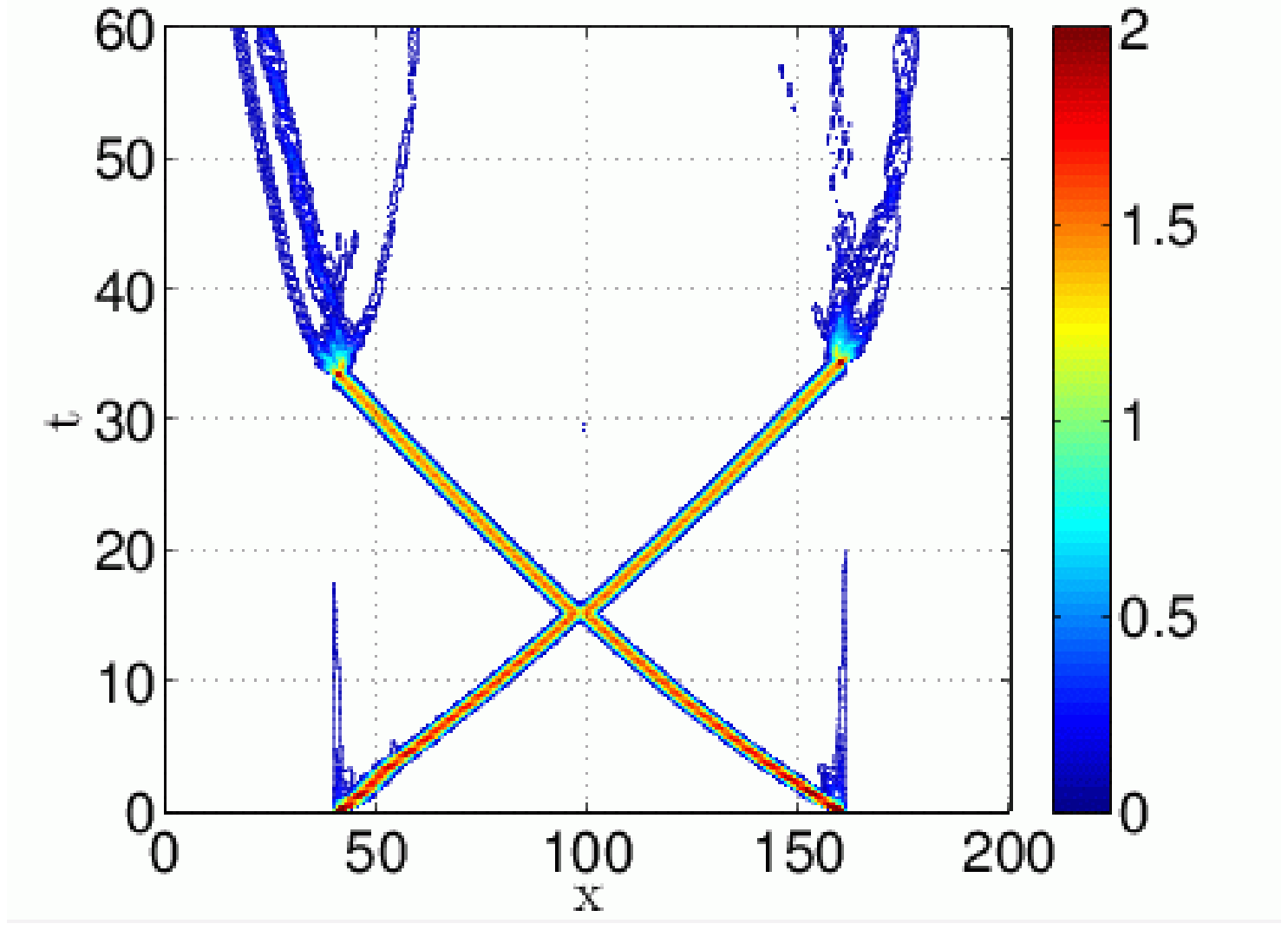}
\includegraphics[width=0.48\textwidth]{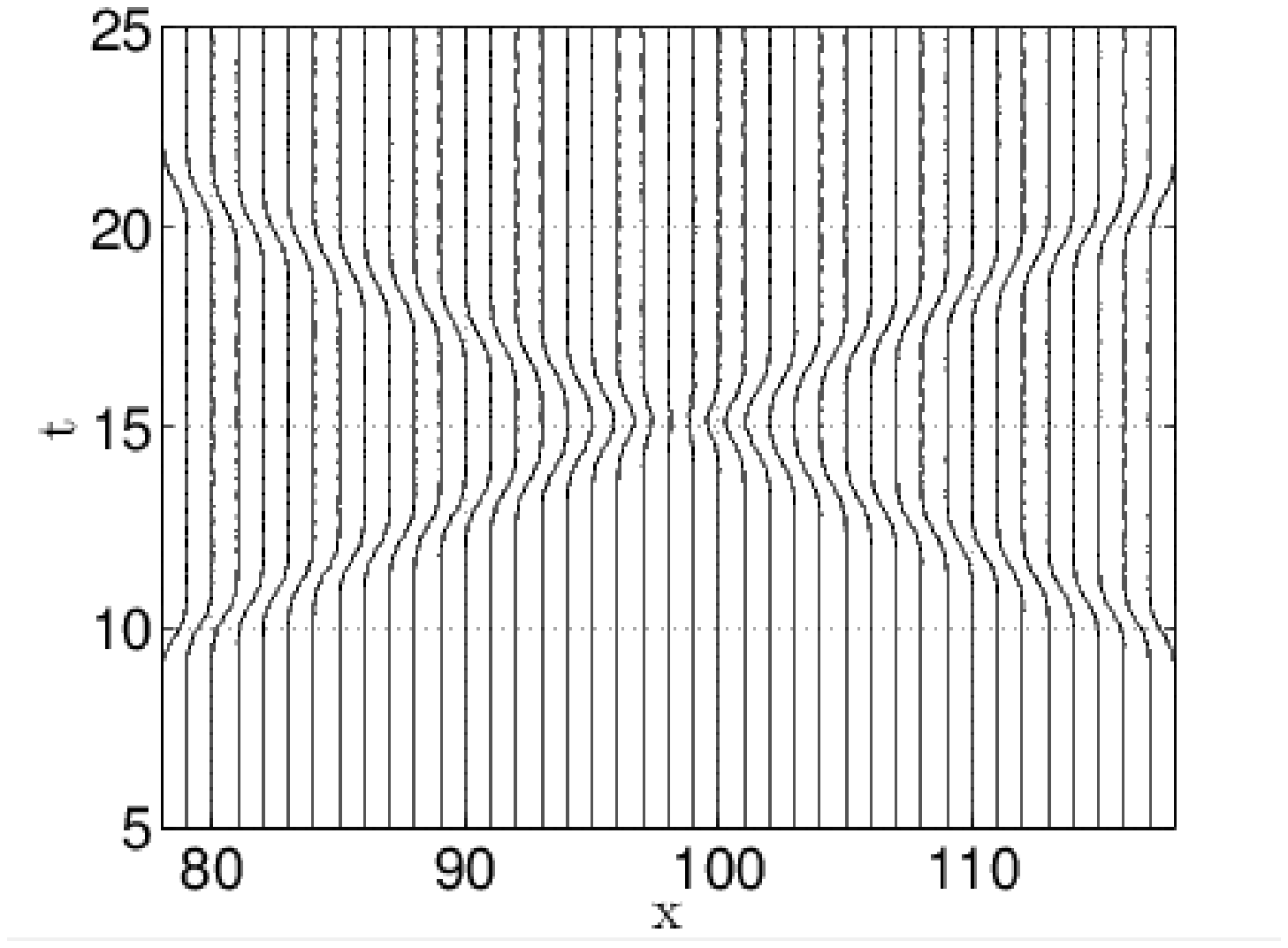}
\caption{In-line collision of two kinks. $N_{x}=200$, $N_{y}=40$,
  $T_{end}=60$, $U_{0}=2$, $u_{y}^{0}=u_{y}^{1}=0$, $u_{x}^{0}=5.25$
  and $u_{x}^{1}=-5.5$. Left: contour plot of the energy density
  function in time on the lattice line. Right: atomic displacement plot
  in the $x$ axis direction on the lattice line.}\label{fig:KKcoll}
\end{figure}

For our second example, we consider in-line kink-kink collisions with
initial kicks of $u_{x}^{0}=5.25$ and $u_{x}^{1}=-5.5$.  Integration
in time is carried out until $T_{end}=60$, see Fig. \ref{fig:KKcoll}.
In the left plot of Fig. \ref{fig:KKcoll} we show the energy density
function in time after each $10$ time steps on the main chain of atoms
where kinks propagate.  Both kicks have produced a kink moving towards
each other.  Around $15$ time units, two kinks collide and re-appear
after the collision, see the displacement plot of Fig.\
\ref{fig:KKcoll} on the right.  Interestingly, when the kinks approach
their initial locations, they fill the vacancies (stationary
anti-kinks) left behind, and this scattering creates breather
solutions.

\begin{figure}
\centering 
\includegraphics[width=0.48\textwidth]{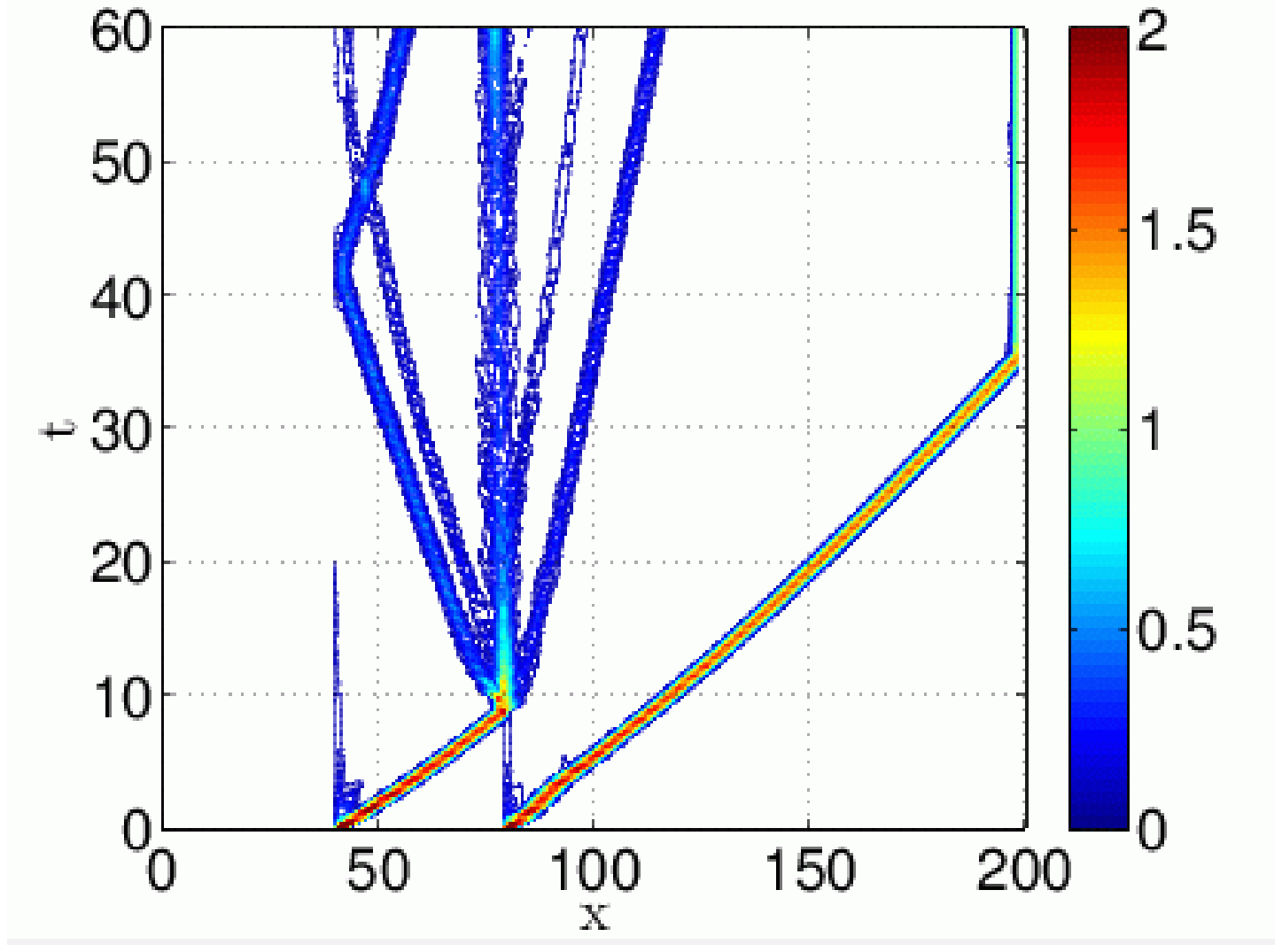}
\includegraphics[width=0.48\textwidth]{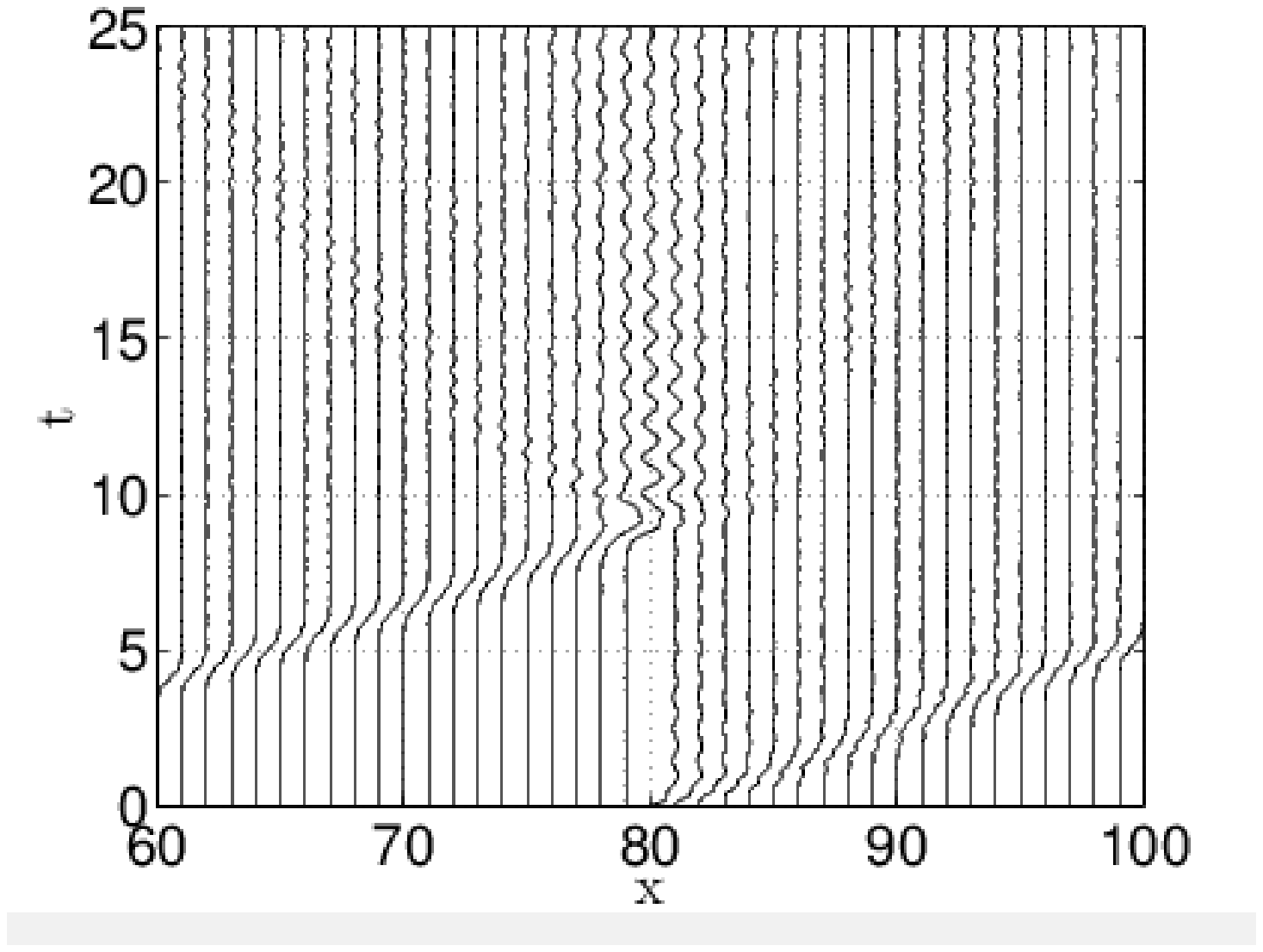} 
\caption{In-line collision of two kinks. $N_{x}=200$, $N_{y}=40$,
  $T_{end}=60$, $U_{0}=2$, $u_{y}^{0}=u_{y}^{1}=0$, $u_{x}^{0}=5.5$
  and $u_{x}^{1}=5.25$. Left: contour plot of the energy density
  function in time on the lattice line. Right: atomic displacement plot
  in the $x$ axis direction on the lattice line.}\label{fig:KKcoll2}
\end{figure}

To illustrate this phenomenon more clearly, we perform additional tests
on the same lattice but with the second atom's initial kick taken
to have opposite sign, i.e.~$u_{x}^{0}=5.5$ and $u_{x}^{1}=5.25$, see
Fig. \ref{fig:KKcoll2}.  Now both kinks propagate in the same
direction.  When the kink on the left approaches the vacancy
(anti-kink) created by the kink on the right, the kink fills the
vacancy and creates a stationary as well as propagating breather
solutions moving in both directions.  The vacancy filling can be
clearly seen in the right plot of Fig. \ref{fig:KKcoll2}, where we
show the displacement of atoms in the $x$ axis direction of the atoms
on the main horizontal lattice chain.  This numerical test shows that
propagating breather solutions can not only be created by the kicks
but also by kink solutions filling vacancies (colliding with
anti-kinks) in the crystal lattice.

\begin{figure}
\centering 
\includegraphics[width=0.48\textwidth]{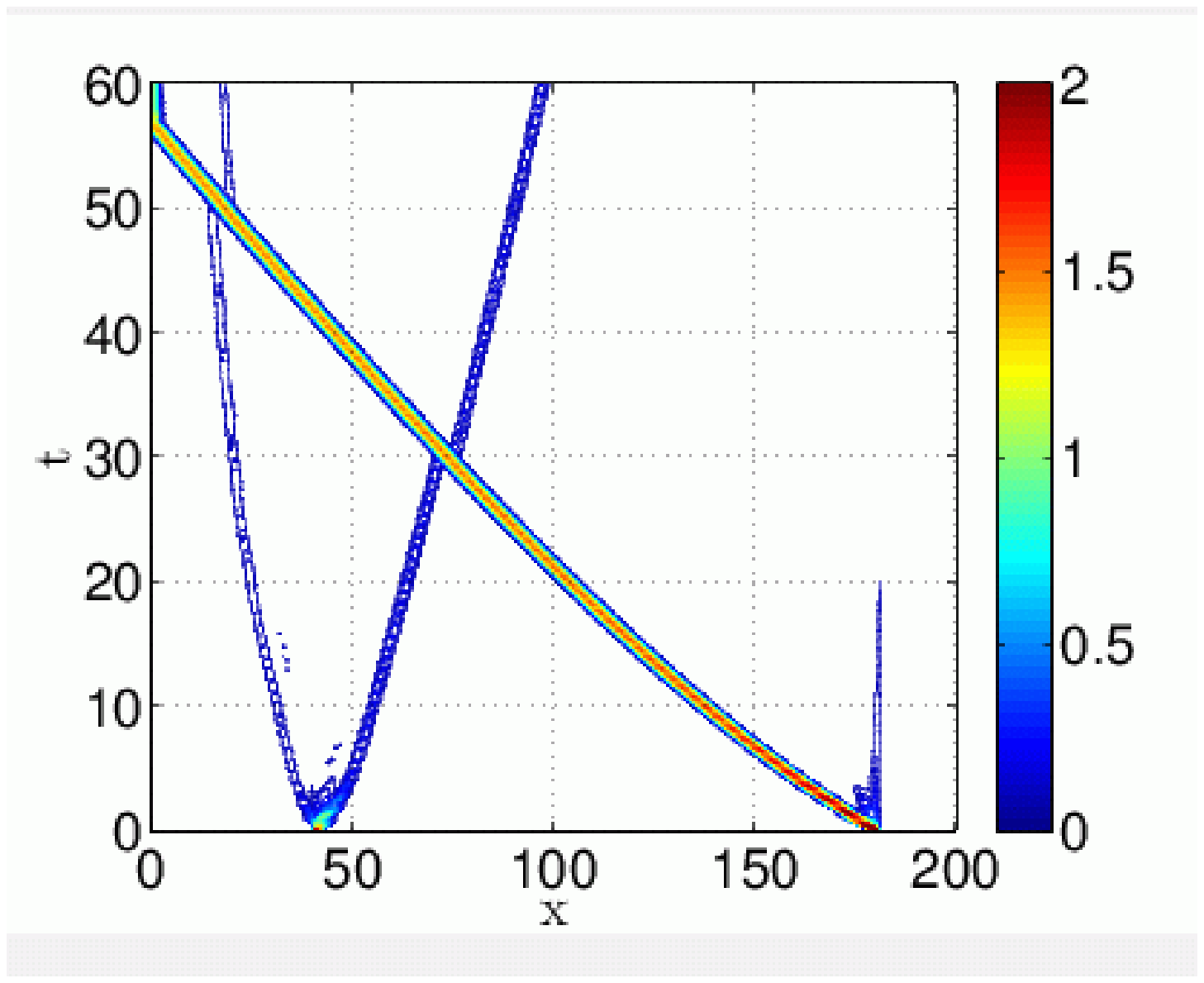}
\includegraphics[width=0.48\textwidth]{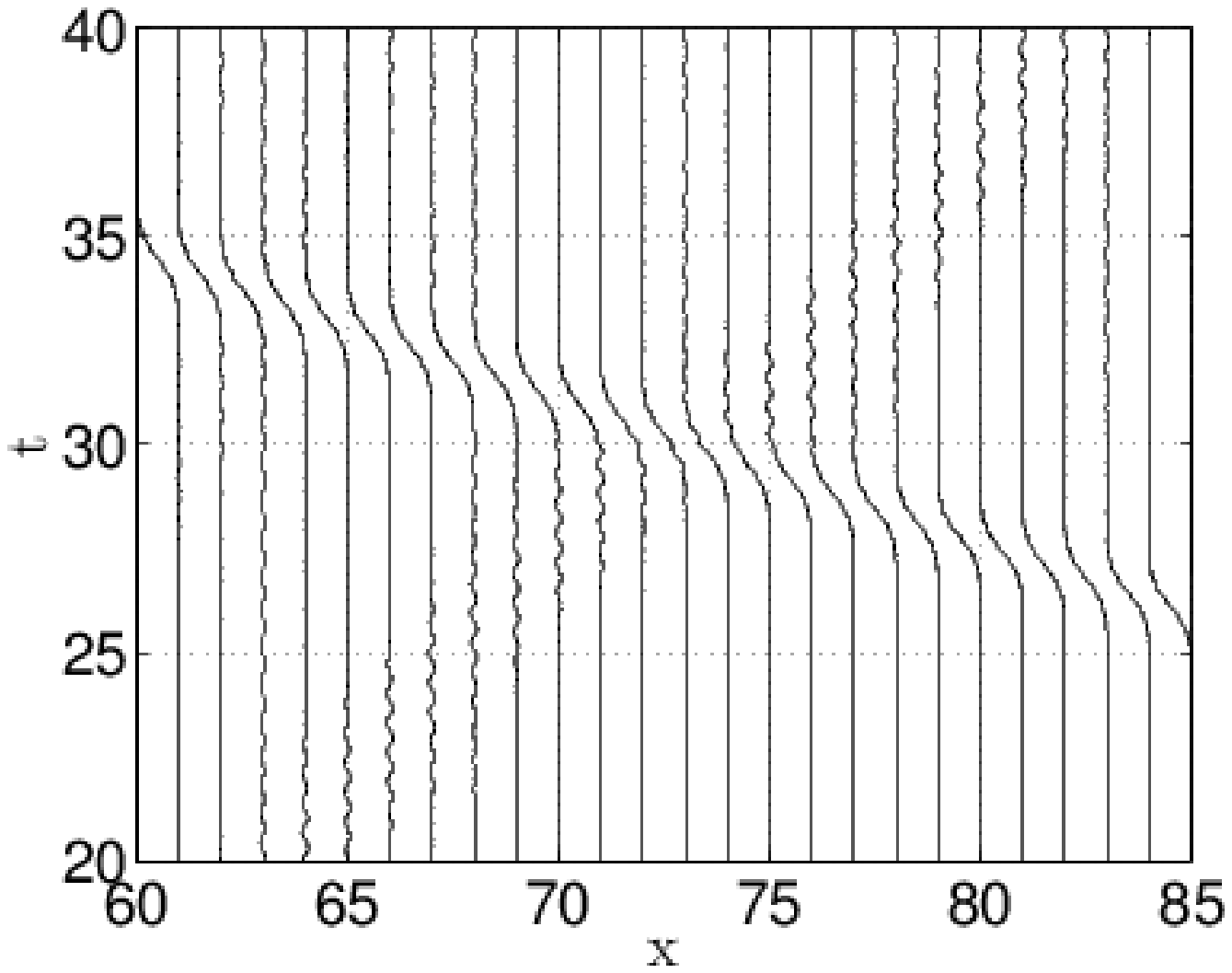}
\caption{In-line collision of breather and kink
  solutions. $N_{x}=200$, $N_{y}=40$, $T_{end}=60$, $U_{0}=2$,
  $u_{y}^{0}=u_{y}^{1}=0$, $u_{x}^{0}=3.5$ and $u_{x}^{1}=-5.5$. Left:
  contour plot of the energy density function in time on the lattice
  line. Right: atom displacement plot in the $x$ axis direction on the
  lattice line.}\label{fig:BKcoll}
\end{figure}

For our final in-line collision experiment, we consider breather-kink
collision with initial velocity kicks $u_{x}^{0}=3.5$ and
$u_{x}^{1}=-5.5$.  We integrate in time until $T_{end}=60$ and
illustrate the numerical results in Fig. \ref{fig:BKcoll}.  In the
left plot of Fig. \ref{fig:BKcoll}, we show the energy density function in
time after each $10$ time steps on the main chain of atoms where the
breather and kink propagate.  The kick on the left has produced two
breather solutions propagating in opposite directions, and the kick on
the right has produced a kink solution moving to the left towards the
breather solutions.  After around $30$ time units, the breather and
kink solutions collide and pass through each other.  Later in time the
kink passes through the second breather solution propagating in the
same direction.  The first collision is also illustrated by the
displacement plot in Fig. \ref{fig:BKcoll} on the right. These results
suggest that breather and kink solutions can easily coexist in our
model of a crystal lattice.

\subsubsection{Numerical results: fully 2D effects}
\label{sec:2Deffects}
So far, except for the horseshoe wave solutions, see
Sec. \ref{sec:FrontSol}, all numerical examples have addressed the
quasi-one-dimensional nature of propagating discrete breather and kink
solutions.  In this section we demonstrate full 2D effects of the
numerical solutions by considering kink-kink and breather-kink
collisions on adjacent chains of atoms, and breather-breather
collision at $60\,^{\circ}$ angle to each other.

\begin{figure} 
\centering 
{\includegraphics[width=0.32\textwidth]{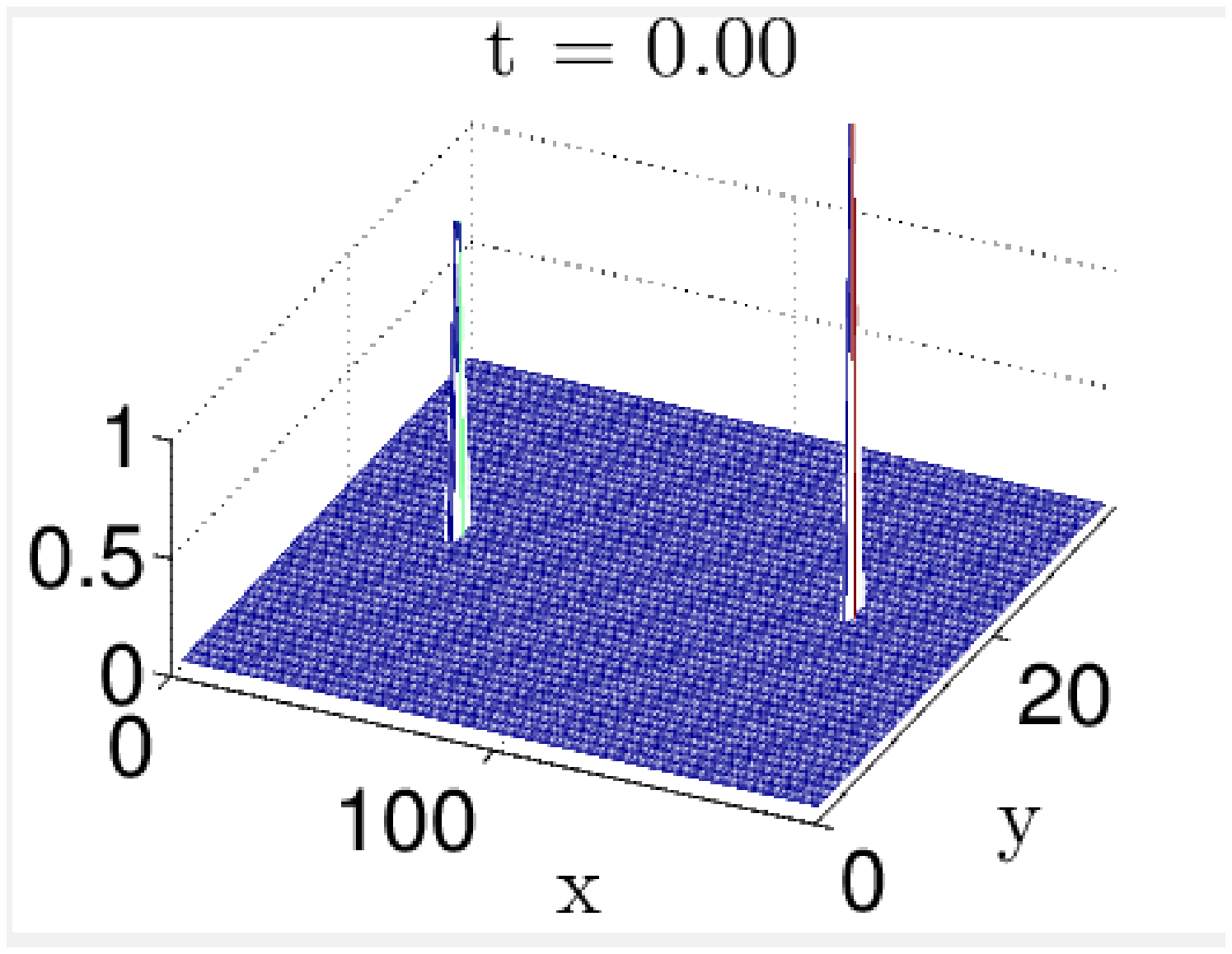}}
{\includegraphics[width=0.32\textwidth]{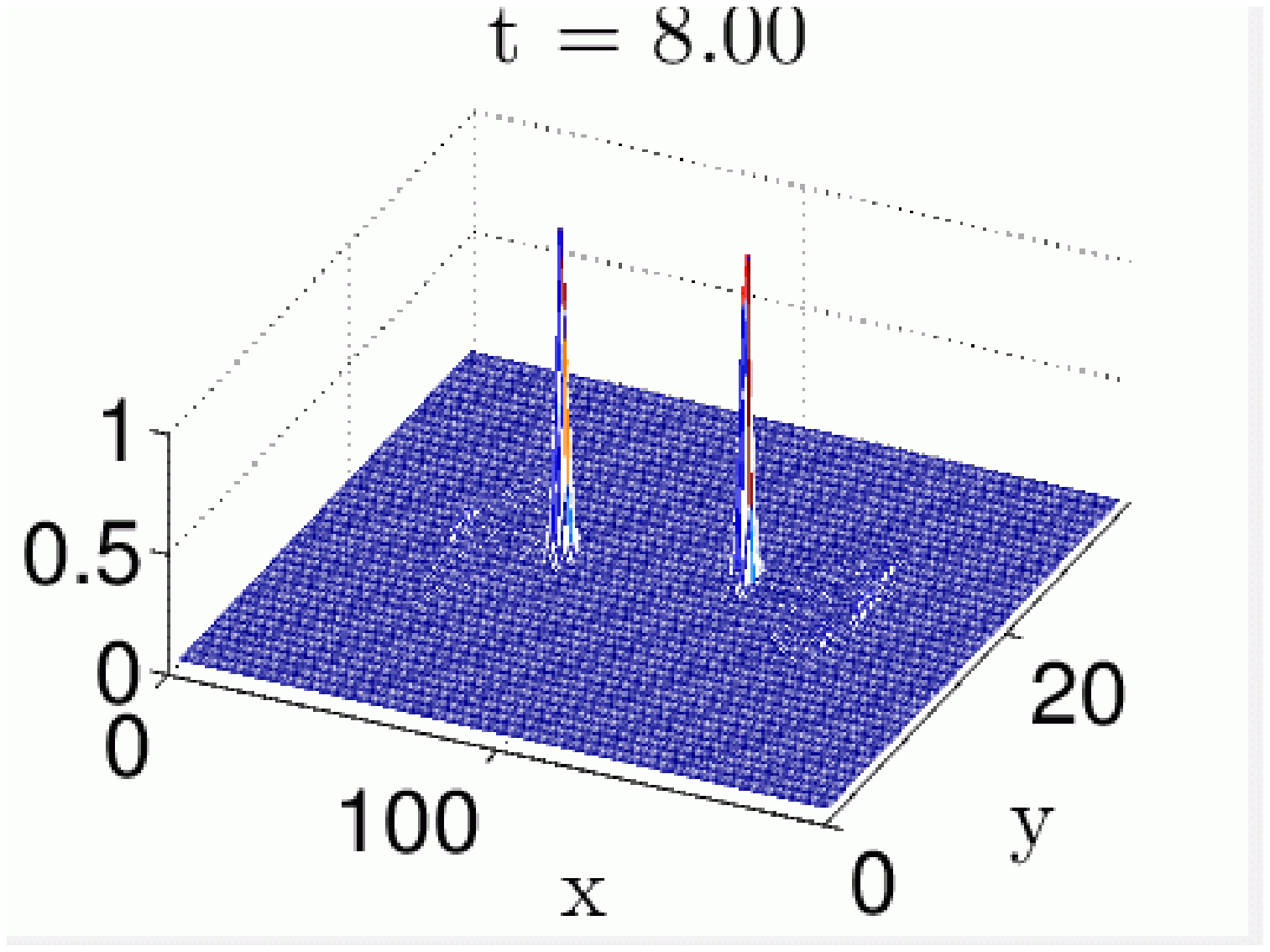}}
{\includegraphics[width=0.32\textwidth]{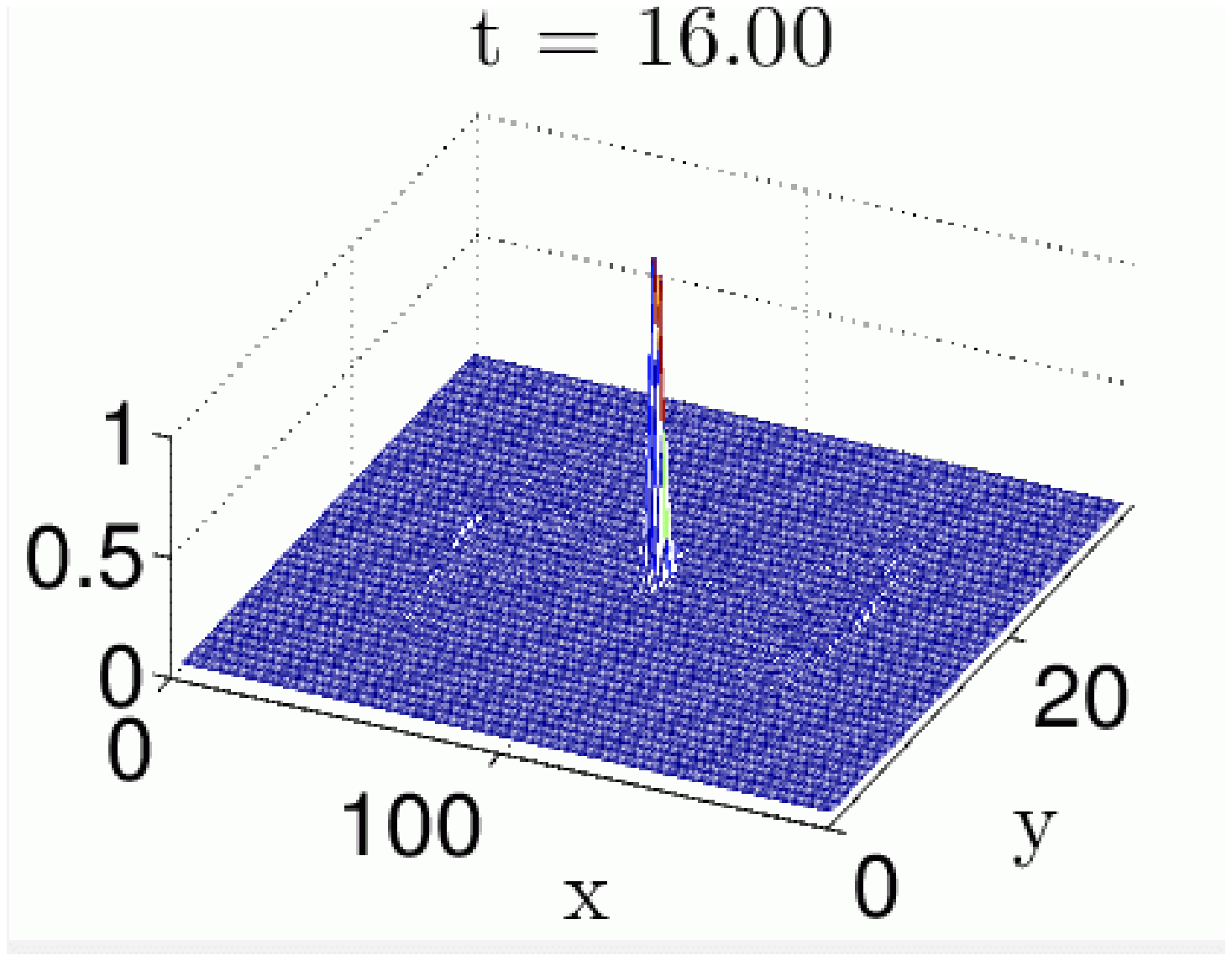}}
{\includegraphics[width=0.32\textwidth]{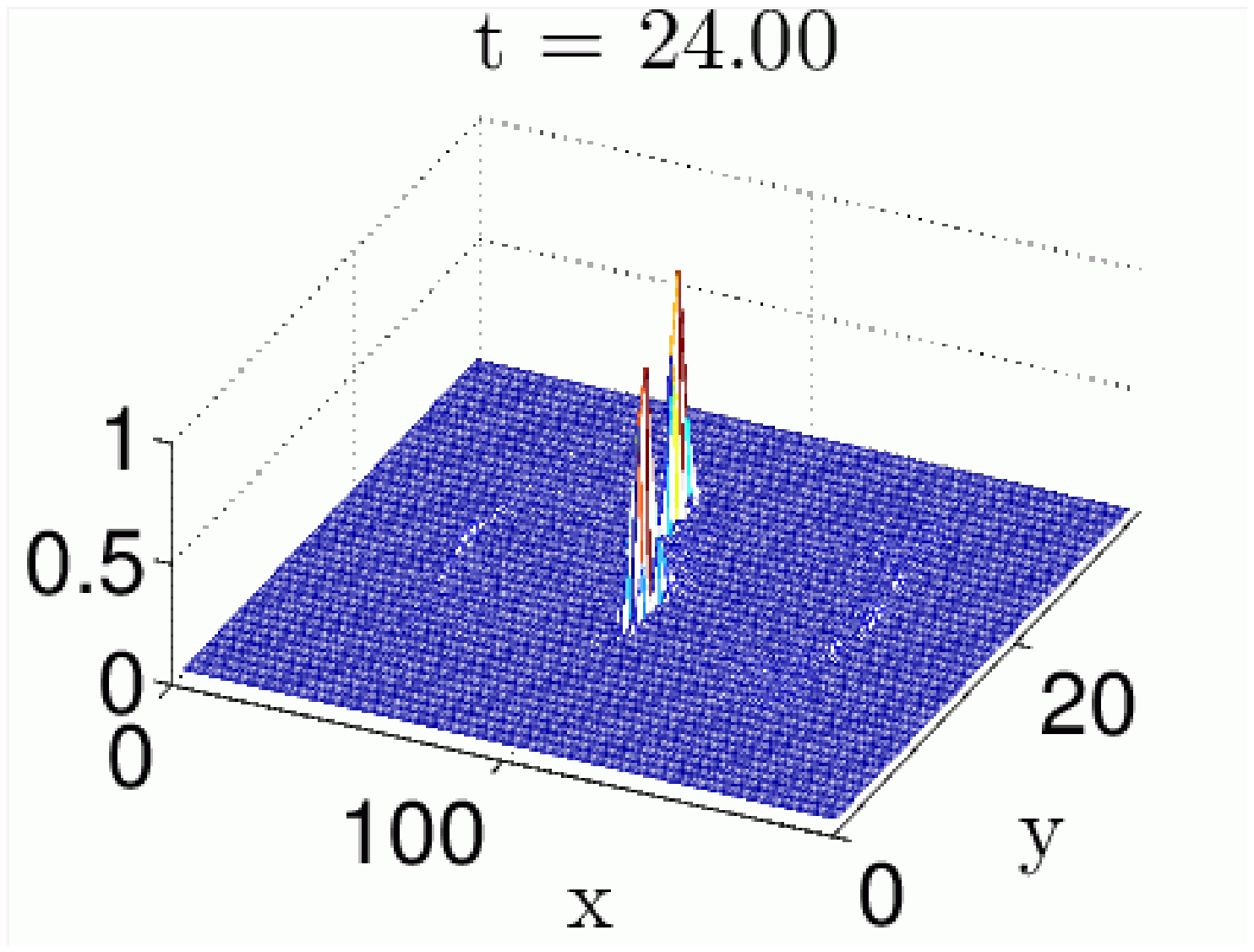}}
{\includegraphics[width=0.32\textwidth]{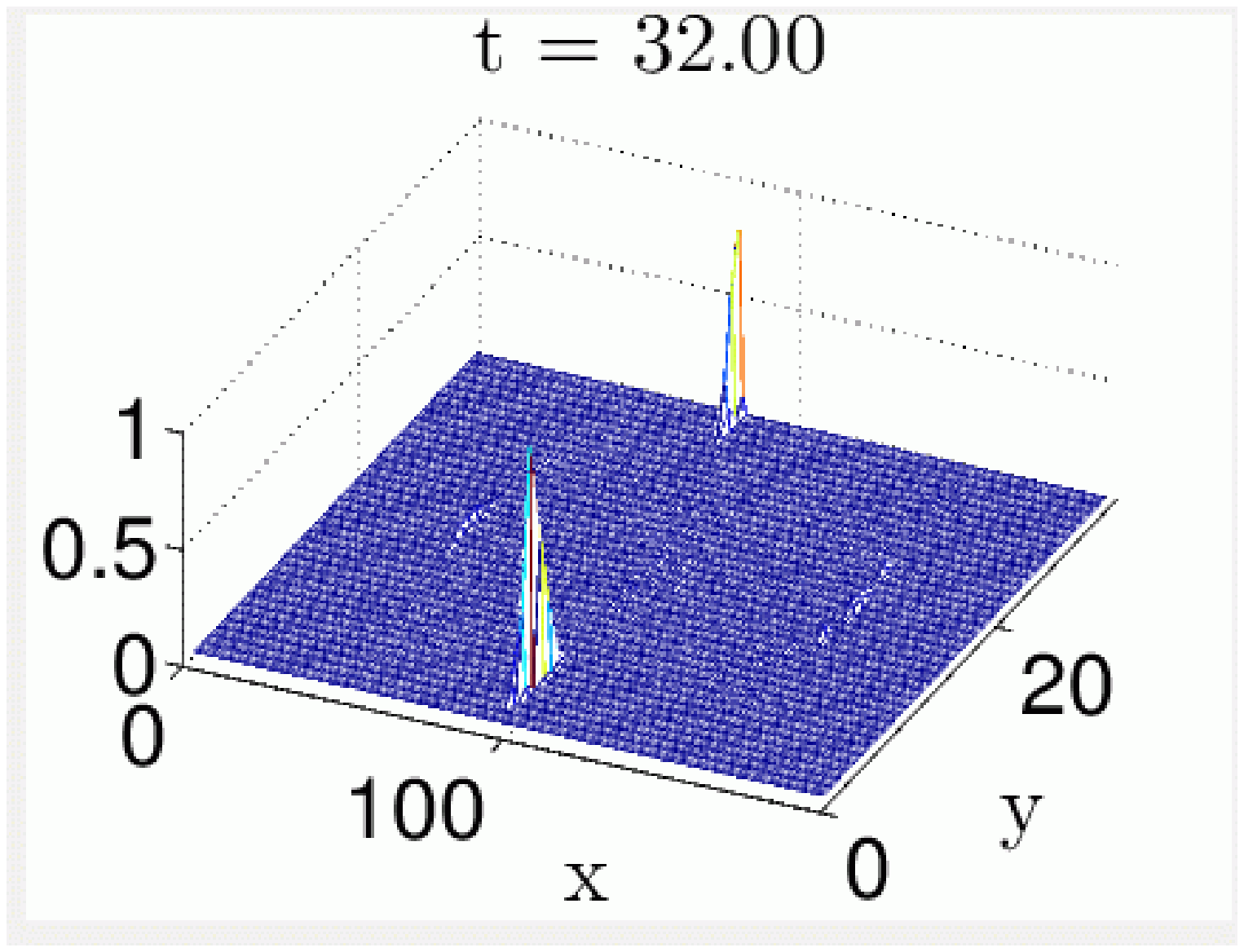}}
{\includegraphics[width=0.32\textwidth]{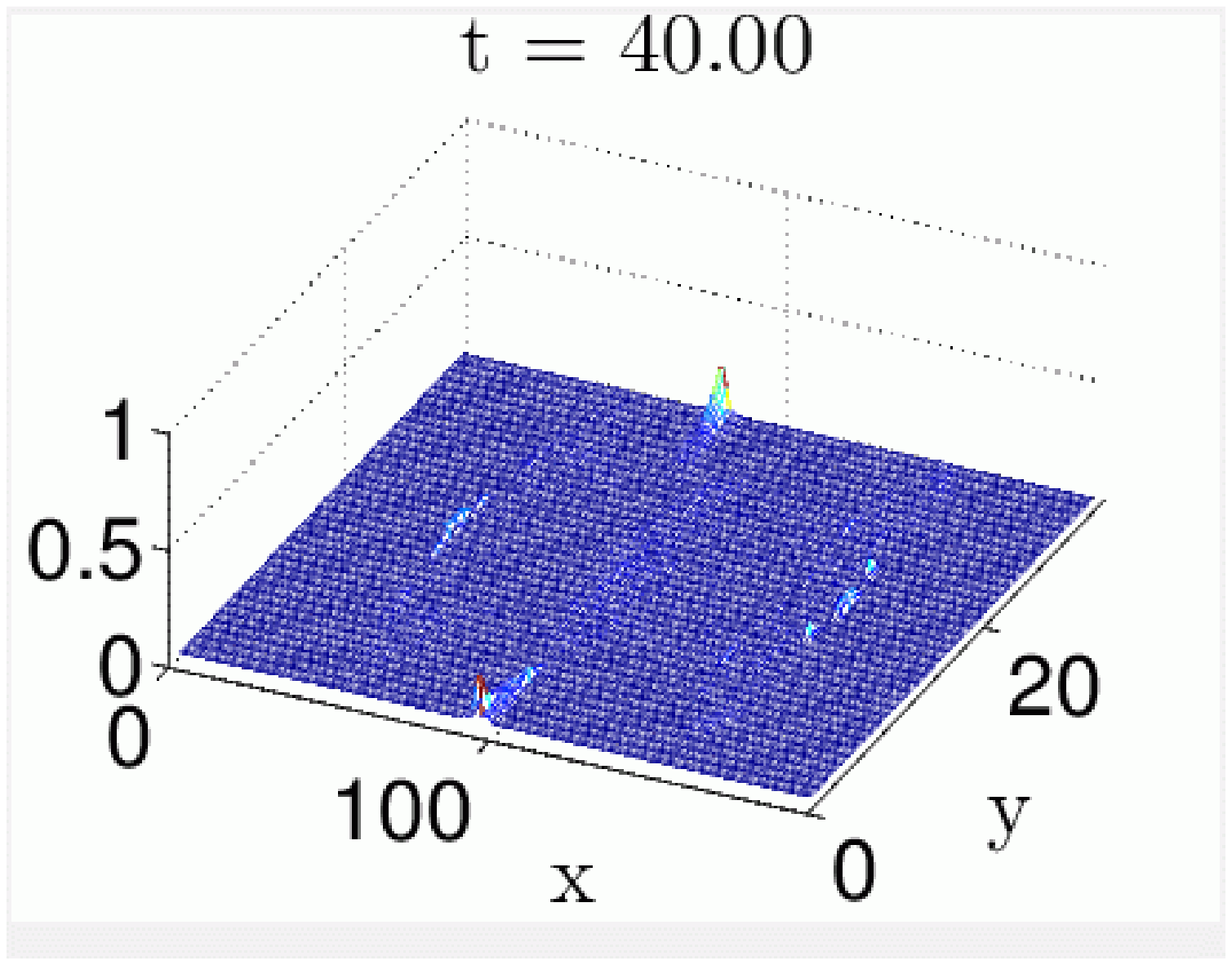}}
\caption{Energy density evolution in time of two kink collision on
  adjacent chains of atoms. $N_{x}=200$, $N_{y}=40$, $T_{end}=40$,
  $U_{0}=2$, $u_{y}^{0}=u_{y}^{1}=0$, $u_{x}^{0}=5.25$ and
  $u_{x}^{1}=-5.5$.}\label{fig:2DKcoll}
\end{figure}

If the kink solutions of our 2D model were truly one-dimensional, we
would expect no interactions between two kink solutions in kink-kink
collisions, with the kinks travelling in opposite directions along
adjacent chains of atoms.  This is not the case, as can be seen in
Fig. \ref{fig:2DKcoll}.  The lattice, parameter values and initial
kicks are identical to the in-line kink-kink collision experiment in
Sec. \ref{sec:Coll}.

In Figure \ref{fig:2DKcoll}, we show the evolution of the energy
density function in time.  The localized energy peaks are associated
with the two kinks propagating towards each other on adjacent lattice
chains.  At $t \approx 16$, the two kinks collide and {\em change
  their propagation directions} after collision.  After a complicated
collision region, the right kink eventually propagates in the
$(1/2,-\sqrt{3}/2)^T$ crystallographic lattice direction, while the
left kink propagates in the $(1/2,\sqrt{3}/2)^T$ crystallographic
lattice direction.  Once each kink has approached the upper or lower
boundary they eject one atom from the lattice.  This example of
collisions shows a new scattering phenomena in a 2D lattice model
which has no counterpart in 1D lattice models.  It shows that there is
at least weak coupling between kink solution and atoms on adjacent
chains.

\begin{figure} 
\centering 
{\includegraphics[width=0.32\textwidth]{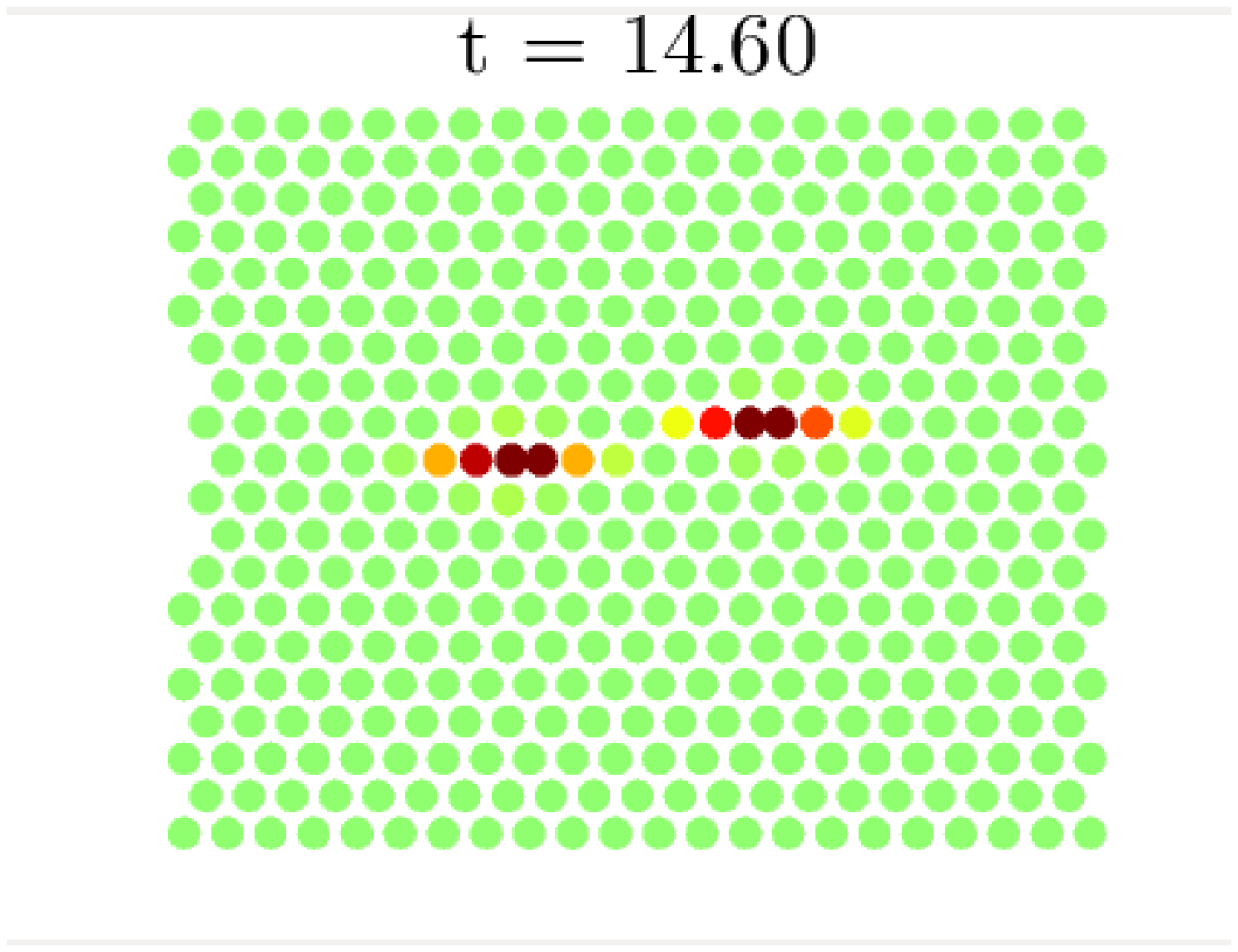}}
{\includegraphics[width=0.32\textwidth]{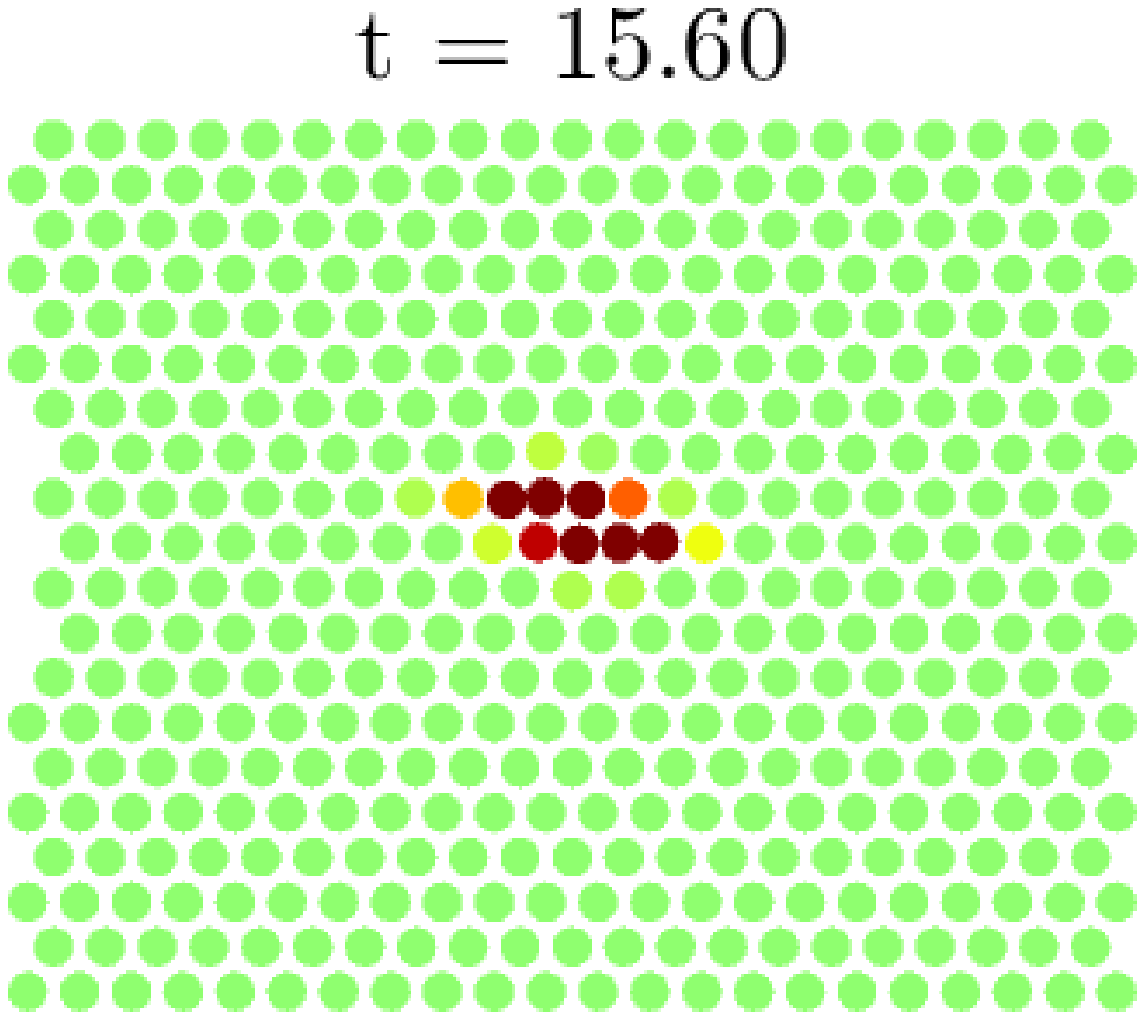}}
{\includegraphics[width=0.32\textwidth]{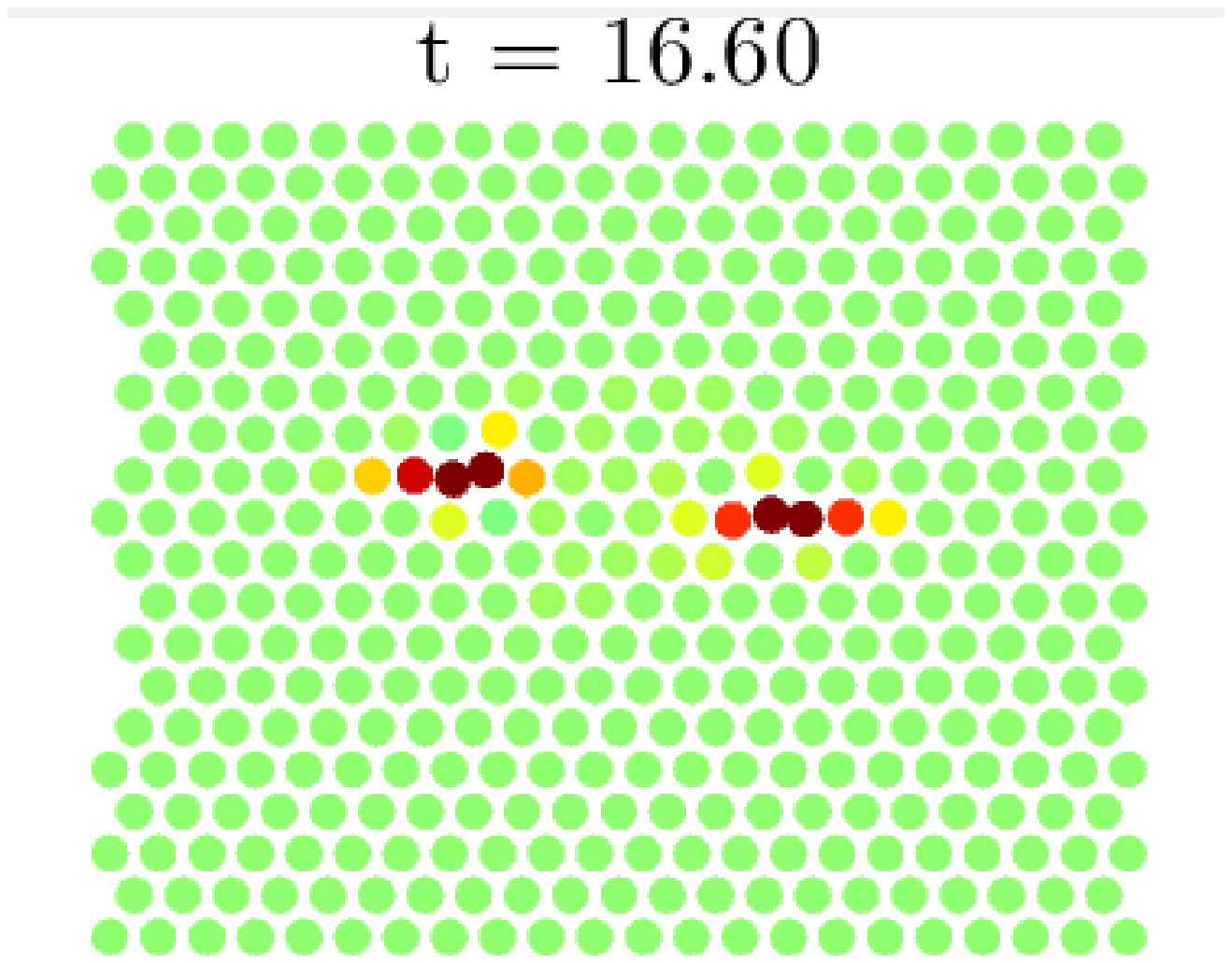}}
{\includegraphics[width=0.32\textwidth]{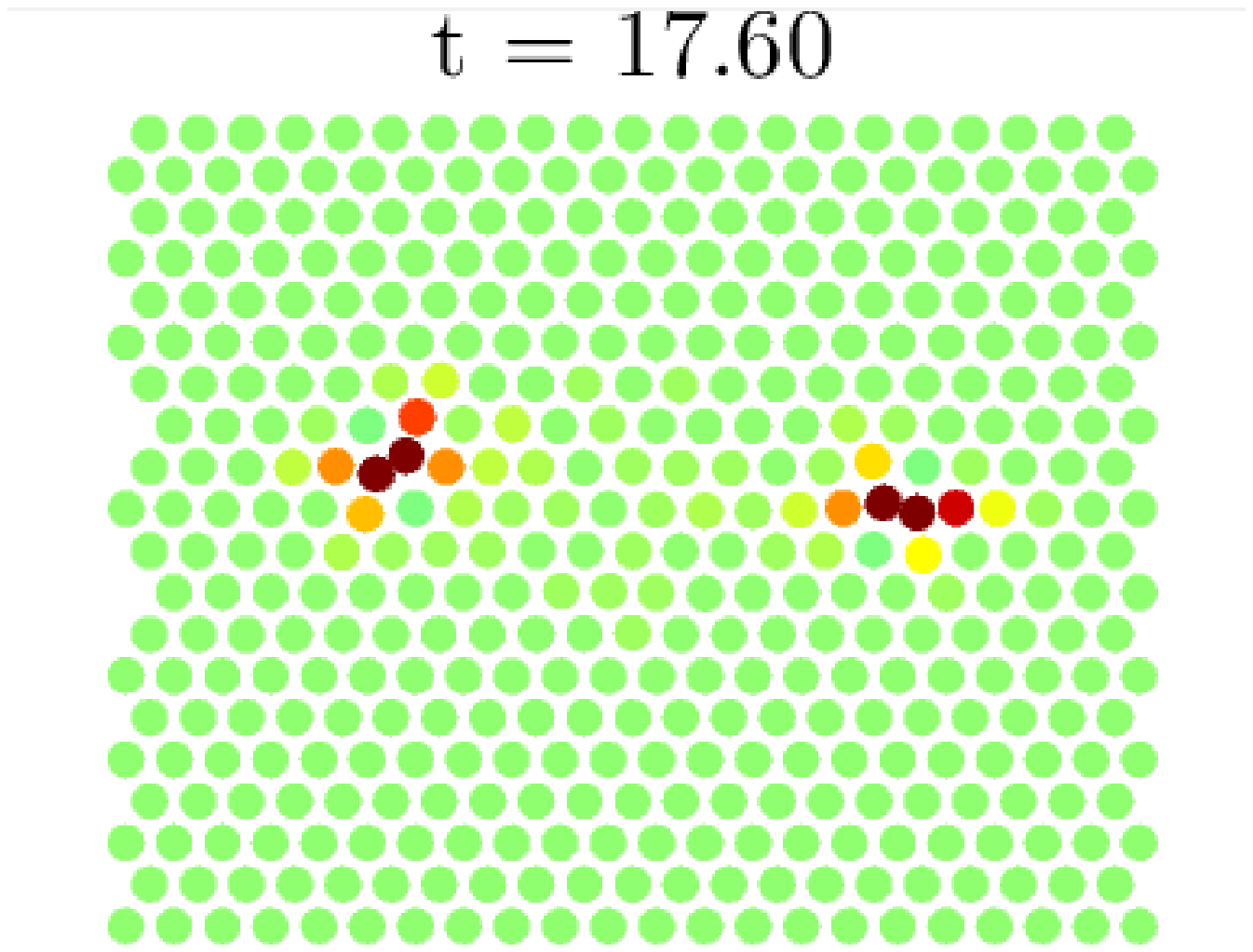}}
{\includegraphics[width=0.32\textwidth]{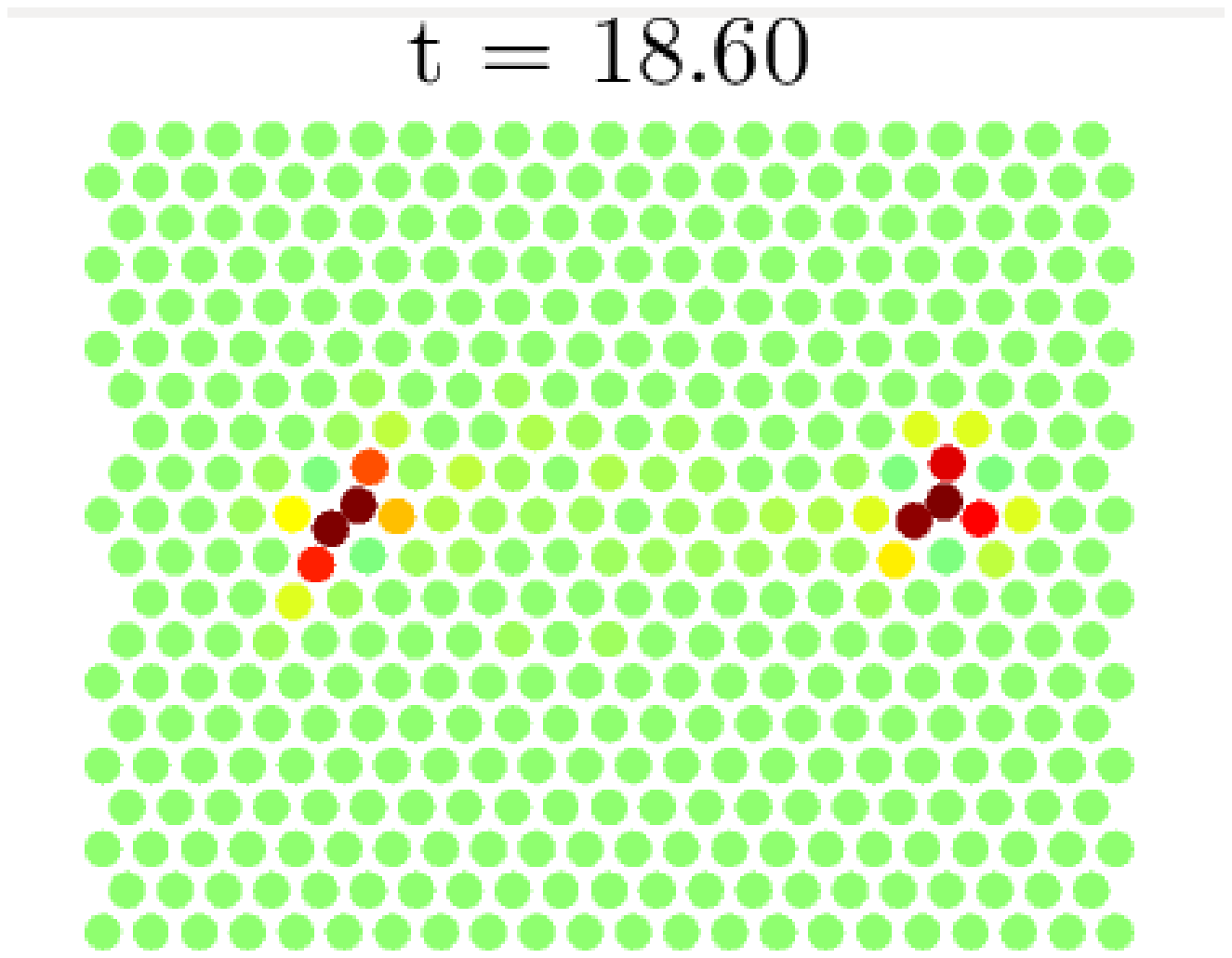}}
{\includegraphics[width=0.32\textwidth]{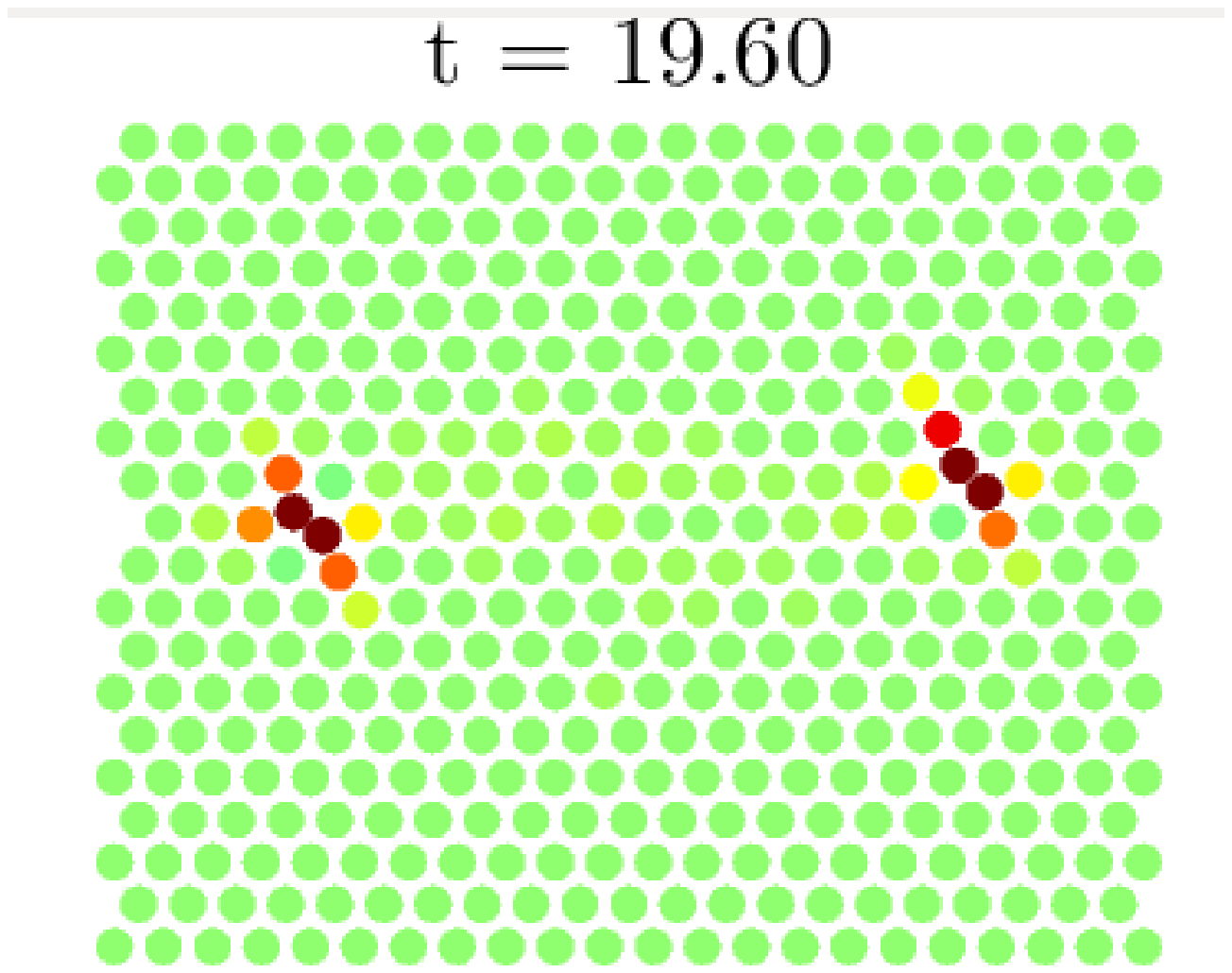}}
{\includegraphics[width=0.32\textwidth]{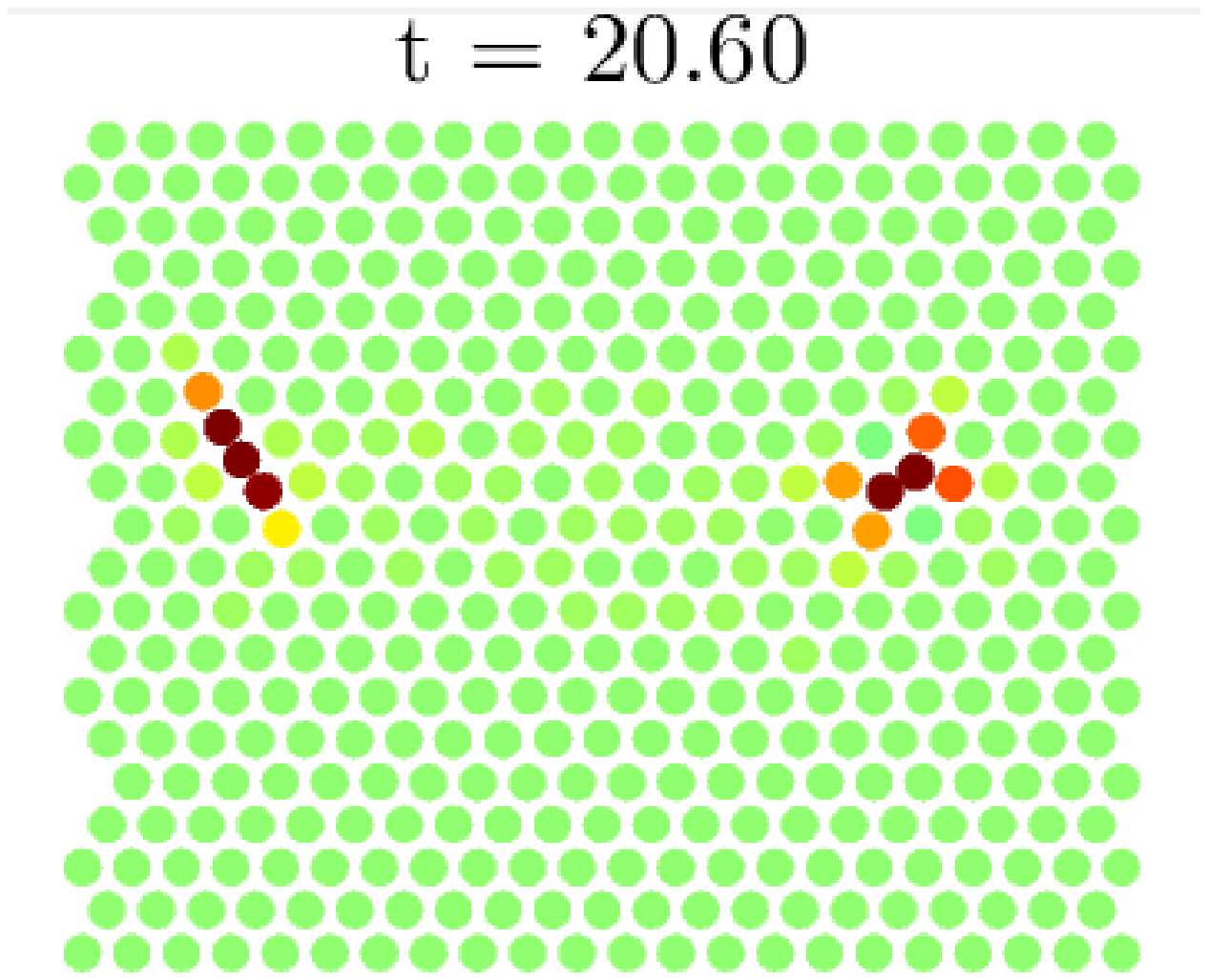}}
{\includegraphics[width=0.32\textwidth]{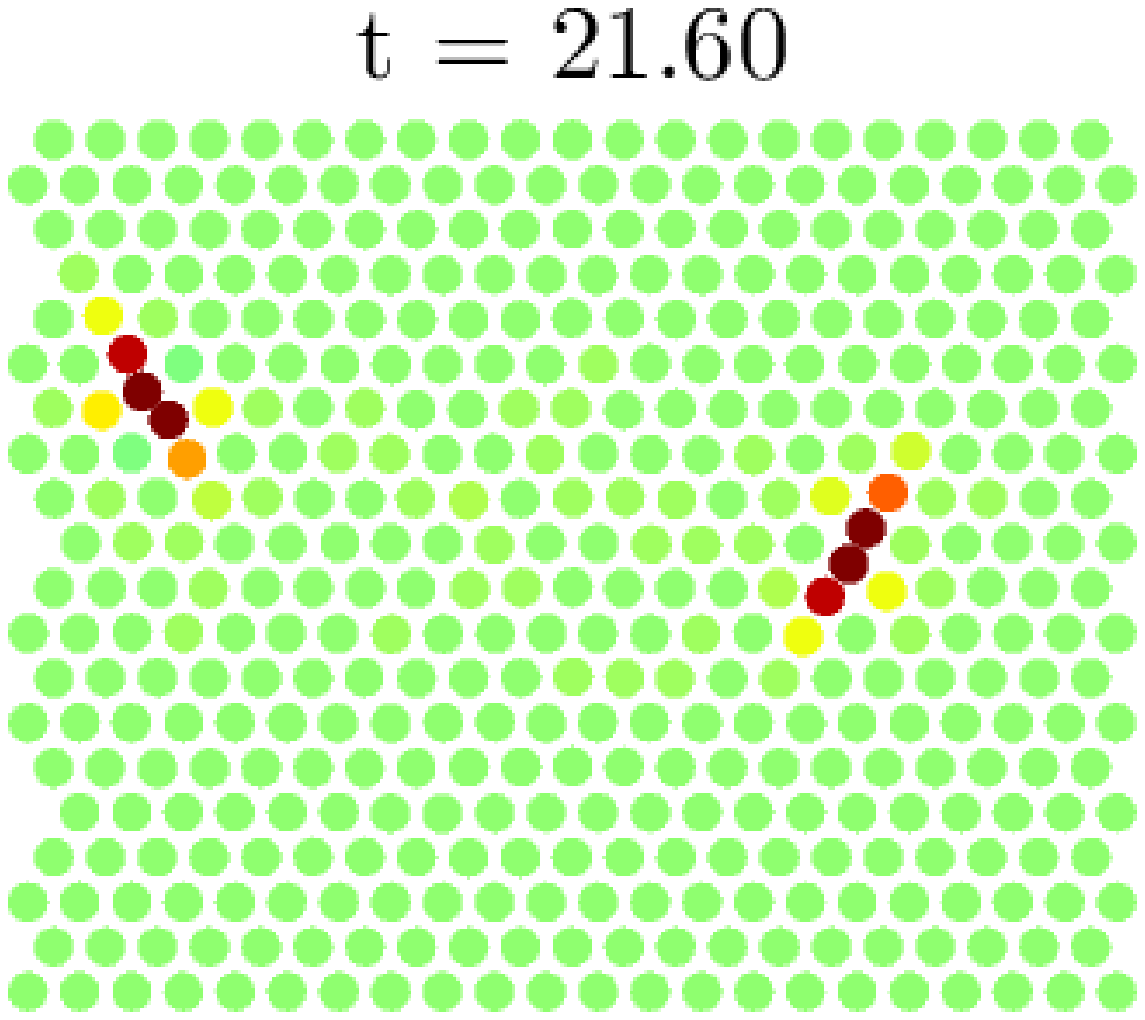}}
{\includegraphics[width=0.32\textwidth]{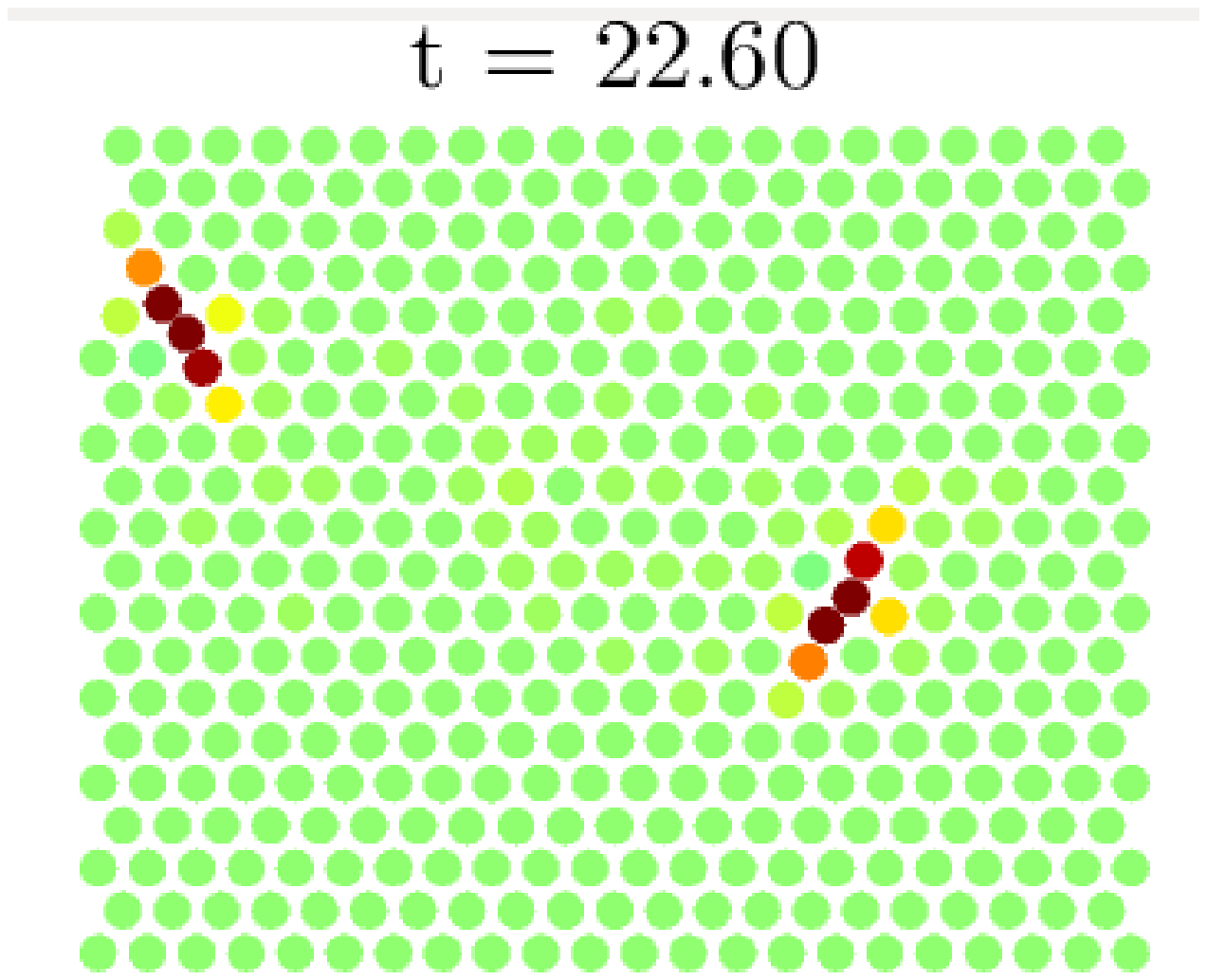}}
\caption{Snapshots of scatter plots of atoms in time of two kink
  collision on adjacent chains of atoms. $N_{x}=200$, $N_{y}=40$,
  $T_{end}=40$, $U_{0}=2$, $u_{y}^{0}=u_{y}^{1}=0$, $u_{x}^{0}=5.25$
  and $u_{x}^{1}=-5.5$.}\label{fig:2DScatt}
\end{figure}

To understand better the events taking place during the kink-kink
collision on adjacent lines, we consider scatter plots of atoms in
time during the collision, see Fig. \ref{fig:2DScatt}.  We zoom into
the lattice area where the collision takes place.  Darker colours
indicate higher energy density function values.  The first plot shows
kinks approaching each other while the final plot shows kink
solutions, already fully developed, propagating in the different
crystallographic lattice directions.  From Figure \ref{fig:2DScatt} it
becomes evident that the two kinks, in fact, passed by each other.  Due
to the weak coupling between kinks on adjacent lines, the collision
has destabilized the kinks by inducing large displacements in the $y$
axis direction.  This induced instability causes the kinks to change
their propagation directions.  This may suggest that long-lived kink
solutions may only exist in completely idealized settings.

\begin{figure}
\centering 
\includegraphics[width=0.48\textwidth]{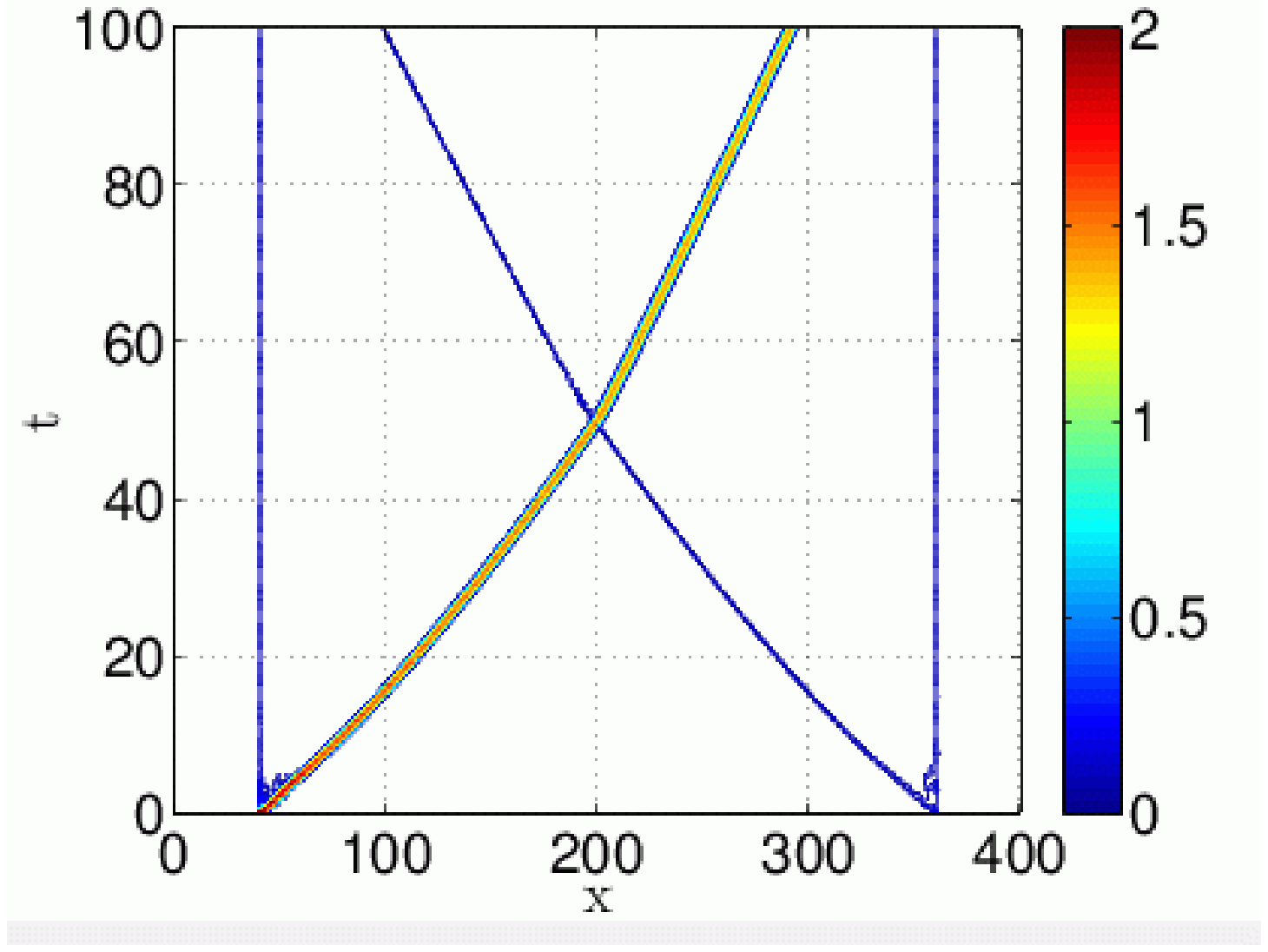}
\includegraphics[width=0.48\textwidth]{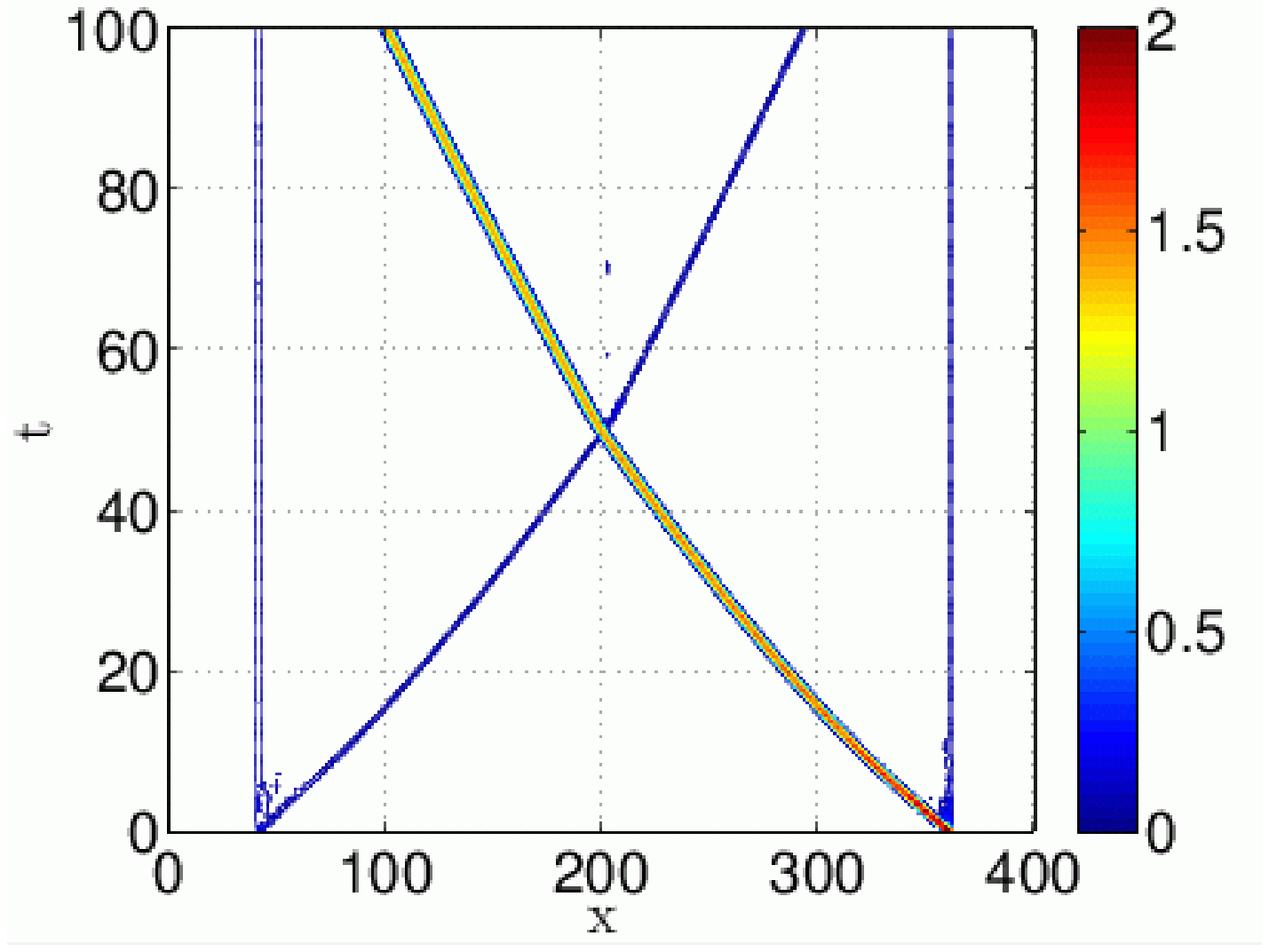}
\caption{Two kink collision on adjacent chains of atoms. $N_{x}=400$,
  $N_{y}=40$, $T_{end}=100$, $U_{0}=2$, $u_{y}^{0}=u_{y}^{1}=0$,
  $u_{x}^{0}=5.3$ and $u_{x}^{1}=-5.4$. Left: contour plot of the
  energy density function in time on the main chain of atoms of the
  kink on the left.  Right: contour plot of the energy density function
  in time on the main chain of atoms of the kink on the
  right.}\label{fig:2DKcoll2}
\end{figure}

In general, results of collisions do not always follow the same
pattern. The outcome will depend on the energy, velocity and phase of
propagating localized modes. We illustrate that with a counter example
of two kink collision on adjacent chains of atoms, see
Fig. \ref{fig:2DKcoll2}. For this experiment we consider a twice
larger lattice: $N_{x}=400$ and $N_{y}=40$, and initial kick values
$u_{x}^{0}=5.3$ and $u_{x}^{1}=-5.4$. In the left plot of
Fig. \ref{fig:2DKcoll2} we show energy density function in time of
atoms on the main chain of the kink moving from the left, and in the
right plot of Fig. \ref{fig:2DKcoll2} we show energy density function
in time of atoms on the main chain of the kink moving from the
right. Integration in time is carried out until $T_{end}=100$ and
results are illustrated after each $20$ time steps.  After around $50$
time units, the two kinks collide, lose some of their initial velocity
and continue to propagate, but slower.  This suggest that both kinks
have lost some energy during the collision to the lattice in the form
of phonons. In addition, plots of Fig. \ref{fig:2DKcoll2} confirm that
there is some energy associated to the kink solutions on adjacent
chains of atoms. Notice the change of the slopes in those energies
after the collision.

The destabilizing effects due to lateral displacements of atoms on the
main chain where the mode propagates is not only present in kink-kink
collisions, but also in breather-kink and breather-breather
collisions.  Recall that there are almost zero lateral displacements
on the main chain of atoms where the breather or kink propagates in an
idealized setting. To support our claims we present numerical
experiments of breather-kink collision on adjacent lines and
breather-breather collision at $60\,^{\circ}$ angles to each other.

\begin{figure} 
\centering 
\includegraphics[width=0.48\textwidth]{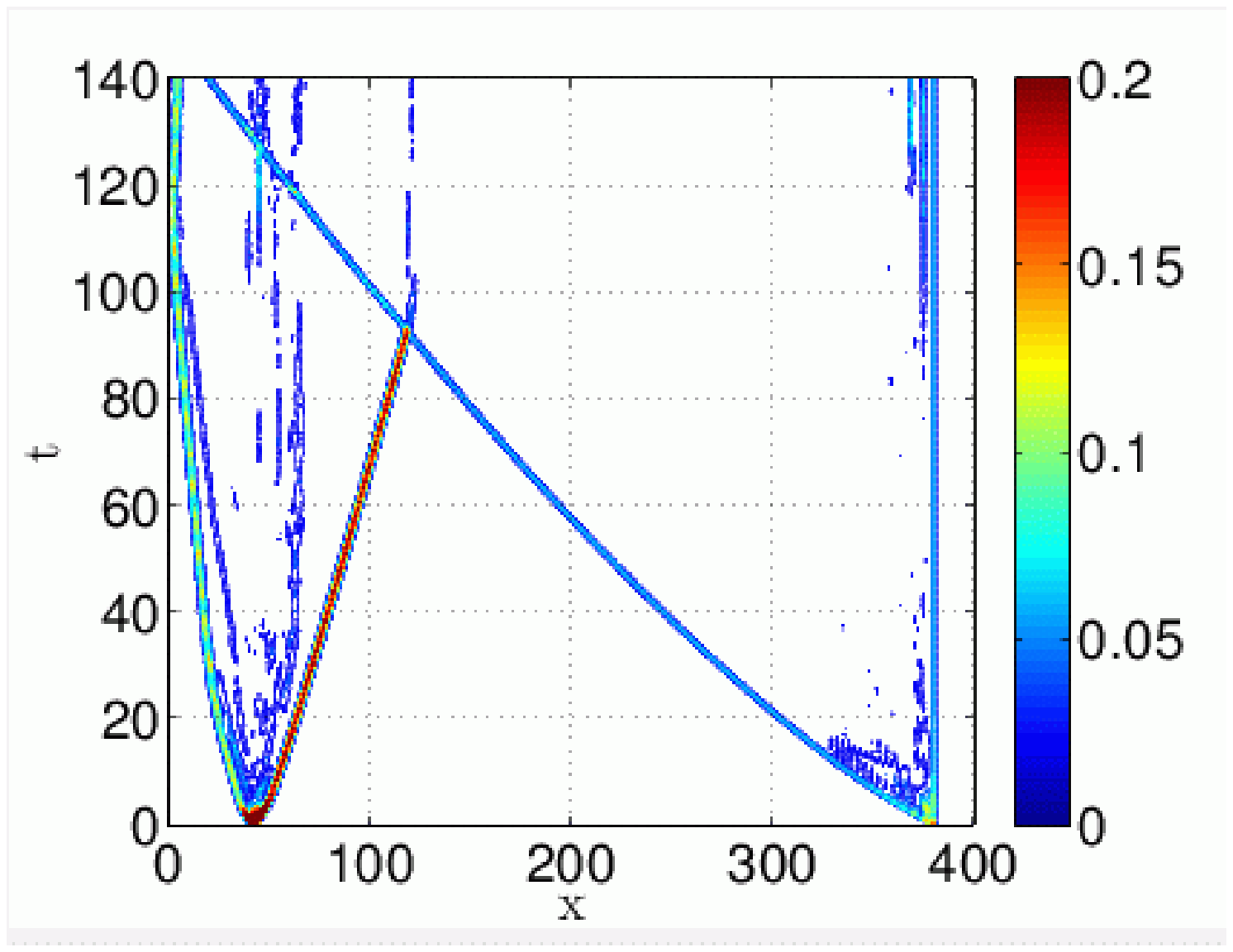}
\includegraphics[width=0.48\textwidth]{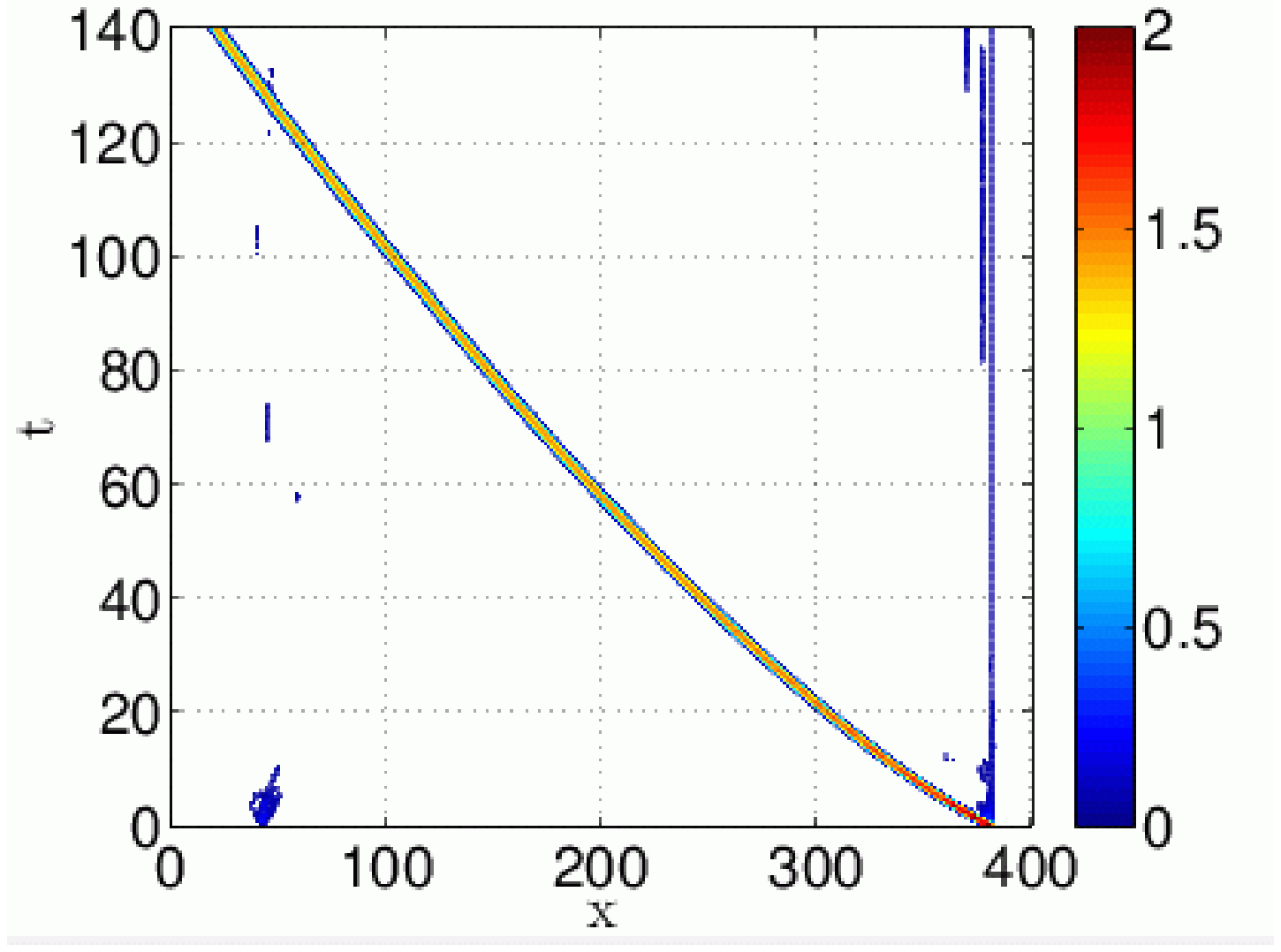}
\caption{Breather-kink collision on adjacent chains of
  atoms. $N_{x}=400$, $N_{y}=40$, $T_{end}=140$, $U_{0}=2$,
  $u_{y}^{0}=u_{y}^{1}=0$, $u_{x}^{0}=3.5$ and $u_{x}^{1}=-5.5$. Left:
  contour plot of the energy density function in time on the main
  chain of the breather solution. Right: contour plot of the energy
  density function in time on the main chain of atoms of the kink
  solution.}\label{fig:2DBKcoll}
\end{figure}

We consider the same breather-kink collision example from
Sec. \ref{sec:Coll}, but on adjacent chains and on the larger (x2)
lattice: $N_{x}=400$ and $N_{y}=40$.  We integrate in time until
$T_{end}=140$ and plot the associated energy density of atoms in both
chains in time after each $20$ time steps in Fig. \ref{fig:2DBKcoll}.
Compare Figs. \ref{fig:2DBKcoll} and \ref{fig:BKcoll}. We find that
the kink has scattered the breather solution during the collision into
the remaining lattice, see the left plot of Fig. \ref{fig:2DBKcoll},
but the collision itself has not affected the kink solution, see the
right plot of Fig. \ref{fig:2DBKcoll}. This example once again
illustrates 2D effects.

\begin{figure} 
\centering 
{\includegraphics[width=0.32\textwidth]{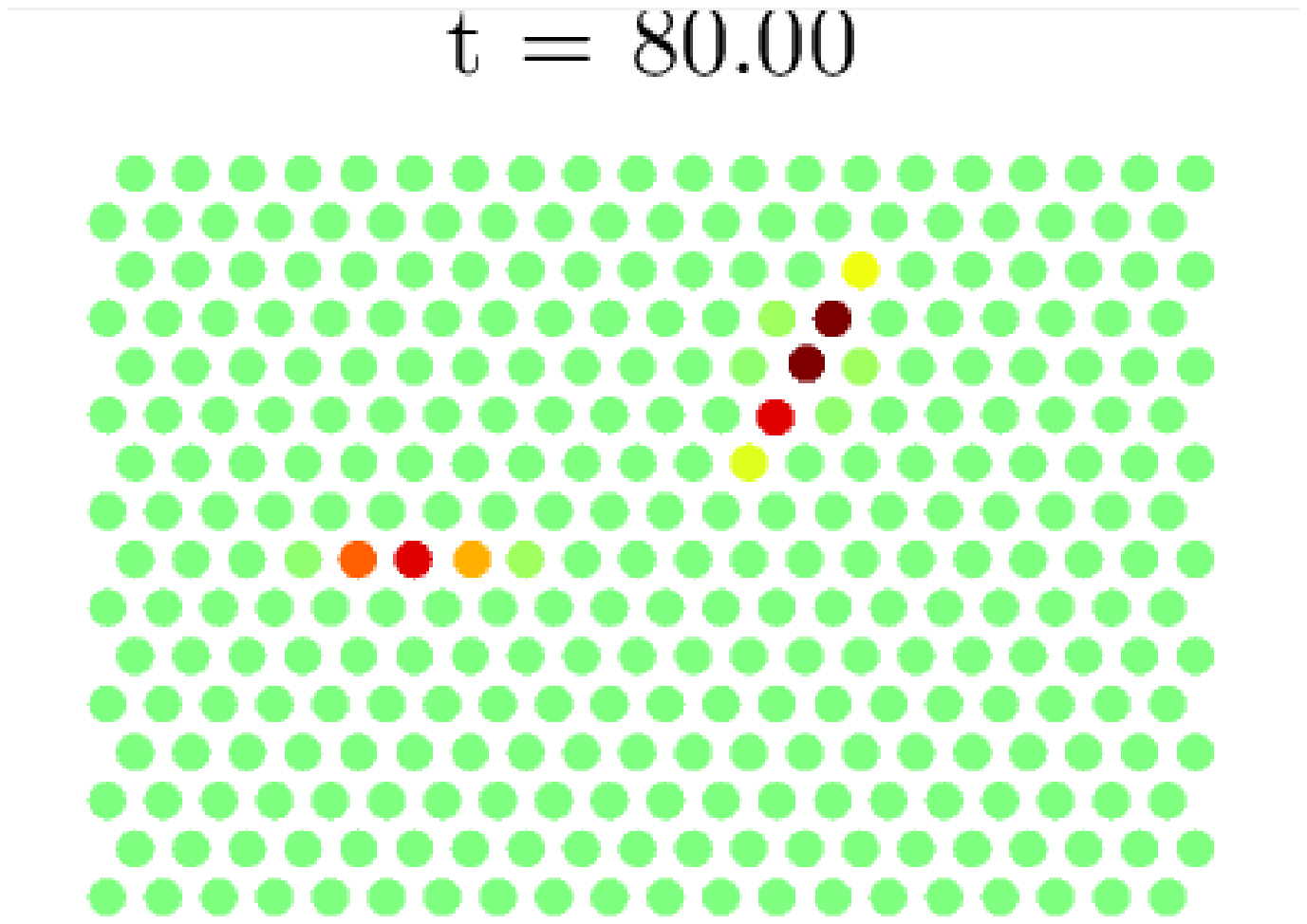}}
{\includegraphics[width=0.32\textwidth]{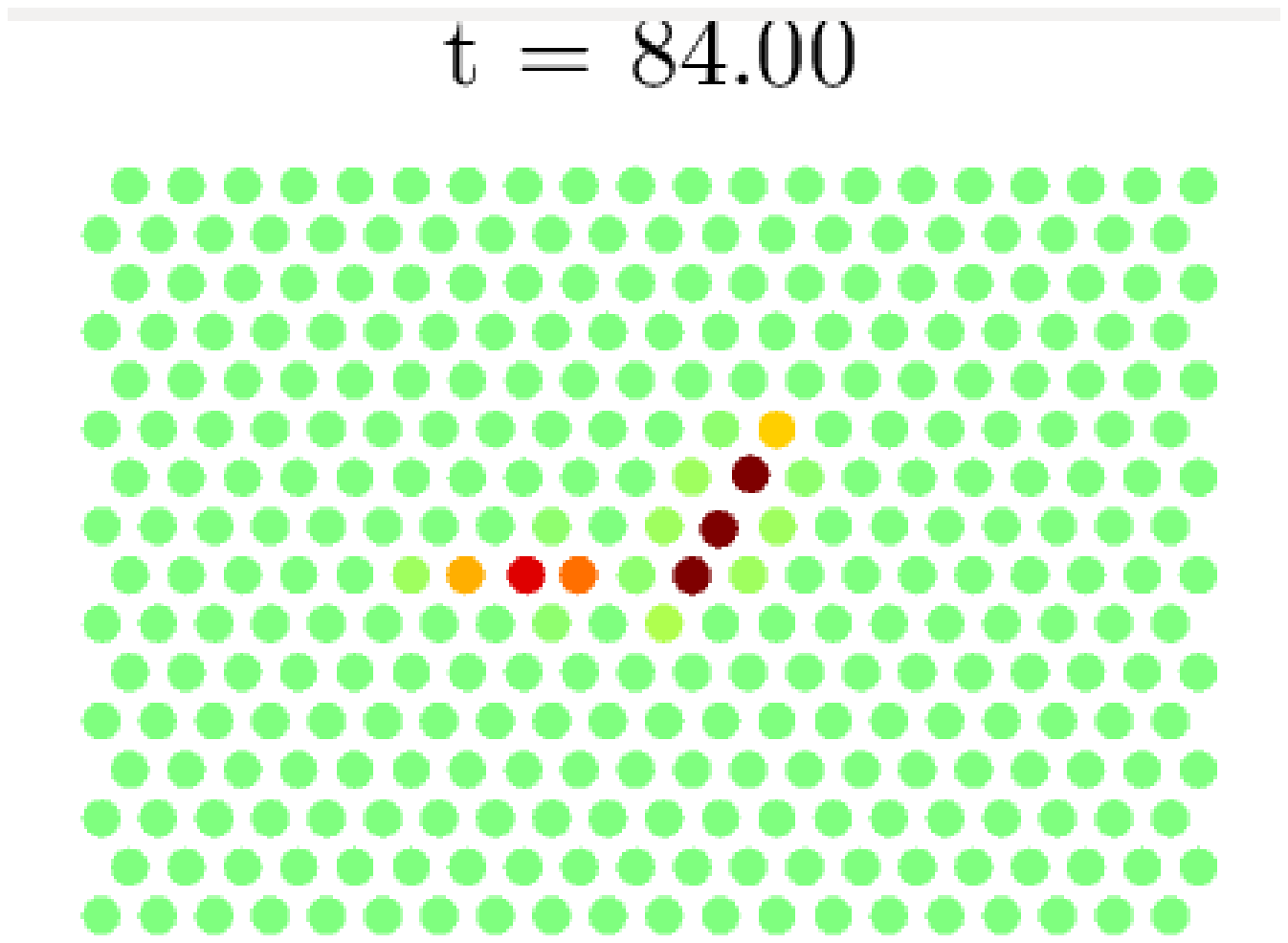}}
{\includegraphics[width=0.32\textwidth]{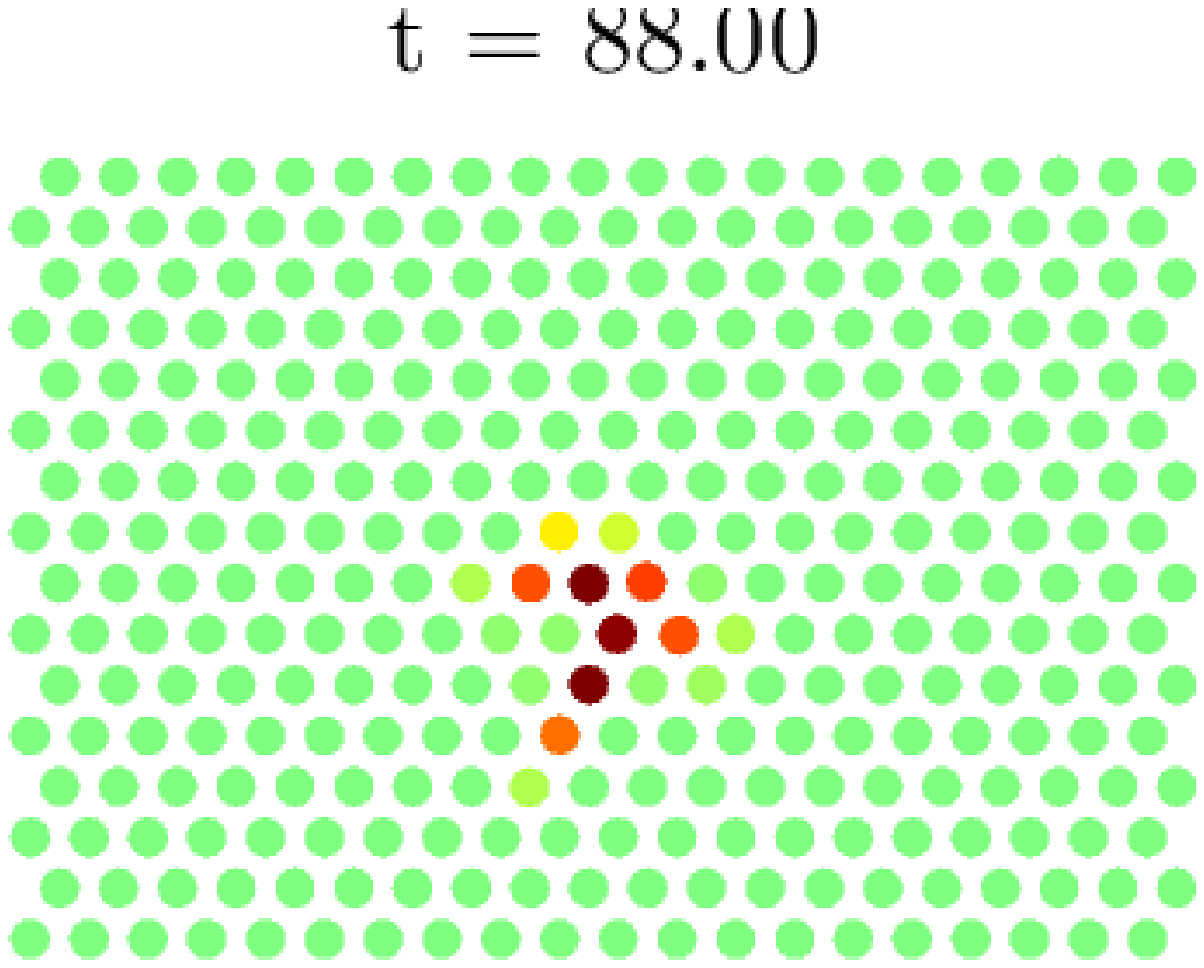}}
{\includegraphics[width=0.32\textwidth]{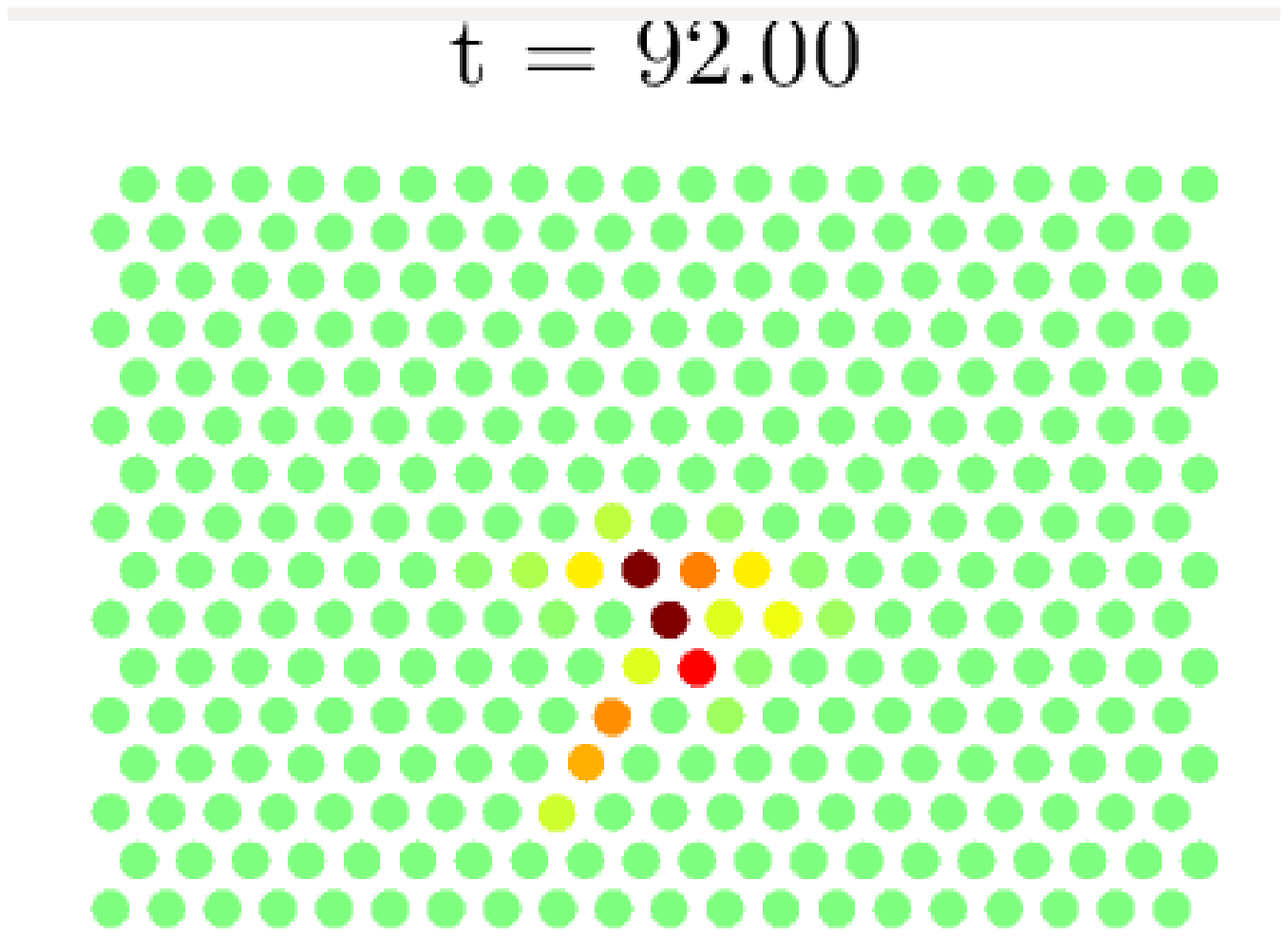}}
{\includegraphics[width=0.32\textwidth]{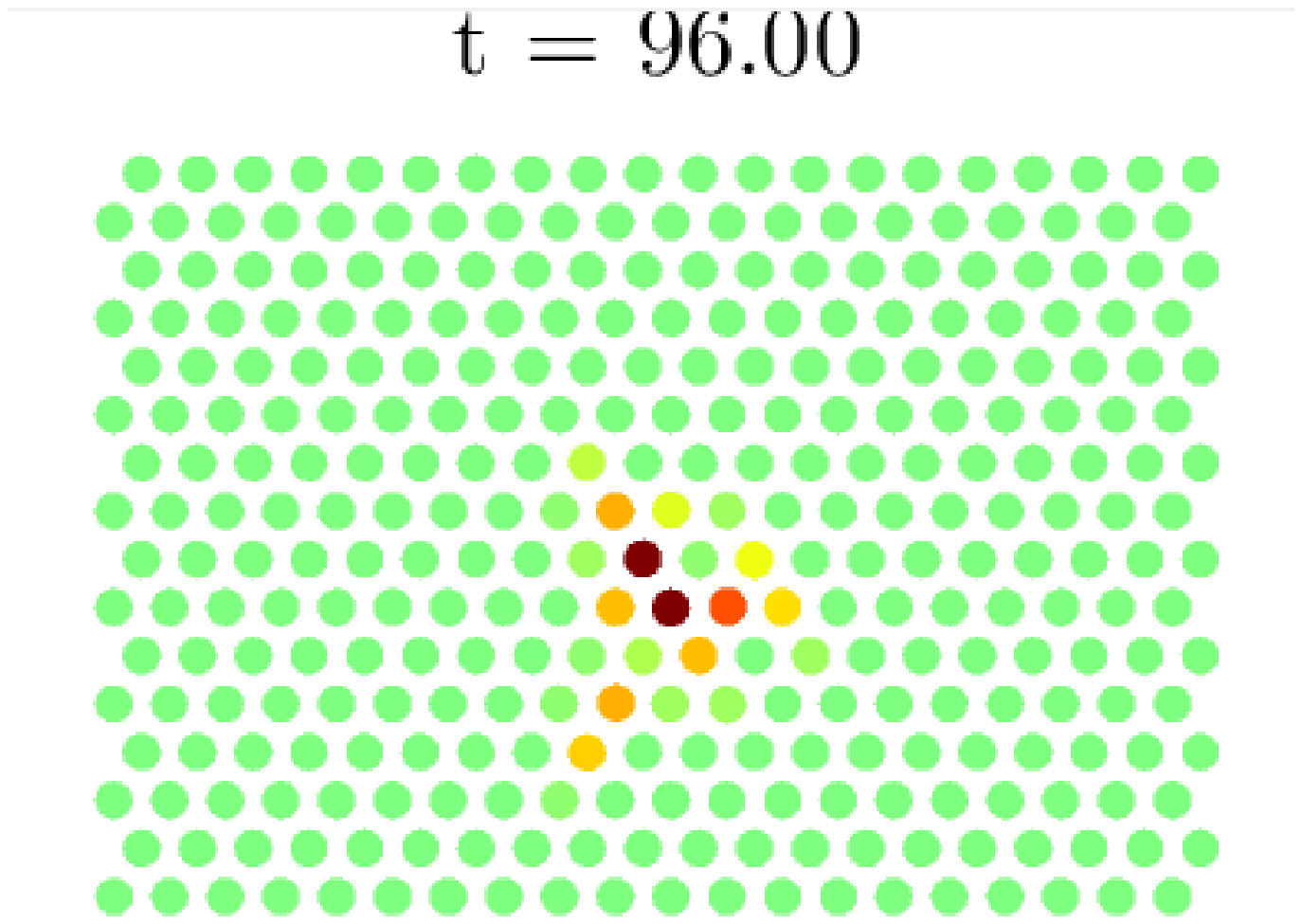}}
{\includegraphics[width=0.32\textwidth]{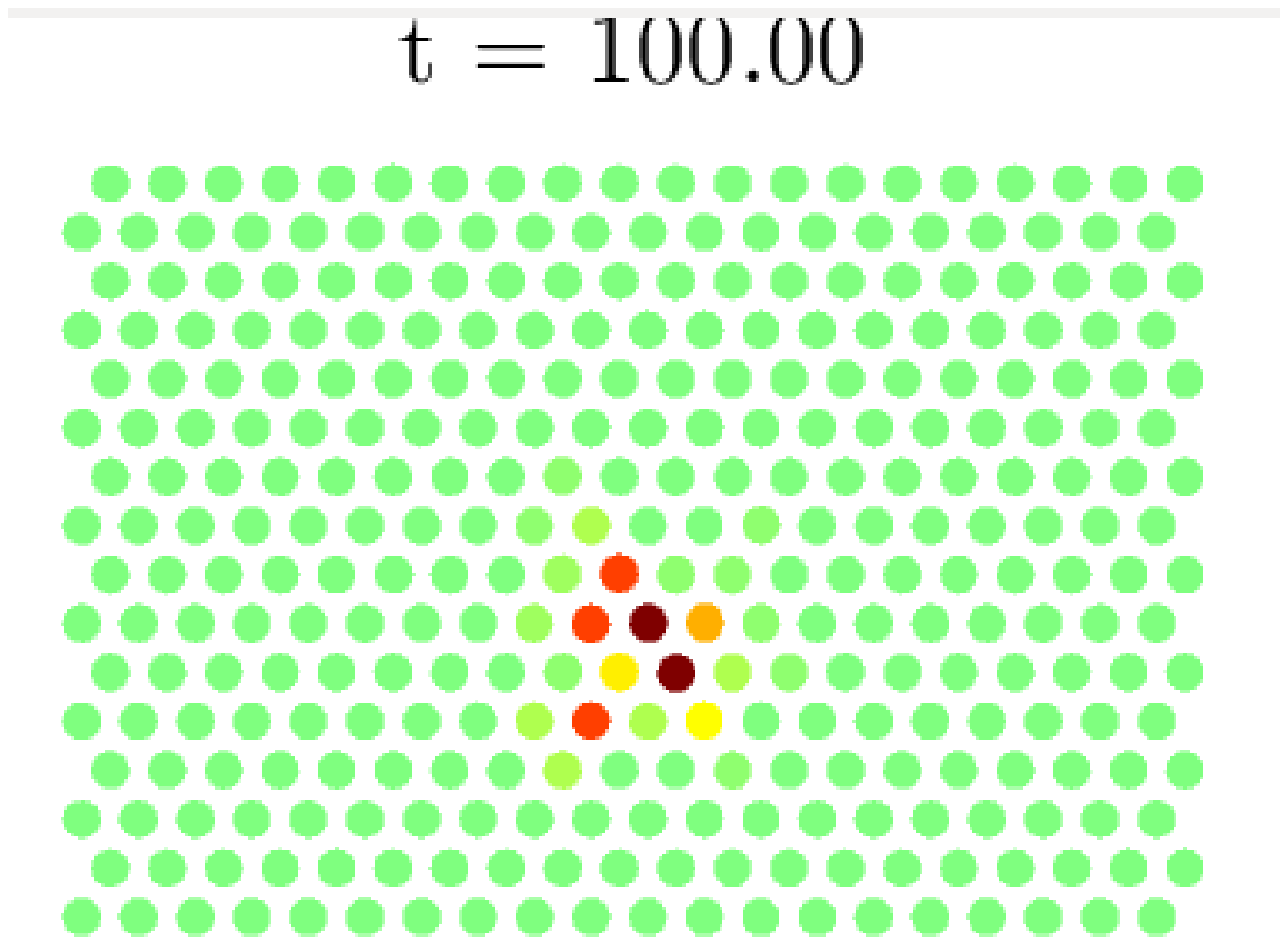}}
{\includegraphics[width=0.32\textwidth]{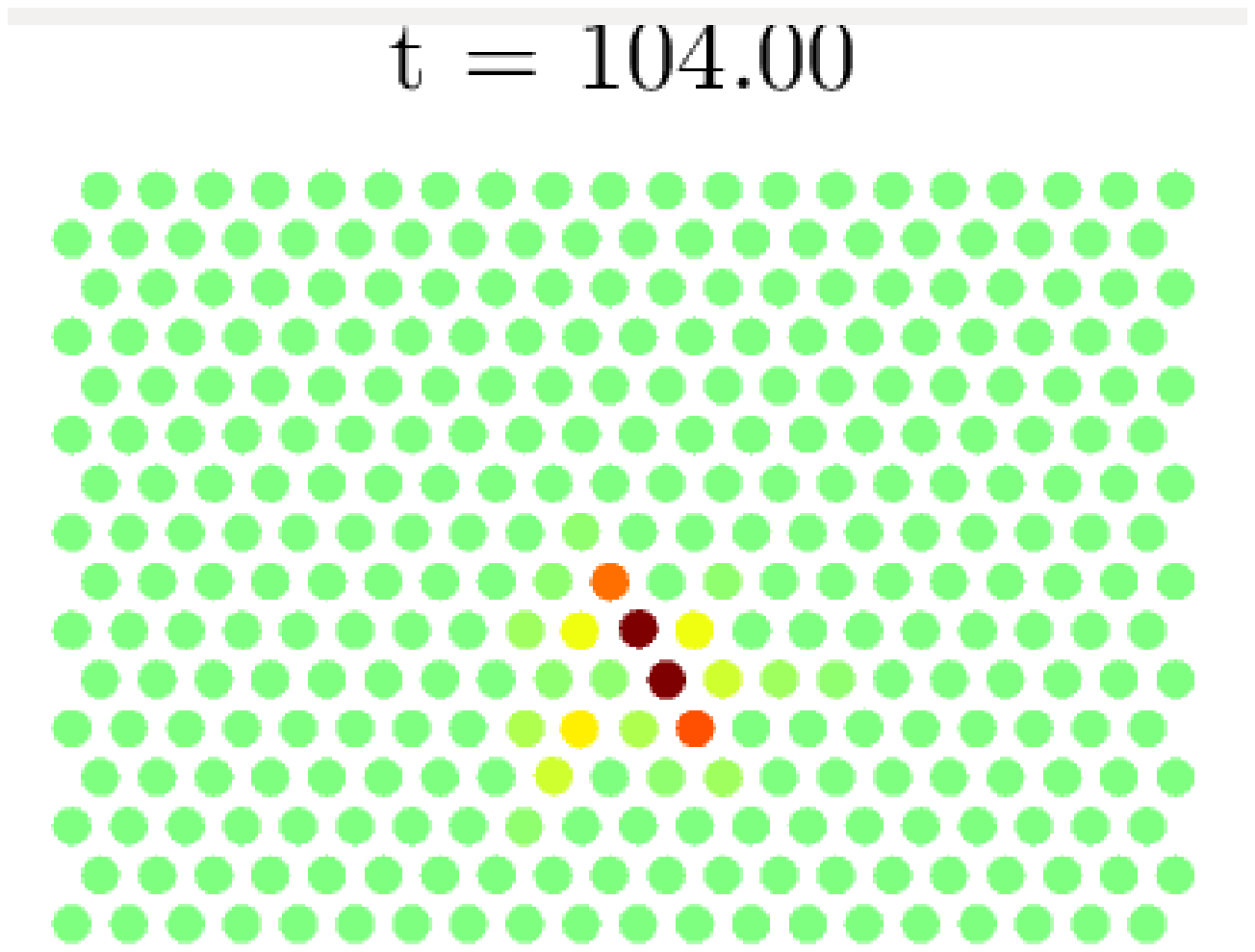}}
{\includegraphics[width=0.32\textwidth]{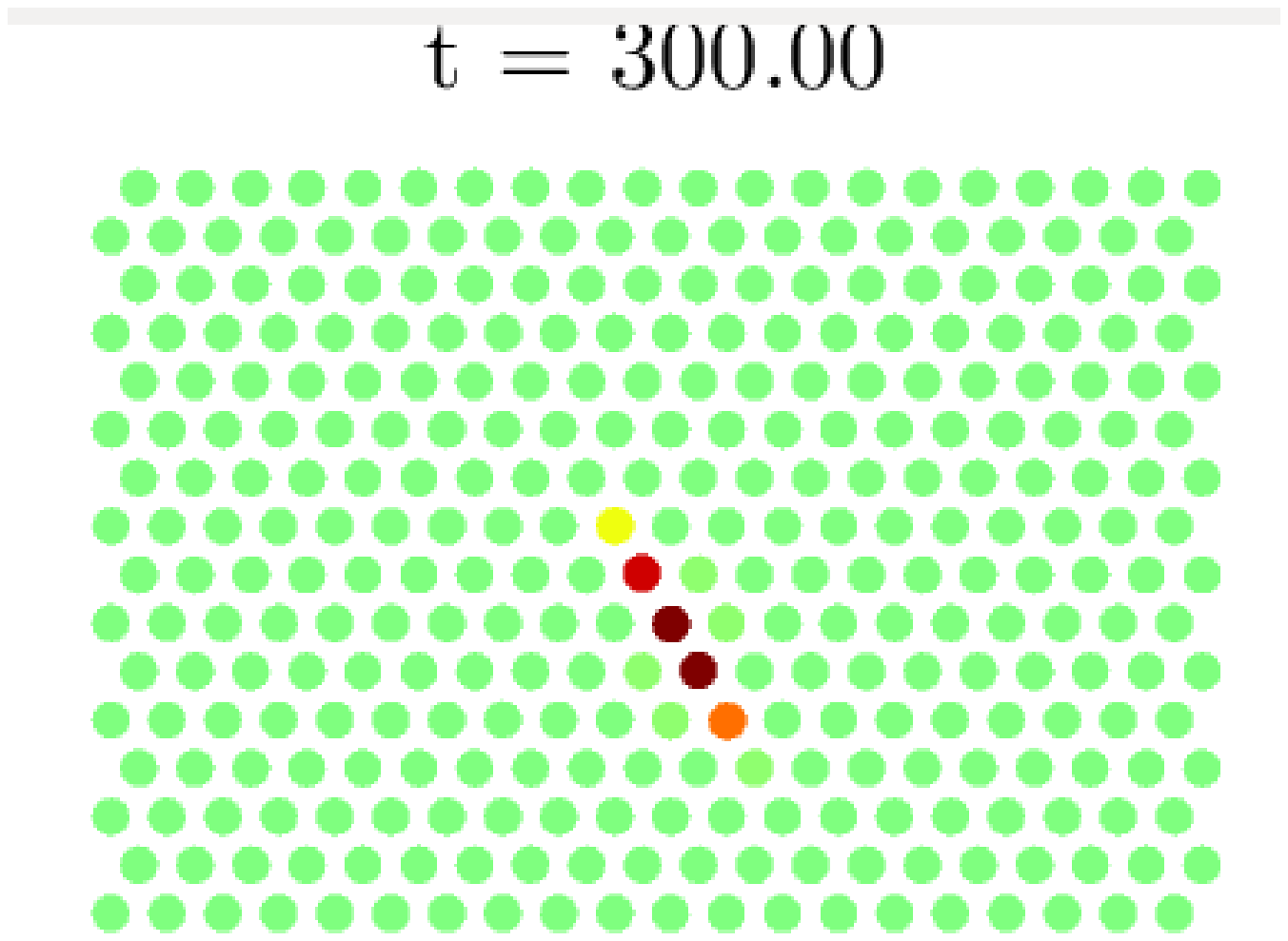}}
{\includegraphics[width=0.32\textwidth]{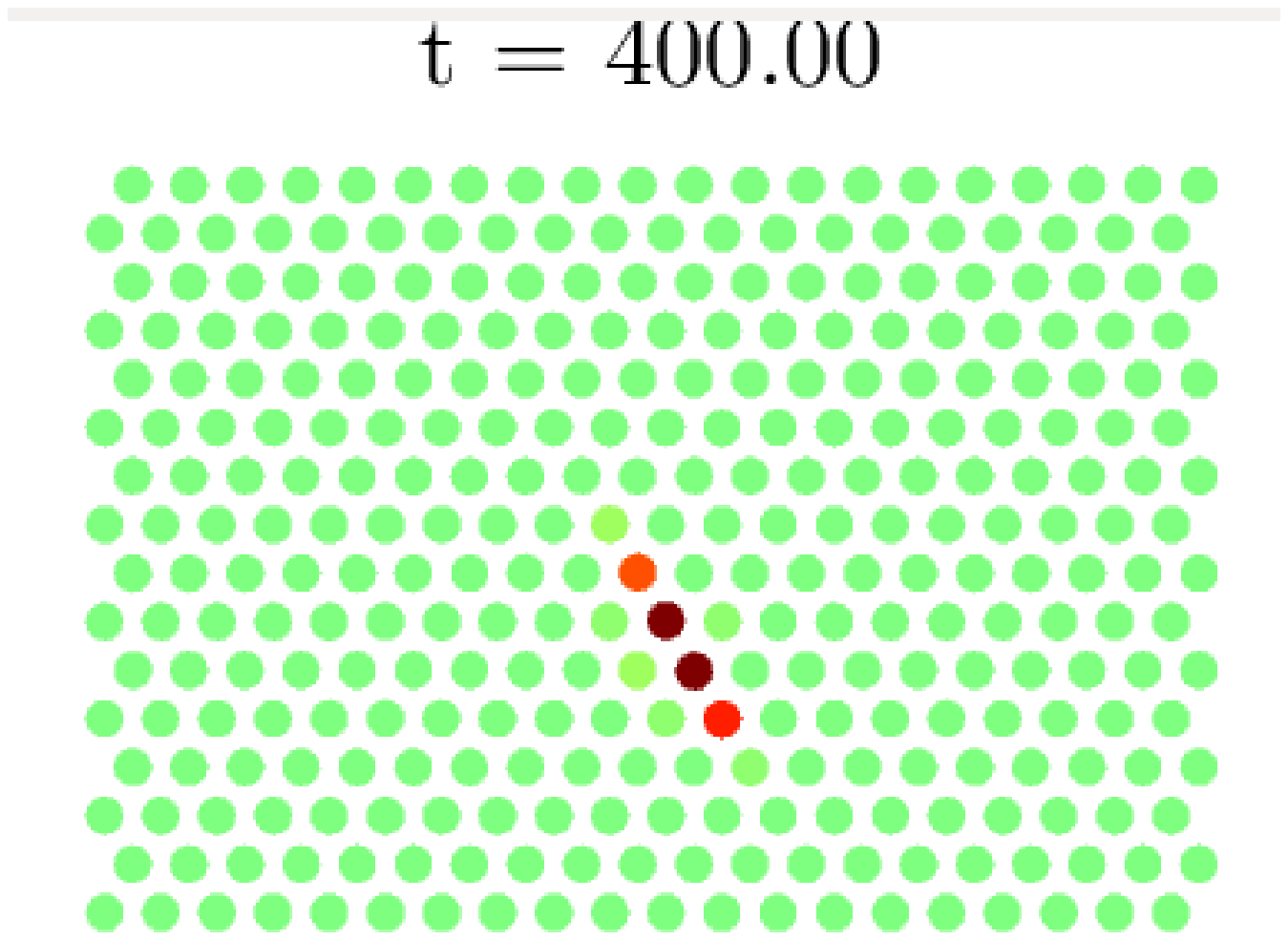}}
\caption{Snapshots of scatter plots of atoms in time of two breather
  collision at $60\,^{\circ}$ angle to each other. $N_{x}=200$,
  $N_{y}=100$, $T_{end}=400$, $U_{0}=2$, $u_{x}^{0}=1$, $u_{y}^{0}=0$,
  $u_{x}^{1}=-2.5\cos(\pi/3)$ and
  $u_{y}^{0}=-2.5\sin(\pi/3)$.}\label{fig:2DBBScatt}
\end{figure}

In the final example of this section we consider a breather-breather
collision at a $60\,^{\circ}$ angles to each other. In this example we
give a kick to one atom in the left lower area and a kick to one atom
in the right upper area of the lattice: $N_{x}=200$ and
$N_{y}=100$. The initial kick values are $u_{x}^{0}=1$ and
$u_{y}^{0}=0$, and $u_{x}^{1}=-2.5\cos(\pi/3)$ and
$u_{y}^{0}=-2.5\sin(\pi/3)$. We carry out integration in time until
$T_{end}=400$. We illustrate the collision area in time with snapshot
scatter plots of atoms in Fig \ref{fig:2DBBScatt}.  Darker colours
indicate higher energy density function values. The first breather
propagates from left to right on the horizontal lattice chain and the
second breather propagates downwards on the $(1/2,\sqrt{3}/2)^T$
crystallographic lattice chain.  During the collision both breathers
merge into one stationary breather localized on the
$(1/2,-\sqrt{3}/2)^T$ crystallographic lattice chain. Depending on the
breather's energies, velocity and phase, we have observed breathers
merging into one stationary or one propagating breather, passing
through each other or changing their propagation directions.

The results presented above can be summarized by one
consideration. The additional degree of freedom introduces three
crystallographic lattice directions, in contrast to 1D models on which
localized modes can travel, thus introducing additional richness into
interaction properties.  Due to the quasi-one-dimensional nature of
travelling modes, lateral displacements of atoms on the main chain
induced through interactions may destabilize propagating modes. At the
same time, the additional degree of freedom allows us to observe new
wave phenomena such as horseshoe wave solutions from
Sec. \ref{sec:FrontSol} and the 2D multi-kink solutions of the
following section.

\subsubsection{Numerical results: 2D multi-kink solution}
\label{sec:MultiKink}
\begin{figure} 
\centering 
{\includegraphics[width=0.48\textwidth]{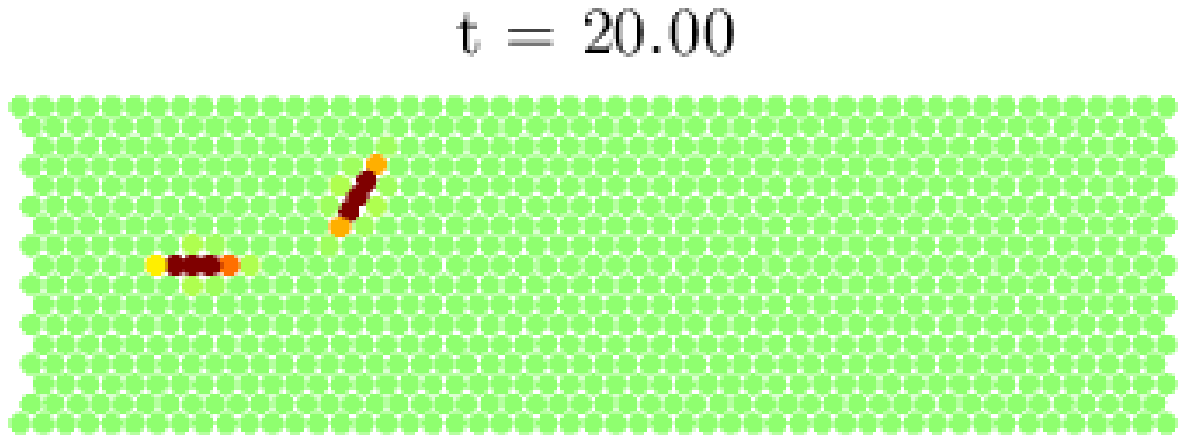}}
{\includegraphics[width=0.48\textwidth]{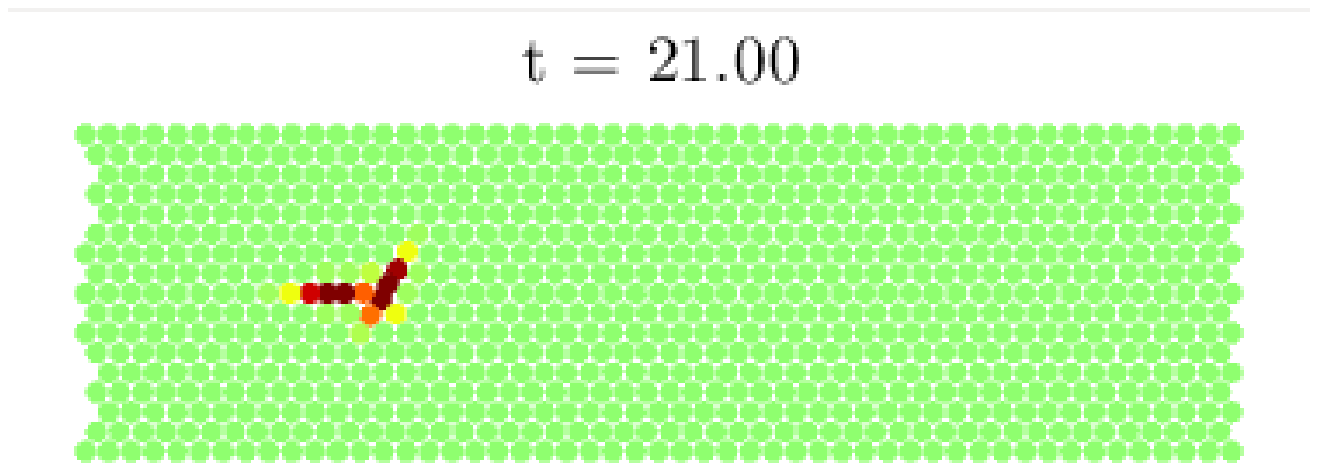}}
{\includegraphics[width=0.48\textwidth]{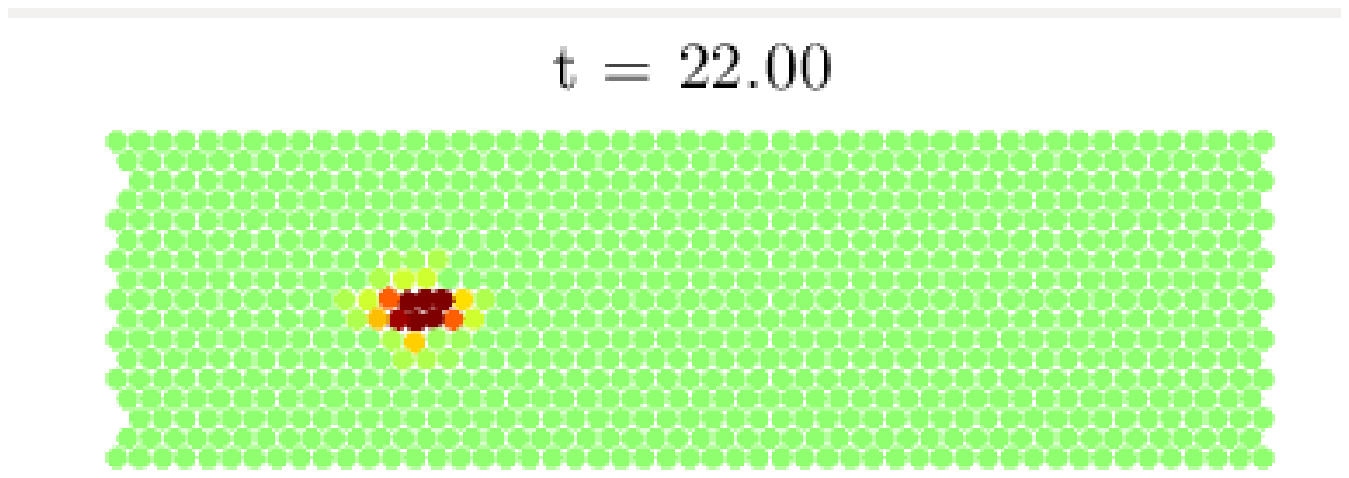}}
{\includegraphics[width=0.48\textwidth]{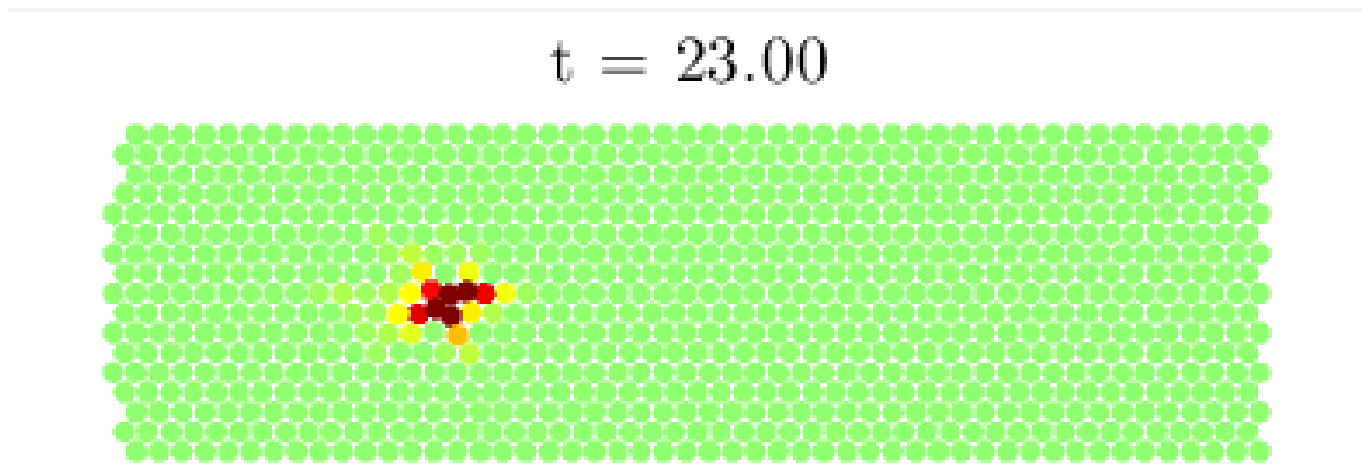}}
{\includegraphics[width=0.48\textwidth]{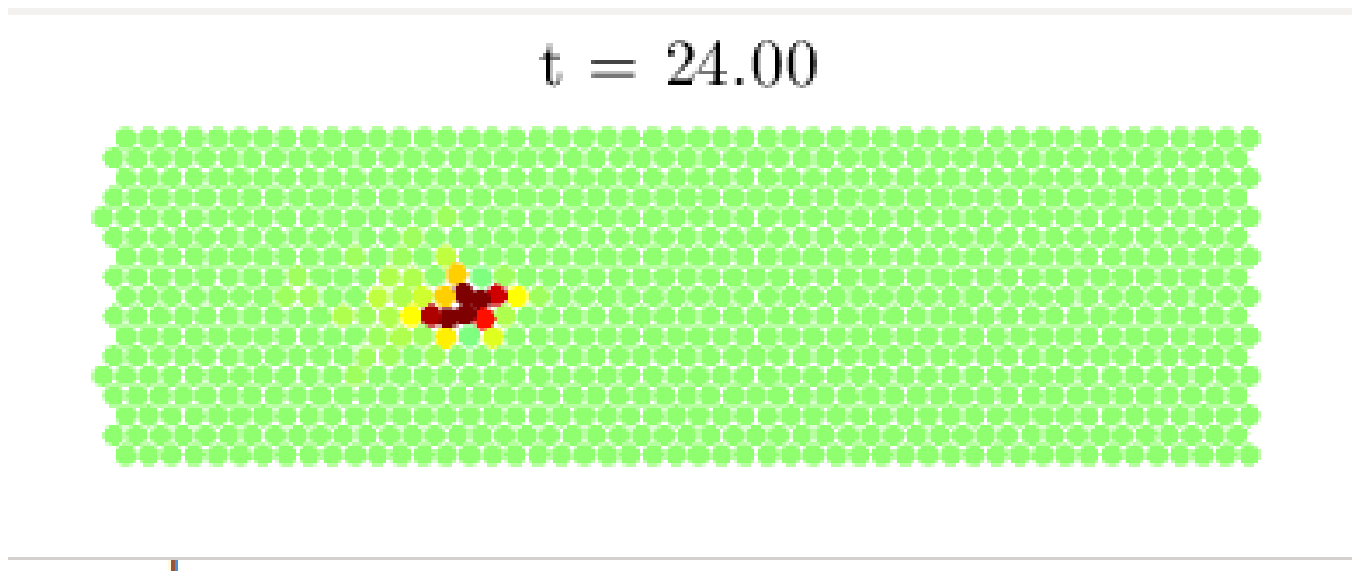}}
{\includegraphics[width=0.48\textwidth]{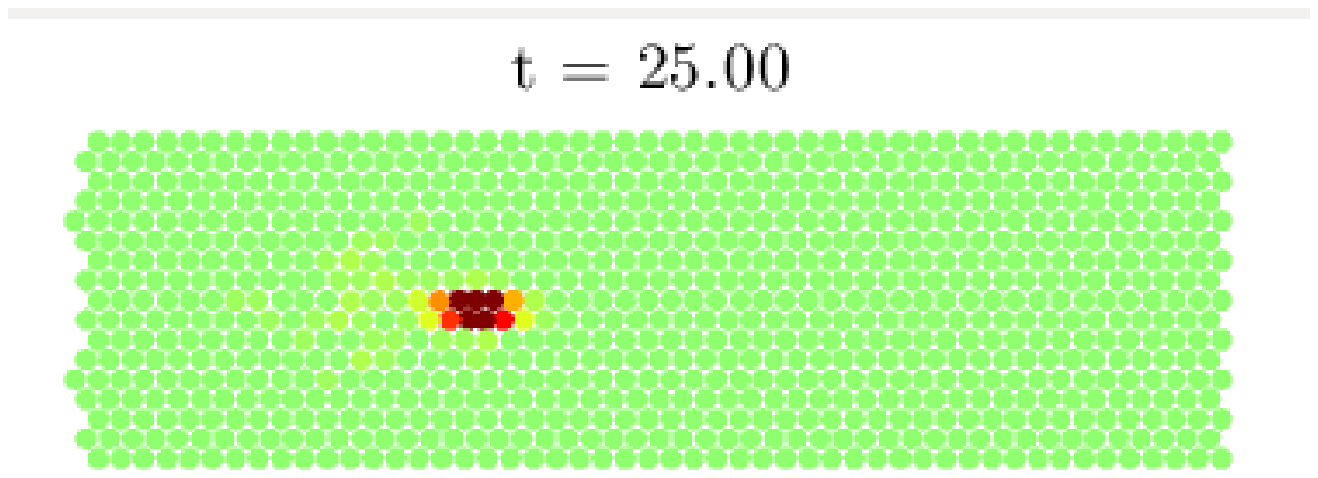}}
{\includegraphics[width=0.48\textwidth]{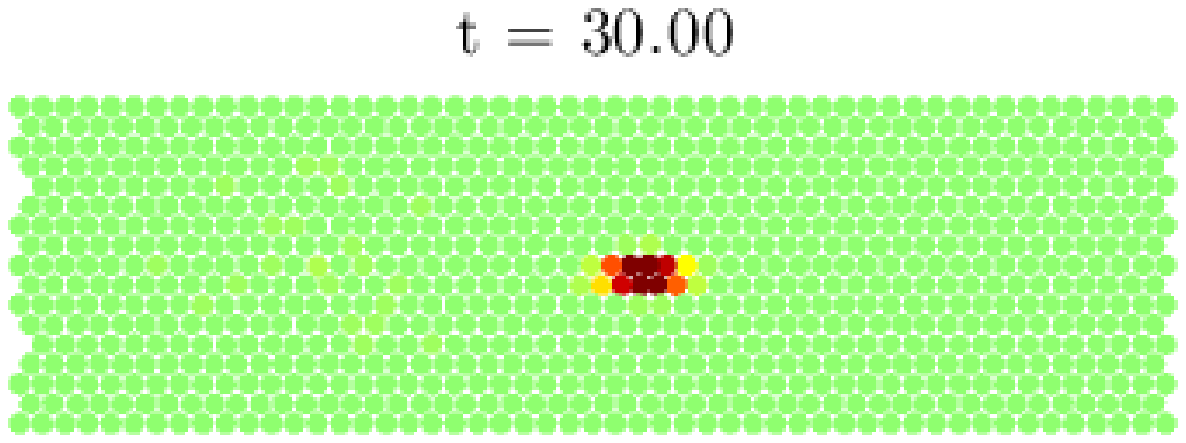}}
{\includegraphics[width=0.48\textwidth]{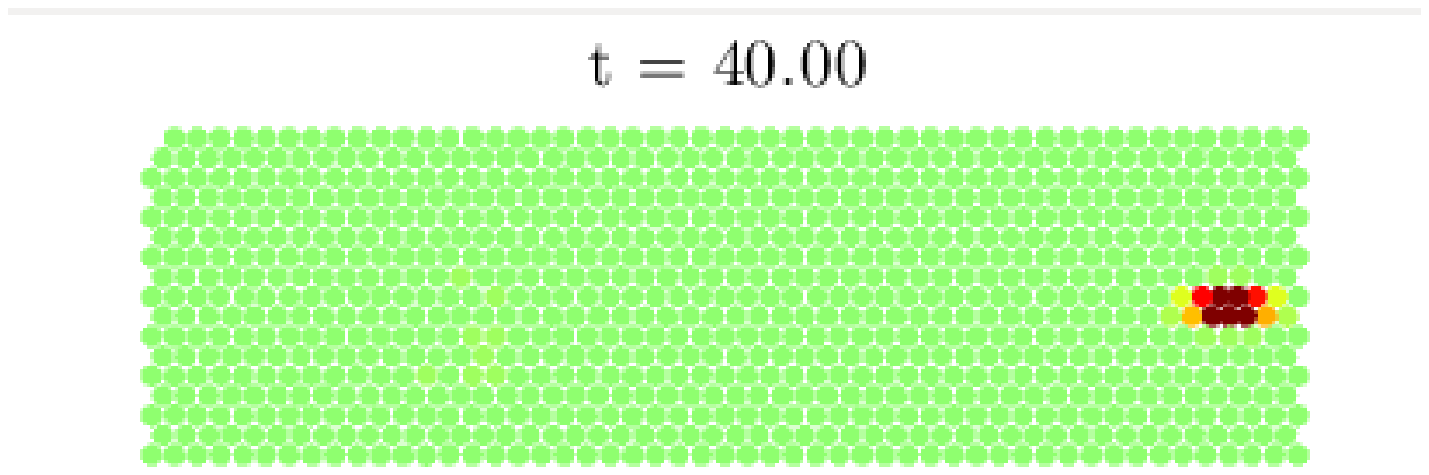}}
\caption{Snapshots of scatter plots of atoms in time of two kink
  collision at $60\,^{\circ}$ angle to each other. $N_{x}=200$,
  $N_{y}=100$, $T_{end}=40$, $U_{0}=2$, $u_{x}^{0}=5.5$,
  $u_{y}^{0}=0$, $u_{x}^{1}=-5.25\cos(\pi/3)$ and
  $u_{y}^{0}=-5.25\sin(\pi/3)$.}\label{fig:2DMKcoll}
\end{figure}

In this section we present a brief example of a 2D coupled-kink
solution.  This is a multiple kink-like mode where two or more kinks
travel together side-by-side with the front perpendicular to the
direction of travel.  The initial formation of such a solution was
observed from a kink-kink collision experiment at $60\,^{\circ}$ angle
to each other which we demonstrate here.  Consider the experiment of
breather-breather collision at $60\,^{\circ}$ angle to each other from
Sec. \ref{sec:2Deffects} but with initial kick values $u_{x}^{0}=5.5$,
$u_{y}^{0}=0$, $u_{x}^{1}=-5.25\cos(\pi/3)$ and
$u_{y}^{0}=-5.25\sin(\pi/3)$.  These particular initial kick values
produce two kink solutions.  The first kink propagates from left to
right on a horizontal lattice chain in $(1,0)^T$ crystallographic
lattice direction and the second kink propagates downwards on the
$(1/2,\sqrt{3}/2)^T$ crystallographic lattice chain, see
Fig. \ref{fig:2DMKcoll}. During the collision both kinks merge
together and form a stable double-kink solution propagating to the
right on two adjacent chains of atoms.
  
\begin{figure} 
\centering 
{\includegraphics[width=0.32\textwidth]{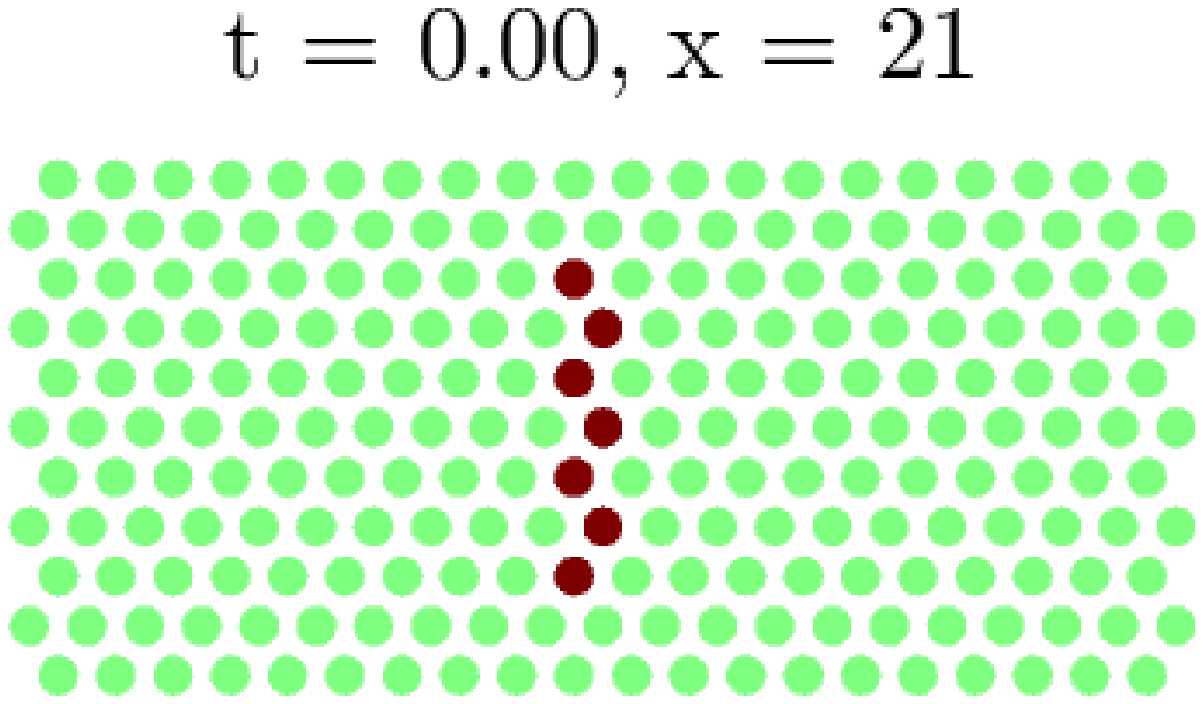}}
{\includegraphics[width=0.32\textwidth]{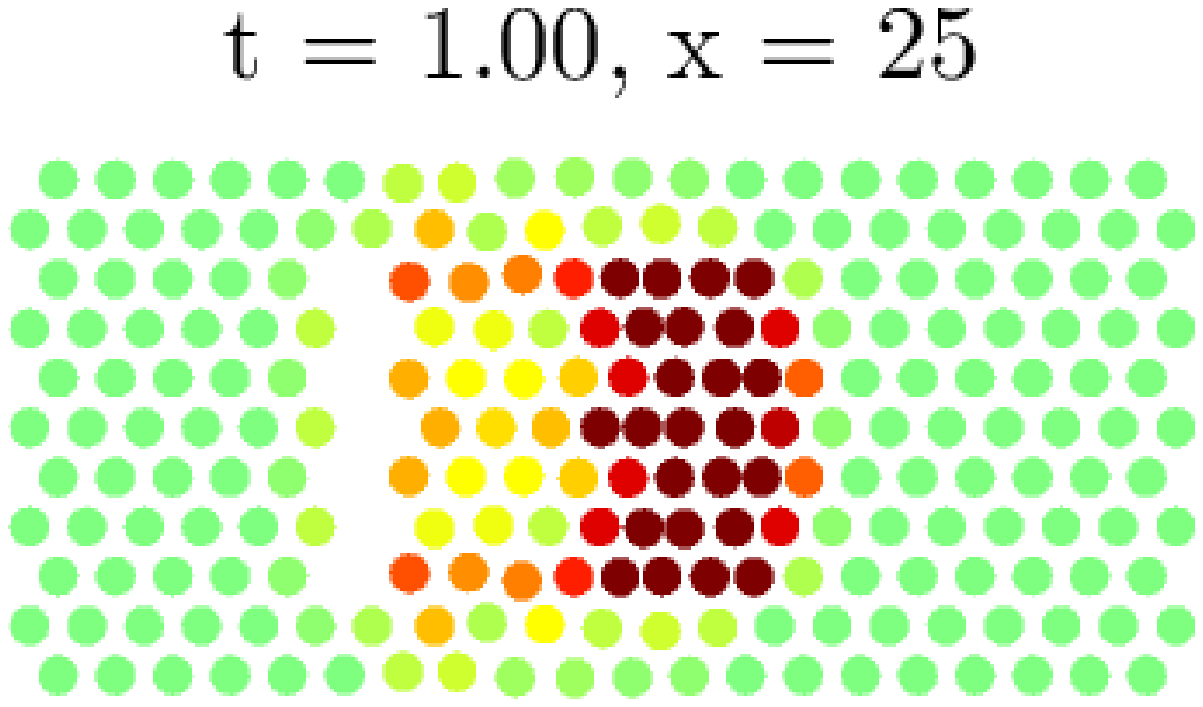}}
{\includegraphics[width=0.32\textwidth]{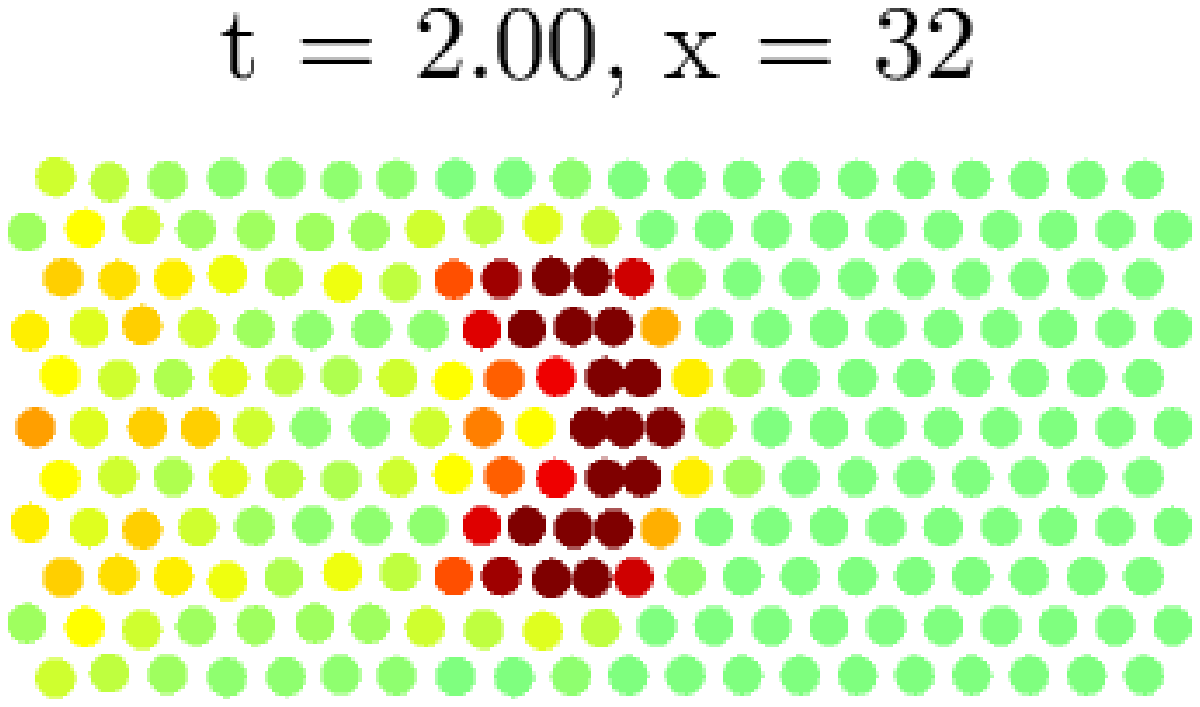}}
{\includegraphics[width=0.32\textwidth]{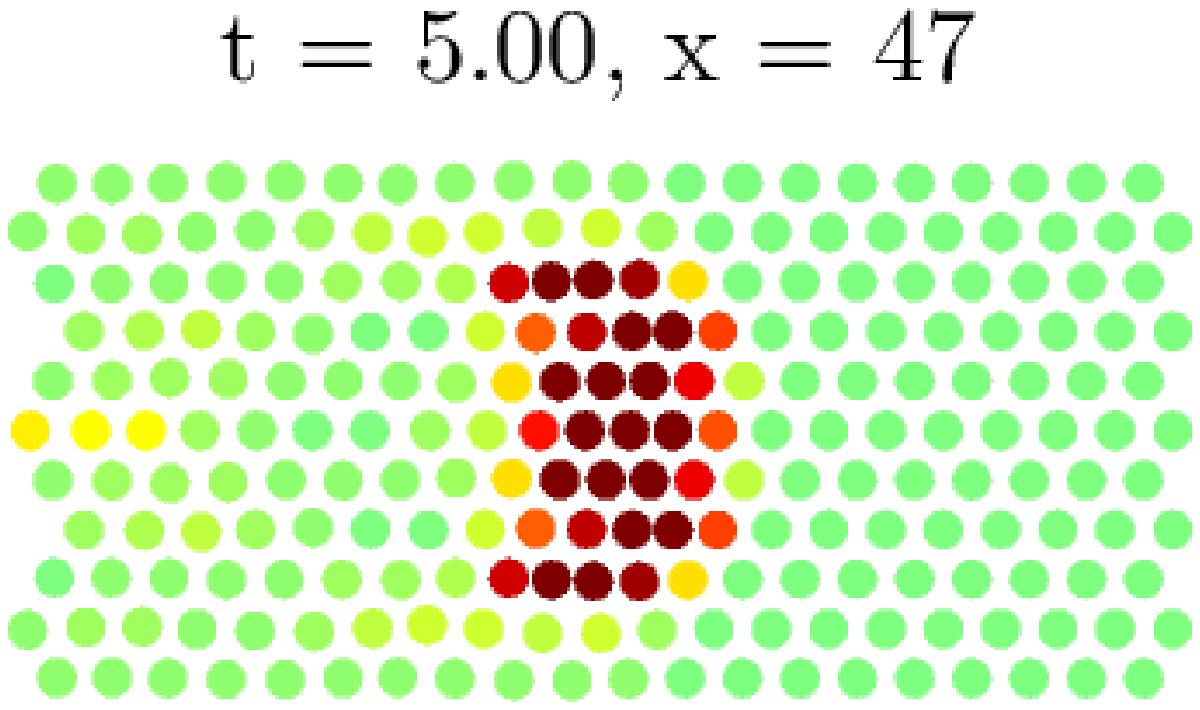}}
{\includegraphics[width=0.32\textwidth]{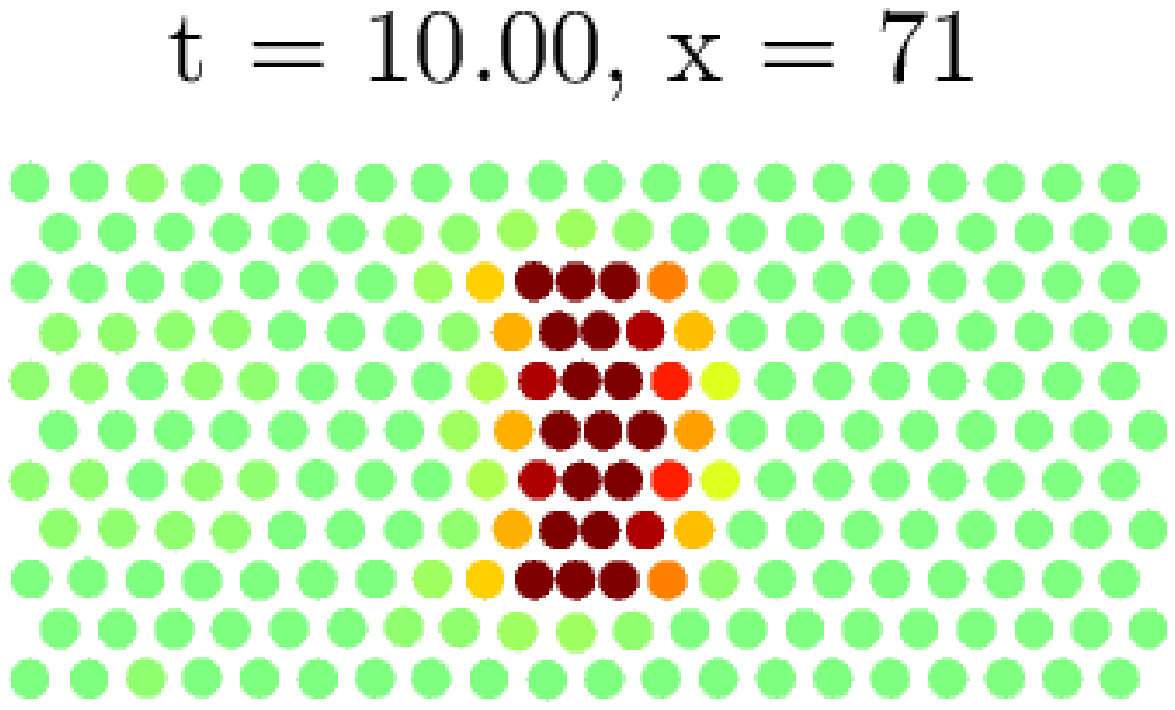}}
{\includegraphics[width=0.32\textwidth]{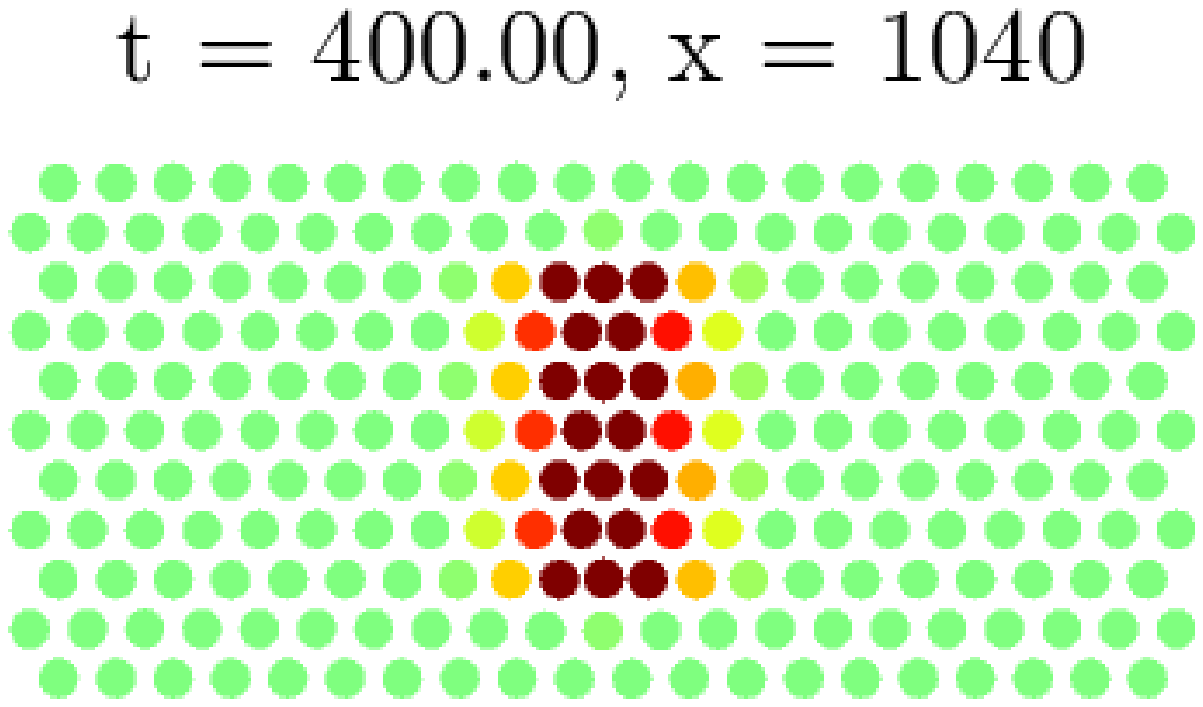}}
\caption{Snapshots of scatter plots of atoms in time of a multi-kink
  solution. $N_{x}=1200$, $N_{y}=40$, $T_{end}=400$, $U_{0}=2$,
  $u_{x,i}^{0}=5.5$ and $u_{y,i}^{0}=0$, where
  $i=1,\dots,{7}$.}\label{fig:MKsol}
\end{figure}

The observation of the stable formation of a double-kink solution,
Fig. \ref{fig:2DMKcoll}, led us to consider coupled multi-kink
simulations, that is, by considering multiple kicks of neighbouring
atoms in the $y$ axis direction.  For this experiment we consider a
lattice: $N_{x}=1200$ and $N_{y}=40$, and equal initial kick values
$u_{x,i}^{0}=5.5$, $u_{y,i}^{0}=0$ on seven atoms,
i.e.~$i=1,\dots,{7}$, see the top left plot of Fig. \ref{fig:MKsol} at
$t=0$.  Importantly, non-equal initial kick values may lead to
scattering of kinks in all three crystallographic lattice directions.
We integrate in time until $T_{end}=400$.  In Figure \ref{fig:MKsol},
we show snapshots of scatter plots of atoms in time at locations of
maximal energy density function in space indicated by the $x$
coordinate.  Numerical results show that the structure of multiple kink
solutions has propagated more than $1000$ lattice sites and suggest
that such type of structures may be long-lived in idealized
settings. Interestingly, the same type of initial kick values did not
lead to the formation of joint breather solutions.

\section{Conclusions and Future Plans}
We have confirmed and much extended the calculations of Mar\'{\i}n et
al.\ showing the existence of long lived quasi-one-dimensional
discrete breathers in hexagonal lattices.  A further paper using a
more conventional particle-particle potential will discuss such
solutions in more detail \cite{bel14}.  The present model also
displays long-lived quasi-one-dimensional discrete kinks in our model
mica lattice.  However as discussed in \cite{bel14}, this type of
solution is more sensitive to the details of the inter-atomic
potentials considered, and other models give much shorter kink
lifetimes.  It remains to be seen if existing or novel materials can
exhibit such kinks in physical situations.
 
We show that the kinks and breathers exhibit a typical rich variety of
phenomena on collision along a mutual line of quasi-one-dimensional
travel.  In addition we demonstrate fully 2D collision phenomena for
the first time, for kinks/breathers travelling on adjacent lines or at
$60^{\circ}$ angles to each other.  Moreover we observe a new type of
spreading shock wave, the horseshoe wave, with a breather profile.  In
view of the many different possible outcomes of such collisions, a
more systematic and quantitative study is required for the future.

We have not discussed thermal or other random perturbations to the
model in the present paper, some brief studies will be reported in
\cite{bel14}.  In at attempt to understand ejection and sputtering in
such models, it would be important to model surface forces properly.
In general a more serious attempt to fit model parameters to real MD
data from mica is required.  A further study should concentrate on the
effects of longer-range forces and how these effect breather and kink
lifetimes.

A multi-core and HPC version of the code will be an important next
step, as this is necessary for long runs, to establish the maximum
lifetimes of breathers and kinks under ideal conditions.  In the full
2D model it would be interesting to investigate scattering of
breathers and kinks with vacancies, dislocations and inclusions, etc.,
to generalise the 1D studies such as \cite{ckaer03}.  Adding
temperature effects to this sort of study would be important.

The present study shows a variety of new interesting phenomena, but
the field of 2D breathers and kinks is still in its infancy, with much
still to be done, in both theoretical and experimental areas. 

\section*{Acknowledgements} 
JB and BJL acknowledge the support of the Engineering and Physical
Sciences Research Council which has funded this work as part of the
Numerical Algorithms and Intelligent Software Centre under Grant
EP/G036136/1.


\begin{thebibliography}{10}

\bibitem{alti89}
M.~P. Allen and D.~J. Tildesley.
\newblock {\em {Computer Simulation of Liquids}}.
\newblock Oxford science publications. Oxford University Press, USA, 1989.

\bibitem{bel14}
Janis Bajars, J.~Chris Eilbeck, and Ben Leimkuhler.
\newblock Nonlinear propagating localized modes in a 2d hexagonal crystal
  lattice.
\newblock preprint, 2014.

\bibitem{ckaer03}
J.~Cuevas, C.~Katerji, J.~F.~R. Archilla, J.~C. Eilbeck, and F.~M. Russell.
\newblock Influence of moving breathers on vacancies migration.
\newblock {\em Phys. Lett. A}, 315:364--371, 2003.

\bibitem{dcer11}
Q.~Dou, J.~Cuevas, J.~C. Eilbeck, and F.~M. Russell.
\newblock Breathers and kinks in a simulated crystal experiment.
\newblock {\em Discrete and Continuous Dynamical Systems - Series S},
  4:1107--1118, 2011.

\bibitem{defw93}
D.~B. Duncan, J.~C. Eilbeck, H.~Feddersen, and J.~A.~D. Wattis.
\newblock Solitons on lattices.
\newblock {\em Physica D}, 68:1--11, 1993.

\bibitem{ei86}
J.~C. Eilbeck.
\newblock Numerical simulations of the dynamics of polypeptide chains and
  proteins.
\newblock In C~Kawabata and A.~R. Bishop, editors, {\em Computer Analysis for
  Life Science}, pages 12--21, Tokyo, 1986. Ohmsha.

\bibitem{els84}
J.~C. Eilbeck, P.~S. Lomdahl, and A.~C. Scott.
\newblock Soliton structure in crystalline acetanilide.
\newblock {\em Phys. Rev. B}, 30:4703--4712, 1984.

\bibitem{els85}
J.~C. Eilbeck, P.~S. Lomdahl, and A.~C. Scott.
\newblock The discrete self-trapping equation.
\newblock {\em Physica D}, 16:318--338, 1985.

\bibitem{fe91}
H.~Feddersen.
\newblock {\em Solitary wave solutions to the discrete nonlinear Schr\"odinger
  equation}, pages 159--167.
\newblock Springer, 1991.

\bibitem{fw98}
S.~Flach and C.R. Willis.
\newblock Discrete breathers.
\newblock {\em Physics Reports}, 295:181--264, 1998.

\bibitem{fl12}
Sergej Flach.
\newblock Discrete breathers in a nutshell.
\newblock {\em Nonlinear Theory and Its Applications, IEICE}, 3:12--26, 2012.

\bibitem{GeGr13}
A.~K. Geim and I.~V. Grigorieva.
\newblock Van der {W}aals heterostructures.
\newblock {\em Nature}, 499:419--425, 2013.

\bibitem{lr04}
B.~Leimkuhler and S.~Reich.
\newblock {\em Simulating Hamiltonian Dynamics}.
\newblock Cambridge University Press, 2004.

\bibitem{ma94}
R.S. MacKay and S.~Aubry.
\newblock Proof of existence of breathers for time-reversible or hamiltonian
  networks of weakly coupled oscillators.
\newblock {\em Nonlinearity}, 7:1623--1643, 1994.

\bibitem{ms04}
N.~Manton and P.~Sutcliffe.
\newblock {\em Topological Solitons}.
\newblock Cambridge Universty Press, 2004.

\bibitem{mer98}
J.~L. Mar\'{\i}n, J.~C. Eilbeck, and F.~M. Russell.
\newblock Localised moving breathers in a 2-d hexagonal lattice.
\newblock {\em Phys. Lett. A}, 248:225--229, 1998.

\bibitem{mer00}
J.~L. Mar\'{\i}n, J.~C. Eilbeck, and F.~M. Russell.
\newblock {\em 2-D Breathers and applications}, pages 293--306.
\newblock Springer, 2000.

\bibitem{mre01}
J.~L. Mar\'{\i}n, F.~M. Russell, and J.~C. Eilbeck.
\newblock Breathers in cuprate superconductor lattices.
\newblock {\em Phys. Letts. A}, 281:225--229, 2001.

\bibitem{ov70}
A.~A. Ovchinnikov.
\newblock Localized long-lived vibrational states in molecular crystals.
\newblock {\em Soviet Physics JETP}, 30:147--150, 1970.

\bibitem{pk84}
Michel Peyrard and Martin~D. Kruskal.
\newblock Kink dynamics in the highly discrete sine-gordon system.
\newblock {\em Physica D: Nonlinear Phenomena}, 14:88--–102, 1984.

\bibitem{rc95}
F.~M. Russell and D.~R. Collins.
\newblock Lattice-solitons and non-linear phenomena in track formation.
\newblock {\em Rad. Meas.}, 25:67--70, 1995.

\bibitem{re07}
F.~M. Russell and J.~C. Eilbeck.
\newblock Evidence for moving breathers in a layered crystal insulator at 300k.
\newblock {\em Phys. Letts. A}, 78:10004, 2007.

\bibitem{sm83}
A.~C. Scott and L.~MacNeil.
\newblock Binding energy versus nonlinearity for a small stationary soliton.
\newblock {\em Phys. Lett. A}, 98:87--88, 1983.

\bibitem{yaduyachli11}
Y.~Yang, W.~S. Duan, L.~Yang, J.~M. Chen, and M.~M. Lin.
\newblock Rectification and phase locking in overdamped two-dimensional
  {F}renkel-{K}ontorova model.
\newblock {\em Europhysics Letters}, 93(1):16001, 2011.

\end{thebibliography}

\end{document}